%% file: main.tex
\documentclass[trackchanges,preprint,astrosymb]{aastex7}

\begin{document}

\title{CHARA Array Observations of the Evolved Components in Symbiotic Star Systems}

\correspondingauthor{Ryan Norris}
\email{ryan.norris@nmt.edu}

\author[0009-0000-6957-8466]{Thomas Martin Gaudin}
\affiliation{ Department of Astronomy and Astrophysics, The Pennsylvania State University, University Park, PA 16802, USA}
\email{tmg6006@psu.edu} 

\author[0000-0002-9120-9728]{Ryan Norris}
\affiliation{New Mexico Institute of Mining and Technology, Workman Center, 801 Leroy Place, Socorro, NM 87801, USA}
\email{ryan.norris@nmt.edu}

\author[0000-0003-1564-7029]{Magdalena Otulakowska-Hypka}
\affiliation{Astronomical Observatory Institute, Faculty of Physics and Astronomy, Adam Mickiewicz University, S\l{}oneczna 36, PL-60286 Poznań, Poland}
\email{magdaot@amu.edu.pl}

\author[0000-0002-9288-3482]{Rachael M.\ Roettenbacher}
\affiliation{Department of Astronomy, University of Michigan, Ann Arbor, MI 48109, USA}
\email{rmroett@umich.edu}

\author{Nirupam Roy}
\affiliation{New Mexico Institute of Mining and Technology, Workman Center, 801 Leroy Place, Socorro, NM 87801, USA}
\email{nirupam.roy@student.nmt.edu}

\author{Yesenia Beltran}
\affiliation{New Mexico Institute of Mining and Technology, Workman Center, 801 Leroy Place, Socorro, NM 87801, USA}
\email{yesenia.beltran@student.nmt.edu}

\author{Cameron Caruso}
\affiliation{New Mexico Institute of Mining and Technology, Workman Center, 801 Leroy Place, Socorro, NM 87801, USA}
\email{cameron.caruso@student.nmt.edu}

\author{Cody Gustafson}
\affiliation{New Mexico Institute of Mining and Technology, Workman Center, 801 Leroy Place, Socorro, NM 87801, USA}
\email{Cody.Gustafson@student.nmt.edu}

\author{Mason Earick}
\affiliation{New Mexico Institute of Mining and Technology, Workman Center, 801 Leroy Place, Socorro, NM 87801, USA}
\email{}

\author{Andrew Kotowski}
\affiliation{New Mexico Institute of Mining and Technology, Workman Center, 801 Leroy Place, Socorro, NM 87801, USA}
\email{}

\author{Rebecca Proni}
\affiliation{New Mexico Institute of Mining and Technology, Workman Center, 801 Leroy Place, Socorro, NM 87801, USA}
\email{}

\author{Jacob Sandusky}
\affiliation{New Mexico Institute of Mining and Technology, Workman Center, 801 Leroy Place, Socorro, NM 87801, USA}
\email{}

\author[0000-0002-8376-8941]{Fabien Baron}
\affiliation{Center for High Angular Resolution Astronomy and Department 
of Physics and Astronomy, Georgia State University, P.O. Box 5060, Atlanta,
GA 30302-5060, USA}
\email{}

\author[0000-0002-8349-9366]{Michelle J. Creech-Eakman}
\affiliation{New Mexico Institute of Mining and Technology, Workman Center, 801 Leroy Place, Socorro, NM 87801, USA}
\email{}

\author[0000-0002-3380-3307]{John D. Monnier}
\affiliation{Department of Astronomy, University of Michigan, Ann Arbor, MI 48109, USA}
\email{monnier@umich.edu}

\author[0000-0001-6017-8773]{Stefan Kraus}
\affiliation{Astrophysics Group, Department of Physics \& Astronomy, University of Exeter, Stocker Road, Exeter, EX4 4QL, UK}
\email{S.Kraus@exeter.ac.uk}

\author[0000-0002-2208-6541]{Narsireddy Anugu}
\affiliation{The CHARA Array of Georgia State University, Mount Wilson Observatory, Mount Wilson, CA 91023, USA}
\email{}

\author[0000-0002-0493-4674]{Jean-Baptiste Le Bouquin}
\affiliation{Institut de Plan\'etologie et d'Astrophysique de Grenoble, 38058 Grenoble, France}
\email{jean-baptiste.lebouquin@obs.ujf-grenoble.fr}

\author[0000-0001-8926-9732]{Sorabh Chhabra}
\affiliation{Astrophysics Group, Department of Physics \& Astronomy, University of Exeter, Stocker Road, Exeter, EX4 4QL, UK}
\email{S.Chhabra@exeter.ac.uk}

\author[0009-0005-8088-0718]{Isabelle Codron}
\affiliation{Astrophysics Group, Department of Physics \& Astronomy, University of Exeter, Stocker Road, Exeter, EX4 4QL, UK}
\email{ic302@exeter.ac.uk}

\author[0000-0001-9764-2357]{Claire Davies}
\affiliation{Astrophysics Group, Department of Physics \& Astronomy, University of Exeter, Stocker Road, Exeter, EX4 4QL, UK}
\email{}

\author[0000-0002-1575-4310]{Jacob Ennis}
\affiliation{Department of Astronomy, University of Michigan, Ann Arbor, MI 48109, USA}
\email{ennisj@umich.edu}

\author[0000-0002-3003-3183]{Tyler Gardner}
\affiliation{Cooperative Institute for Research in Environmental Sciences at the University of Colorado Boulder, Boulder, CO, 80309 USA}
\email{}
\author[0009-0006-0225-4444]{Mayra Gutierrez}
\affiliation{Department of Astronomy, University of Michigan, Ann Arbor, MI 48109, USA}
\email{mgutie60@ucsc.edu}

\author[0000-0002-1788-9366]{Noura Ibrahim}
\affiliation{Department of Astronomy, University of Michigan, Ann Arbor, MI 48109, USA}
\email{inoura@umich.edu}

\author[0000-0001-9745-5834]{Cyprien Lanthermann}
\affiliation{The CHARA Array of Georgia State University, Mount Wilson Observatory, Mount Wilson, CA 91023, USA}
\email{}

\author[0000-0001-5980-0246]{Benjamin R. Setterholm}
\affiliation{Max-Planck-Institut für Astronomie, Heidelberg, Germany}
\email{}
\author[0000-0001-9939-2830]{Christopher D. Farrington}
\affiliation{The CHARA Array of Georgia State University, Mount Wilson Observatory, Mount Wilson, CA 91023, USA}
\email{}
\author{Olli Majoinen}
\affil{The CHARA Array of Georgia State University, Mount Wilson Observatory, Mount Wilson, CA 91023, USA}
\email{}
\author{Norm Vargas}
\affil{The CHARA Array of Georgia State University, Mount Wilson Observatory, Mount Wilson, CA 91023, USA}
\email{}
\author[0000-0001-8537-3583]{Douglas R. Gies}
\affiliation{Center for High Angular Resolution Astronomy and Department 
of Physics and Astronomy, Georgia State University, P.O. Box 5060, Atlanta,
GA 30302-5060, USA}
\email{}
\author[0000-0002-0114-7915]{Theo ten Brummelaar}
\affiliation{The CHARA Array of Georgia State University, Mount Wilson Observatory, Mount Wilson, CA 91023, USA}
\email{}
\author[0000-0001-5415-9189]{Gail H. Schaefer}
\affiliation{The CHARA Array of Georgia State University, Mount Wilson Observatory, Mount Wilson, CA 91023, USA}
\email{}

\begin{abstract}

The nature of the mechanisms that drive mass transfer in symbiotic stars remains an area of active research in stellar astronomy. Constraining the role that both stellar winds and Roche-lobe overflow play in this process is crucial to improving our understanding of these binaries and connecting them to important transient events such as recurrent novae and Type Ia supernovae. The high-resolution capabilities of an optical interferometer can resolve the geometric structure of the red giant in symbiotic stars and help answer this question. This work presents the results of an optical interferometric study using the Center for High Angular Resolution Astronomy (CHARA) Array for the purpose of measuring the angular diameter of and imaging the cool giant in four symbiotic and related systems.  Here we report \textit{H} band observations collected with MIRC-X. Model fitting and image reconstruction are used to test for Roche-lobe-filling geometries. Near-simultaneous infrared spectroscopy taken using the NASA InfraRed Telescope Facility (IRTF) is used to determine the fundamental stellar parameters of the cool giant in each system. The parametric fits reported here favor circularly symmetric disk models over elongated geometries, while imaging suggests the presence of surface features on three of these stars. We find that the three systems  with constrained orbits have inferred time-averaged filling factors below unity.

\end{abstract}


\section{Introduction} \label{sec:intro}

Symbiotic stars are the widest type of interacting stellar binary system, consisting of a red giant (RG) star or asymptotic giant branch star (AGB) and a white dwarf (WD) or other compact object \citep[and references therein]{Munari2019,2025Galax..13...49M}.\footnote{In this paper, we will generally use the term RG or giant to refer to the donor star in these systems, although these objects could be more evolved.} These systems are characterized by orbital periods on the order of hundreds of days to years, a wide separation between component stars, and a circumstellar medium (CSM) that surrounds the system. Generally, they are grouped into two classes, S-types, which contain an RG, and D-types, which contain an AGB and are named after their dusty circumstellar envelope \citep{2019Akras}. Measurements of the temperature and luminosity of the WD, photoionization of the wind of the RG, X-ray emission, flickering, and in some cases outbursts or novae, are among the evidence pointing to interaction in the system \citep{2025Galax..13...49M}. Symbiotic stars are thought to be connected to several important late-stage stellar evolution objects, including X-ray binaries, cataclysmic variable stars, and recurrent novae. They have been proposed as progenitor systems of Type Ia supernovae \citep{1992ApJ...397L..87M,1999ApJ...522..487H} but simulations \citep{2025A&A...698A.155L} and observations \citep{2025ApJ...991..111S} suggest that they may be the source of a small fraction of Galactic Type Ia supernovae.

The key to determining the role that symbiotic stars play in late-stage binary stellar evolution lies in improving our understanding of mass transfer in these systems. In wide interacting binary systems like many symbiotic stars, the donor loses mass through a stellar wind, a fraction of which is accreted by the companion \citep{2025MNRAS.544.2387M}. The earliest model of wind accretion was the Bondi-Hoyle-Lyttleton (BHL) prescription (\citealp{bh1, bh2}; see \citealp{2025ApJ...980..226T} for a brief review of BHL implementation in accreting binary systems). It was later found that the BHL model over-predicts the mass-accretion rate in systems where the wind velocity is smaller than the orbital velocity, such as in symbiotic systems, apart from those with the largest orbital periods \citep[and references therein]{Boffin2015, 2016A&A...588A...3H, 2010ApJ...723.1188S, 2025ApJ...980..226T}. However, \citet{2025ApJ...980..226T} found that modifying the BHL model by the addition of a geometric projection factor better predicts the observed accretion efficiency. Simulations of the evolution of symbiotic systems using this modified BHL model found agreement with the orbital properties and bolometric luminosities of several wide-orbit S-type systems \citep{2025ApJ...989..108M}. 

While studying mass transfer in wide symbiotic systems containing Mira variables, \citet{wrlo1} found that for cases where the RG's wind velocity was less than the escape velocity of the star, material flows through the L$_1$ point as it is transferred toward the accreting companion, a process known as Wind Roche-Lobe Overflow (WRLOF). In subsequent simulations, \citet{wrlo2} found that WRLOF can also take place when the dust-formation radius is comparable to the RG's Roche-lobe radius. Notably, the simulations presented in these works found that WRLOF resulted in larger accretion rates than BHL mass transfer. Similarly, hydrodynamic simulations by \citet{2009ApJ...700.1148D}, and \citet{2022ApJ...931..142L} found that gravitational focusing of BHL winds plays an important role in mass transfer in interacting systems, resulting in a stream similar to Roche-lobe overflow (RLOF) and the formation of an accretion disk. Moreover, simulations by \citet{2025MNRAS.544.2387M} and \citet{2025ApJ...980..224V} found that systems with more evolved donor stars can alternate between wind accretion and WRLOF.

In some close symbiotic systems, it is possible that mass transfer may take place via RLOF. Many symbiotic stars exhibit ellipsoidal variations that have been interpreted as evidence of tidal deformation \citep[e.g. EG Andromedae in ][]{1997Wilson}. As a result, some systems have been proposed to experience mass transfer via RLOF \citep{Munari2019}. In this process, any material from the distorted RG that crosses this equipotential boundary is no longer bound to the RG and is free to be accreted by the hot companion or become part of the surrounding CSM.  However, mass transfer driven by RLOF may be too unstable to support the long lifetime expected for the symbiotic star binary stellar evolution phase \citep{2007podsiadlowski}. \citet{2012Mikolajewska} noted that there are large discrepancies in the Roche-lobe filling factors derived from radial velocities and light curve analysis. Moreover, \citet{skopal_windtransfer} and \citet{2021shagatova} suggested that a focused wind could mimic the appearance of a tidally distorted RG and lead to ellipsoidal variation in the light curves.

 \citet{2007apn4.confE..45C} and \citet{2012Mikolajewska} recommended the use of optical interferometry for studying the RG in these systems. If the RG can be shown to be completely filling its Roche lobe, we would expect the system to be undergoing RLOF. Conversely, if the RG is underfilling its Roche lobe, mass transfer would be taking place via a wind-based mechanism such as those described above. 

The Precision Integrated-Optics Near-infrared Imaging ExpeRiment (PIONIER) at the Very Large Telescope Interferometer (VLTI) has been used to observe 13 symbiotic stars and three related interacting systems, searching for radial asymmetry to test methods of mass transfer. Although not a symbiotic system, the study of the interacting binary system SS Lep, which consists of an M-type RG and an evolved A-type companion, by \citet{2011blind} paved the way for future optical interferometric studies of symbiotic star systems. Observations at two different epochs during 2010 indicated evidence of mass transfer from the RG to the A-type companion, but suggested that the Roche lobe was only partially filled. This study was followed by a larger observational study from \citet{2014Boffin} that observed six new targets. Two of these targets in the 2014 study were found to be filling or almost filling their Roche lobes while the other four were measured to be underfilling their Roche lobes. \citet{2014Boffin} noted that observation of these systems at a higher angular resolution than could be achieved with the VLTI at that time is needed to resolve the RGs in their targets.  \citet{2024A&A...692A.218M} reported that a model image of a Roche lobe best fit quadrature observations of HD 352 taken with the VLTI,  showing that optical interferometry can be used to distinguish between binaries filling their Roche lobes and those which are not. Following this, \citet{Merc2025I} and \citet{Merc2025II} presented observations of a total of 13 symbiotic star systems, including those previously reported in \citet{2014Boffin}, of which only one (ZZ CMi) was close to filling its Roche lobe. In addition to these VLTI observations, \citet{2026ApJ..1004L..13N} used the Center for High Angular Resolution Astronomy (CHARA) Array to measure the angular diameter of the RG in T Coronae Borealis at different orbital phases, finding that the derived stellar radius supported Roche lobe filling.

In light of the need for higher angular resolution observations of symbiotic star systems noted in \citet{2014Boffin}, we began a survey of four nearby systems using the Center for High Angular Resolution Astronomy (CHARA) Array in 2021. The CHARA Array is a six-telescope optical interferometer array with a longest baseline of ~330 meters, capable of resolving objects with angular sizes as small as 0.5 milliarcseconds (mas) in the \textit{H}-band \citep{ten_brummelaar_first_2005}. Here we report on the results of observations of four interacting star systems, three of which are symbiotic stars, made with the facility. We observed V1472 Aql (HD 190658; M2.5 III; \citealt{1997samus}), EG And (HD 4174; M2.4 III; \citealt{egandtype}), BD Cam (HD 22649; S3.5/2; \citealt{bdcamtype}), and SU Lyn (HD 47648; M5.8 III; \citealt{2016mukai}) with the goals of analyzing the shape of each binary system’s RG component and of imaging the giant. We also report the results of near-infrared spectroscopic observations, which provide additional stellar parameters for the giants in these systems. The parametric fits reported here reject Roche-lobe-filling geometries and imaging suggests the presence of surface features on most of these stars. 

\section{Observations} \label{sec:obs}

\subsection{Target Selection}

We choose targets by querying SIMBAD \citep{2000Simbad} and the New Online Database of Symbiotic Stars \citep{2019Merc,2019AN....340..598M} for symbiotic systems having a bright \textit{H}-band magnitude, a distance within $\sim$1 kpc from Earth, a right ascension between 07 hours and 21 hours, and a declination greater than -20 degrees. These parameters ensured that targets were bright enough to be observed by CHARA, large enough to be resolved and imaged, and observable during the semester in which these data were collected. Of those stars fitting these constraints, we selected V1472 Aql, due to its inclusion in the sample of \citet{2014Boffin}, although we note that \citet{Merc2025I} excluded it from their recent revisit of symbiotic stars as it does not display the characteristics of a typical symbiotic system. We chose EG And due to its use as a prototypical example of a symbiotic system experiencing wind-based mass transfer \citep{2016shagatova,2021shagatova}. We selected BD Cam because its distance suggested a large angular diameter, which would enable more detailed imaging. Finally, we chose to observe SU Lyn because it had been recently identified as a symbiotic system \citep{2016mukai} and it is thought to be the prototype for a hidden population of symbiotic stars \citep[for a review, see][] {Munari2019}.  Table \ref{tab:int_log} reports a log of all observations made for each target.

There do not seem to be any publications that present evidence of a WD or hot companion in V1472 Aql, hence its removal from the symbiotic systems revisited by \citet{Merc2025I}. However, \citet{1982Lucke} noted that it may be filling its Roche lobe and \citet{1997samus} found evidence of an eclipse by an unseen companion. Due to its past inclusion in lists of symbiotic binaries, we include it in this study with the caveat that the companion may be some other faint object. EG And, BD Cam, and SU Lyn all indicate the presence of WDs via ultraviolet emission (EG And: \citealt{egandwd1}; BD Cam: \citealt{bdcamwd1,bdcamwd2}; SU Lyn: \citealt{2016mukai,2021kumar}).  Using EG And as an example, given a WD of effective temperature $T_{\text{eff}}\approx 7.5 \times 10^{4}$ K and radius $R = 0.018~R_{\odot}$ \citep{1991A&A...249..173V} and giant of $T_{\rm eff}\approx3630$ K and
$R\approx77\,R_{\odot}$ (see Section \ref{sec:fundamental}), we can expect $\Delta H\approx 13.4~\text{mag}$, and thus do not expect to detect the WD with the observations reported here.

\input{observations}

\subsection{Interferometric Observations} \label{subsec: interferometry observations}

We observed the four systems using Georgia State University's CHARA Array in Mount Wilson, CA in a three-night window from UTC dates 2021 September 20 to 2021 September 22 using the six-beam combiner MIRC-X \citep{2018Anugu, 2018Kraus, 2020Anugu} in high spectral resolution (R $\sim$ 190) mode, operating in the \textit{H} band (1.5 - 1.72$~\mu$m). Simultaneous observations made with the six-beam  \textit{K}-band combiner ($\lambda=2.0 - 2.39 ~\mu$m) MYSTIC \citep{Setterholm2023} will be described in a later publication. We were able to observe three of our four targets using all six telescopes, but due to BD Cam's high declination, we could not observe the system with the S1 telescope. We did not expect any of the targets to change appreciably in angular diameter within the time span of the observations,  so we combined multi-night data collected on BD Cam and V1472 Aql into a single data set for each star.

Observations of each star consisted of standard ``calibrator-target-calibrator" brackets. We selected calibrator stars within 10 degrees of the target on the sky, with a similar \textit{H}-band magnitude to the target ($\pm2$ magnitudes), without reported binarity, and with an angular diameter smaller than 0.75 mas. For targets in frequently-observed areas of the sky, we selected calibrators by querying the literature for calibrator stars that had previously been utilized in CHARA observations. In the case of targets that had no suitable previous calibrators in close proximity, we used SearchCal \citep{2006Bonneau,searchcal} to identify appropriate calibrators. Table \ref{tab:int_log} reports the nights on which each calibrator was observed and in conjunction with which target. Table \ref{tab:int_log_cal} provides information about each calibrator.
\input{cals}

We reduced data using the MIRC-X Data Reduction Pipeline version 1.3.5 described in \citet{2020Anugu}\footnote{\url{https://gitlab.chara.gsu.edu/lebouquj/mircx_pipeline}}. Calibrated OIFITS files used in this paper are available in the JMMC Optical Interferometry Database\footnote{\url{https://oidb.jmmc.fr/index.html}}. As an example, Figure \ref{fig:su lyn} presents the squared visibilities and closure phase of SU Lyn.  Plots of  $(u,v)$ coverage, squared visibilities, and closure phases for all targets are available in Figures \ref{fig:uv_coverage}-\ref{fig:t3phiall} in Appendix \ref{sec:all plots}.

\input{sulynobs}

\input{irtfobs}

\subsection{Spectroscopic Observations} \label{subsec: spectroscopy observations}
 We observed each of our four stars using SpeX \citep{2003Rayner} on the NASA Infrared Telescope Facility (IRTF) at Mauna Kea, Hawaii on UTC 2021 September 02 providing near-simultaneous spectral information about each interferometry target. We also observed three of our four targets on UTC November 26, 2021. We did not observe V1472 Aql a second time as it was too near the horizon during that observation. The data collected on SU Lyn on 2021 September 02 was of poor quality as the star was observed close to the end of the night and is not reported here. We used both the SXD and LXD-long modes using a slit width of $0.3''\times15''$ with resolution R $\sim$ 2500 covering a wavelength range of 0.7 $\mu$m to 5.3 $\mu$m. 

During observations, we used an ``AB" beam pattern, nodding the telescope to observe the target on different parts of the slit. This helped to reduce systematic error such as variations in the sky brightness that may be introduced during observation. We collected at least 4 AB pairs on each target in each mode in order to average out seeing and guiding variations and to build a sufficiently high signal-to-noise ratio (SNR) for each target. We observed standard stars of spectral type A0V in both SXD (0.7 $\mu$m - 2.55 $\mu$m) and LXD-long (2.25 $\mu$m - 5.3 $\mu$m) modes to allow for telluric correction of target spectra. We selected these stars by querying the ``Locator for Nearby A0V and G2V stars" form\footnote{\url{http://irtfweb.ifa.hawaii.edu/~spex/find_a0v/}} for standards that were near targets both in hour angle and angular separation. This ensured that the difference in airmass between standard and target was minimal at the time of observation. A log of observations is reported in Table \ref{tab:spec log}.

We reduced data using the standard IRTF data reduction pipeline, Spextool \citep{2004Cushing,2003Vacca}, written in the Interactive Data Language (IDL). First, we created calibration frames from observations of flat field and argon arc lamps to extract the AB pair spectra from the raw FITS files generated during observations. After separately merging individual spectral data files to create single target and standard spectra, we used a combination of the standard spectrum and a convolved high-resolution model of the star Vega to generate a telluric spectrum. We then used the telluric spectrum to correct the target spectrum for atmospheric telluric absorption. Finally, we scaled and merged the individual orders to create a final target spectrum. We repeated this process for both SXD and LXD spectral data separately before merging the final SXD and LXD spectra into a single file and smoothing out any false flux spikes generated in areas of low SNR. Figures \ref{fig:sept_spectra} and \ref{fig:nov_spectra} in the Appendix present spectra observed at each epoch.\footnote{Calibrated spectra used in this paper are available on Zenodo \citep{Gaudin_Zenodo_2026}.}

\section{Angular Diameters and Shape} \label{sec:models}
\subsection{Model Fits} \label{sec:fits}
\input{modelfits}
We performed model fitting to the interferometric observations using OITOOLS\footnote{\url{https://fabienbaron.github.io/OITOOLS.jl/dev/}} \citep{Martinez2021, Anugu2023} and verified these using PMOIRED\footnote{\url{https://github.com/amerand/PMOIRED}} \citep{2022SPIE12183E..1NM}. Within OITOOLS.jl we used Bayesian inference with the nested sampling Monte Carlo algorithm MLFriends \citep{2016S&C....26..383B,2019PASP..131j8005B} supplied by UltraNest \citep{,2021JOSS....6.3001B}\footnote{\url{https://johannesbuchner.github.io/UltraNest/}}. This allowed us to calculate the Bayesian evidence, log(Z), presenting the probability of the data given the model. PMOIRED uses gradient descent for $\chi_{\nu}^{2}$ minimization with the option for bootstrapping. Because the models we fit were all centrosymmetric, we fit to squared visibilities only, so as to remove the possible influence of non-centrosymmetric surface features, which are expected in evolved stars on the closure phases. The results from PMOIRED and OITOOLS were within $1\sigma$ of each other, so only the results from OITOOLS are reported and used here. Following the fitting process, the angular diameters were divided  by $1.0054\pm0.0006$ \citep{2018Gardner} to account for wavelength calibration, as specified by the MIRC-X-MYSTIC Pipeline User Manual.

Following the methods described in \citet{2007verhoelst} and \citet{2014Boffin}, we fitted four models: a uniform disk (UD), a limb-darkened (LD) disk, an elongated uniform disk (ED), and a multi-component model with a smaller disk superimposed on top of a larger elongated disk. The UD model is given by the complex visibility function:
\begin{equation}
    V(\theta_{UD}) = \frac{2J_1(\pi \nu \theta_{UD})}{\pi \nu \theta_{UD}} \label{eq:ud}
\end{equation}
where $J_1$ is a Bessel function of the first kind, $\nu$ is the sampled frequency range, and $\theta_{UD}$ is the angular diameter of the target. This model is included to best simulate a spherical star that is well within its Roche lobe with no radial asymmetries present on its visible disk. For the LD disk model, we used the Hestroffer law \citep{hestlaw}:
\begin{equation}
    I_{\lambda}(\mu)/I(1)=\mu^{\alpha} \, ; \, \alpha \in \mathbb{R}^{+} \label{eq:heflaw}
\end{equation}
where $I$ is intensity, $\mu=\sqrt{1-(2r/\theta_{LD})^{2}}$ with $r$ the angular distance from the center of the star, $\theta_{LD}$ the angular diameter of the photosphere, and $\alpha$ is the limb-darkening parameter. In the ED model, the equation without limb darkening is modified to take into account semi-major axis orientation  ($\phi$) and axis ratio ($\epsilon$) of the ellipse. In this model, we adopt an elliptical spatial
frequency:

\begin{equation}
\rho_{\rm ell} =
\left[
\epsilon^2
\left(u\cos\phi-v\sin\phi\right)^2
+
\left(v\cos\phi+u\sin\phi\right)^2
\right]^{1/2}.
\label{eq:ellipse_rho}
\end{equation}

\noindent
The corresponding complex visibility is:

\begin{equation}
V(\theta_{\rm ED}) =
\frac{2 J_1\left(\pi \theta_{\rm ED}\rho_{\rm ell}\right)}
     {\pi \theta_{\rm ED}\rho_{\rm ell}},
\label{eq:ellipse_visibility}
\end{equation}

\noindent
where $\theta_{\rm ED}$ is the angular diameter in radians.

 The multi-component (hybrid) model employs both the UD and ED models. The diameter of the UD is fixed at 50 percent of the best-fitting diameter from the UD fit and the ED diameter is given a lower limit of that same value and given no upper limit. By allowing the ED to have a larger diameter, the model combines a compact circular component with a larger elongated component, allowing a simple non-circular extended structure. In this model, we fixed the contribution of each component, with the UD contribution $70\%$ of the flux. 

The parameters of the best-fitting models and the resulting $\chi_{\nu}^2$ and $\log(Z)$ are reported in Table \ref{tab:model_fits}. Note that the position angle of the ellipse in OITOOLS is measured counterclockwise from the $u$-axis (East) in the ($u$,$v$) plane, not from the axis of rotation. The $\chi_{\nu}^2$ is lowest for the LD disk models V1472 Aql, BD Cam, and SU Lyn. Likewise, for these three objects the LD disk model is consistently preferred by Bayesian model comparison over UD models (V1472 Aql: $\Delta \log Z = 2.82$; BD Cam: $\Delta \log Z = 1.50$ ; SU Lyn: $\Delta \log Z = 11.64$). There is an even stronger preference over the elongated (V1472 Aql: $\Delta \log Z = 5.03$; BD Cam: $\Delta \log Z = 3.23$; SU Lyn: $\Delta \log Z = 14.72$) and hybrid models (V1472 Aql: $\Delta \log Z = 15.18$; BD Cam: $\Delta \log Z = 13.15$; SU Lyn: $\Delta \log Z = 39.69$). For EG And, the lowest $\chi_{\nu}^{2}$ is for the fit to elongated model and the Bayesian evidence is nearly identical for the UD and LD models, with a slight preference for the UD model ($\Delta \log Z = 0.22$). The UD model is more strongly preferred over the ED model ($\Delta \log Z = 2.47$) and very strongly preferred over the hybrid model ($\Delta\log Z = 19.55$). This suggests that a disk is a better model of the star, with the fitting routine having trouble distinguishing between uniform and LD disks due to the smaller angular diameter of the star. Thus, the model fits suggest that the projected disks of the stars are consistent with circular symmetry rather than elongation. Nonetheless, the relatively high $\chi_{\nu}^{2}$ values of the best fitting models for each star, apart from EG And, suggest that additional parameters, such as surface features, are needed to properly model these stars.

\subsection{Simulations}\label{sec:sims}
\input{modelfits_sim}
\input{simulations}
To test the ability of our model fitting approach to distinguish between spherical and Roche filling stars, we used ROTIR\footnote{https://github.com/fabienbaron/ROTIR.jl} \citep{Martinez2021} to simulate observations of LD stars of various radii and stars with various potential-based fillout factors ($f = 0.5$ to $1.0$), where $f=1.0$ represents a star whose equipotential surface is at the L$_{1}$ Lagrange point \citep{leahy2015}. The simulated RG had stellar parameters of $T_{\text{eff}} = 3597$ K, limb darkening coefficient $\alpha = 0.47$, and (for the Roche filling stars only) gravity darkening coefficient $\beta = 0.08$. The Roche filling star was modeled as a member of a binary with a non-visible companion in a system with mass ratio $q = \frac{M_{\text{G}}}{M_{\text{2}}}=0.5$, binary separation $a=8.496$ mas, inclination $i = 90\degr$, distance = 235.43 pc, and synchronous rotation. Thus the Roche lobe volume radius of the modeled system was 2.73 mas. Because the simulated images are flux normalized, distance only impacts the resulting angular diameter. We copied the $(u,v)$ coverage and error characteristics of the SU Lyn observations when using OITOOLS to make simulated observations, adding Gaussian noise scaled to the observed error bars. We then used the fitting procedure described in Section \ref{sec:fits}. We also modeled spherical LD stars of radii r=1.00 mas, r= 1.50 mas, r=1.85 mas, and r=1.98 mas to verify that the fitting procedure does not identify a spherical star as elongated.

Table \ref{tab:modelfitssim}  presents the results of the model fits to the simulations and Figure \ref{fig:simimages} presents images of the simulated stars. Although the $\chi_{\nu}^{2}$ for the fits of the ellipse model to the Roche filling star is 1.0 for many fillout factors, the axis ratios of these ellipses are very close to 1 until a fillout factor of 0.8. Moreover, it is only around a fillout factor of 0.9 that the Bayesian evidence difference becomes small enough for the Roche geometry to be distinguishable. This suggests that our modeling procedure is capable of distinguishing between spherical and Roche filling geometries for high fillout factors, reinforcing our finding that the target stars are likely underfilling their Roche lobes.

\section{Fundamental Stellar Parameters} \label{sec:fundamental}
\input{parameters}
\input{orbitalparams}
\input{priors}
\input{masstab}

In order to determine the likeliest fit for the temperature, metallicity, and reddening of each RG, we used the spectral inference package Starfish \citep{2015Czekala,2018Czekala}.  To do this, we trained a spectral emulator covering the parameter ranges of $T_{\text{eff}} = [2300, 5000]$ K, $\log{g} = [0.0, 3.0]$ dex, and [Fe/H] = [-1.0, 0.5] using synthetic high-resolution spectra generated from PHOENIX spherical model atmospheres \citep{2013Husser}. As a first solution, we obtained a maximum a posteriori (MAP) solution using Nelder-Mead optimization, which we used as starting point for a Markov Chain Monte Carlo (MCMC) fit. We fixed  Starfish's Gaussian process (GP) covariance hyperparameters because allowing them to vary freely prevented convergence. We used the MAP solution to initialize MCMC walkers, with convergence determined using the integrated autocorrelation time $\tau$, stopping when $N>50\tau$ and $|\Delta \tau|/\tau<0.01$. We used 16 walkers, a maximum of 3000 iterations, a burn-in of $3\tau_{\text{max}}$, and thinned the chains by $0.3\tau_{\text{min}}$.   

For each star, we fit the effective temperature ($T_{\text{eff}}$), visual extinction, $A_{V}$, and unless otherwise mentioned in the following subsections, metallicity, [Fe/H]. After comparing model spectra with different $\log$ g for given effective temperatures and metallicities, we kept $\log$~g constant at values reported in literature or estimated from the spectral type and class of each RG, because we found that the model spectra were insensitive to changes in $\log$~g at the spectral range and effective temperatures used in our measurements. For SU Lyn, we used $\log$~g=0.0 due to its evolutionary status \citep{2022Ilkiewicz}, and for all other targets we used $\log$ g=0.5.  The priors we used were: a Gaussian centered on literature values for $T_{\text{eff}}$, a half-normal prior on $A_{V}$ informed by values of Galactic reddening determined using the Bayestar19 dust map \citep{bayestar19} at each target's \textit{Gaia} Early Data Release 3 (EDR3) \citep{gaiamain,2021A&A...649A...1G} geometric distance as reported in \citet{gaiadist}, and a uniform prior for [Fe/H] over [-0.5,+0.5]~dex, unless otherwise stated.

In our fits, we used the \textit{H} and \textit{K} band because literature shows that the $^{12}$CO lines in both regions are sensitive to the effective temperature of M giants \citep{2022Ghosh,2016Schultheis}. This region also lacks significant contamination from the accretion disk or WD because the giant is responsible for the majority of the flux of the system at these wavelengths. Within these bands, we fit over 1.50--2.38$~\mu$m in order to avoid an artifact in the emulator boundary near 2.40$~\mu$m and we masked the telluric band at 1.82--1.91$~\mu$m. The observed spectra were thinned by a factor of three to one pixel per resolution element and flux normalized by the median.

As noted, we used a dust map as a prior for $A_{V}$ based on Galactic dust maps but we left $A_{V}$ free. In Figure \ref{fig:cornerstarfish}, one can see the correlation between derived effective temperature and visual extinction. There is a moderate positive correlation between the two, and higher extinction values lead to fitting slightly higher temperatures. As described in Section \ref{sec:res}, the color excesses we derive from the visual extinctions we fit largely agree with literature values. 

The metallicity was poorly constrained in our fits. To test the impact of metallicity on the reported parameters, we ran the fitting routine for EG And spectra using a free uniform prior as described above and also with a fixed Gaussian [Fe/H]=$-0.54\pm0.10$ from \citet{2023MNRAS.526..918G}, which was determined using higher resolution spectra ($R\sim50,000$). We found no significant difference in resulting parameters: the free uniform prior yielded $T_{\text{eff}}=3630^{+120}_{-110}~\text{K}$ and $A_{V}=0.11^{+0.10}_{-0.08}$ with [Fe/H]=$-0.05^{+0.28}_{-0.27}$ ; the fixed [Fe/H] from \citet{2023MNRAS.526..918G} yielded $T_{\text{eff}}=3680\pm80~\text{K}$ and 
$A_{V}=0.12^{+0.11}_{-0.08}$. 

To obtain physical radii from the angular diameters of V1472 Aql, BD Cam, and SU Lyn, we use the geometric posterior distance estimates inferred by \citet{gaiadist}, who used  \textit{Gaia} EDR3 parallaxes and a direction-dependent Galactic prior. As described in Section \ref{subsubsec:And params}, for EG And we adopt $d=400\pm20$ pc
\citep{1992Vogel,2016AJ....152....1K}.

We chose to use the geometric rather than photogeometric distance estimates because the latter incorporates a color-absolute magnitude prior calibrated on single stars, which would be inappropriate for symbiotic systems. To evaluate the quality of the parallax used in this distance determination, we can use several metrics. The re-normalized unit weight error (RUWE) is a diagnostic of the astrometric solution quality. For an isolated point source RUWE $\approx 1$. \citet{2022MNRAS.513.2437P} proposed $\text{RUWE}>1.25$ as a threshold for identifying objects in \textit{Gaia} DR3 with a poor single-star astrometric solution, as would be expected when binary orbital motion perturbs the photocenter of the source, although \citet{2024A&A...688A...1C} note that this threshold varies somewhat with sky position. The RUWE values for the targets reported in this paper are: V1472 Aql, RUWE = 2.028; EG And, RUWE = 1.814; BD Cam, RUWE = 2.772; SU Lyn, RUWE = 0.912. Thus, all systems apart from SU Lyn may be affected by binary-induced photocenter motion. However, for all targets, the fractional parallax uncertainty (fpu, $\sigma \varpi/\varpi$) meets the criteria (fpu $< 0.1$) suggested in \citet{gaiadist} for which the inverse parallax can be considered a reliable distance estimate and the posterior is parallax rather than prior dominated (V1472 Aql: fpu = 0.018; EG And: fpu = 0.020; BD Cam: fpu =0.061; SU Lyn: fpu =0.046). The geometric distances in \citet{gaiadist} differ from the inverse-parallax distance by 2.5\% or less for all four targets. None of the four targets appears in the DR3 \citep{gaiadr3} non-single star solution catalog \citep{gaiabinary}.

We computed physical radii using the LD angular diameters in Table~\ref{tab:model_fits} and calculated luminosities using the effective temperatures from the spectral fits and the physical radii. Table \ref{tab:stellarparams} lists parameters determined from the fits of the synthetic spectra to observed spectra.  Radius and luminosity are derived from Monte Carlo propagation of the angular diameter, distance, and effective temperature uncertainties, with asymmetric
intervals reported as 16th/84th posterior percentiles. Corner plots and spectral-model fits are available in Figures \ref{fig:cornerstarfish} and \ref{fig:spectrastarfish} in the Appendix.

\section{Masses and Filling Factors\label{sec:masses}}

With a star's radius and luminosity, it is possible to infer its mass from evolutionary tracks, as was done by \citet{2014Boffin} for the RGs in V1472 Aql and for the RG in SU Lyn by \citet{2022Ilkiewicz}. However, the mass function, if available, also constrains the RG's mass $(M_{\text{G}}$), as well as that of the companion star ($M_{2}$). Moreover, to measure the extent to which each RG fills its Roche lobe, the mass of the companion is needed, since the filling factor is described by the equation $f=R_{\text{g}}/R_{L}$ where $R_{L}$ is the Roche-lobe radius determined with the approximation of \citet{1983eggleton}:

\begin{equation}
\label{eq:RL}
r_{\text{L}} = \frac{0.49 q^{2/3}}{0.6 q^{2/3} + \ln{(1 + q^{1/3})}} a, \quad \text{where } q = \frac{M_{\text{G}}}{M_{\mathrm{2}}}, \quad \text{and } a = \left[\frac{G\,(M_{\text{G}} + M_2)\,P^2}{4\pi^2}\right]^{1/3}
\end{equation}

Because $M_{2}$ is correlated with $M_{\text{G}}$ in the mass function equation, we used a joint inference approach as described in Appendix \ref{app:mass}\footnote{Code available at: https://github.com/norrisryan/fitsyms}. This method incorporated both the position of the RG on the MESA Isochrones and Stellar Tracks \citep[MIST;][]{MIST,MIST1} and the mass function of each system as presented in Table~\ref{tab:orbital}. We used the current stellar mass $M_\star$ from the MIST evolutionary tracks rather than the initial (ZAMS) mass $M_\mathrm{init}$, because the mass function constrains the present gravitational mass. MIST uses cumulative RGB and AGB wind mass loss from
\citet{1975MSRSL...8..369R} and \citet{1995A&A...297..727B} in its modeling of $M_{*}$. We found that the resulting shift from $M_\mathrm{init}$ to $M_\star$ decreased fit masses less than the reported uncertainties. For each sample, we drew $\theta_{\rm LD}$, $d$, and $T_{\rm eff}$ from their adopted measurement distributions and $[\mathrm{Fe/H}]$ from the target-specific prior, and determined $M_G$ using a kernel-weighted interpolation over the nearest MIST grid points in
$\log T_{\rm eff}$, $\log L$, and $[\mathrm{Fe/H}]$. Using this value, we then applied the mass function as an importance weight and resampled the complete set of quantities to obtain a joint posterior. The companion-mass, $[\mathrm{Fe/H}]$, and inclination priors were target-specific; all priors are presented in Table \ref{tab:priors}. The resulting values for $M_G$, $M_2$, $\langle r\rangle$, $R_L$, and $f$ are presented in Table \ref{tab:masstab}. For systems with eccentric orbits, we use the orbit-averaged separation $\langle r\rangle$ when calculating the reported Roche-lobe radii and filling factors, as described in Appendix \ref{app:mass}. However the eccentricity for the targets in this sample for which the orbit is known is small. Corner plots and Hertzsprung-Russell (HR) diagrams for the fits are presented in Figures \ref{fig:masscorner} and \ref{fig:hrdiag}.

Note that although the filling factor calculated in this manner does not take into account the changing shape of the star in the way that the fillout factors used in the simulations in Section \ref{sec:sims} do, it enables us to estimate how close the star is to filling its Roche lobe. As shown by the simulations, changes to the shape of the star are not apparent to our fitting routines until fillout factors approach $f=0.9$.

\section{Imaging}
\input{surfing}

\input{images_sphere}

The closure phases we obtained for the targets presented here deviate from zero as seen in Figure \ref{fig:t3phiall} in Appendix \ref{sec:all plots}.  Many past papers on interacting systems \citep{2014Boffin,Merc2025I,Merc2025II} reported closure phases consistent with zero, including for V1472 Aql \citep{2014Boffin}. This is largely due to the shorter baselines of the VLTI in comparison to the CHARA Array. The data collected in those studies spanned only the first lobe of the visibility curve for those objects, whereas the data collected here span additional lobes and therefore provide data on features smaller than the stellar disk. For comparison, consider the closure phases collected by \citet{2018Natur.553..310P} of $\pi^{1}$ Gruis, an AGB star of significantly larger angular diameter than the RGs in symbiotic systems that have been observed by the VLTI. In this case, the asymmetric features were of angular size such that they could be resolved by VLTI and therefore the closure phases deviate from zero.

Because we recorded closure phases that deviate from zero, we also performed image reconstruction using the data we collected. We used SURFING  \citep{2016Natur.533..217R} for image reconstruction, with results presented in Figure \ref{fig:imagessphere}. Prior to imaging, SURFING was used to find best-fitting parameters, as listed in Table \ref{tab:surfing}. Following this, an image for each star was obtained using SURFING for image reconstruction, averaging 50 instances of each surface.

The best fitting angular diameters and limb darkening coefficients obtained with SURFING match with those found by OITOOLS within $1\sigma$ of the parametric fits. With the exception of EG And, the reconstructed images indicate the presence of a varied surface on these stars. These asymmetric structures are possibly the result of large surface convective fluctuations that are characteristic of stars that are in the later red giant or asymptotic giant branch (AGB) phase of their evolution.

\section{Results for Individual Objects} \label{sec:res}

The methods described above enabled us to obtain the fundamental parameters of the RG in each system, as well as measure its filling factor. In the following section, we put these results into the context of past work and also describe the impact of these results on understanding mass transfer in these systems.

\subsection{V1472 Aql} \label{subsubsec:Aql Params}
V1472 Aql lacks emission lines and thus is not a symbiotic star system. Rather, it is classified as an SS-Lep-type binary in the \textit{New Online Database of Symbiotic Variables} \citep{2019Merc}. However, it has the shortest period of the stars in the sample presented here (excluding SU Lyn for which there is no published period). The system displays ellipsoidal variations in its light curve \citep{1997samus,2010MNRAS.409..777T} and many authors have suggested that the RG may be filling its Roche lobe \citep[e.g.][]{2009A&A...498..489J}. A third, possibly gravitationally-related component is 2.3'' from the inner pair and beyond the field of view of our observations \citep{1997samus, 1997A&AS..124...75T,2007A&A...464..377F}.  \citet{2014Boffin} included the system in a study of mass-transferring giants using the VLTI. These authors ultimately concluded that the RG was filling between 43\% and 100\% of its Roche lobe. Note that they came to this determination using a reprocessed Hipparcos parallax that took orbital motion into account; using the standard Hipparcos parallax led to the RG underfilling its Roche lobe. 

We measured an LD angular diameter of $\theta_{\text{LD}}=2.43^{+0.05}_{-0.05}$ mas. \citet{2014Boffin} reported uniform-disk angular diameters instead of LD angular diameters and their measurement of $\theta_{\text{UD}}=2.33\pm0.03$ mas is close to the $\theta_{\text{UD}}=2.36^{+0.04}_{-0.03}$ mas we obtained. However, the \textit{Gaia} EDR3 parallax that was used by \citet{gaiadist} to obtain the distance we used is $\varpi=3.99\pm0.07$ mas, which is larger than the $\varpi=2.4\pm1.0$ used by \citet{2014Boffin}, so we end up with a smaller physical radius than reported by these authors.

Using a distance $d=247.6^{+5}_{
-4}~\text{pc}$ \citep{gaiadist}, we obtain a radius $R=65^{+2}_{-2}~R_{\odot}$. We measured an effective temperature of $T_{\text{eff}} = 3620^{+120}_{-110}~\text{K}$ and derived a luminosity $L=650^{+100}_{-80}~L_{\odot}$. The RG is classified as an M2.5III in \citet{1981ApJS...45..437A} and the effective temperature we measure agrees well with the values reported by \citet{2021ApJ...922..163V} for an M2III-M3III star ($T_{\text{eff}}=3617\pm92~\text{K}$ for M2III, $T_{\text{eff}}=3559\pm120~\text{K}$ for M3III). However, the radius we obtain falls toward the smaller end of those found by these authors for stars of those spectral types ($79.30\pm14.15~R_{\odot}$ for M2III and $75.87\pm20.28~R_{\odot}$ for M3III). The authors of that study tended to use Hipparcos parallaxes over \textit{Gaia} DR2 parallaxes due to the brightness of their targets, and much of the sample they used for M2III and M3III stars were not binaries. So, it is possible that the radius we derive is smaller than the actual physical radius due to the impact of binary motion on the \textit{Gaia} parallax which is the source of the distance we use.

The reddening obtained from the Bayestar19 map \citep{bayestar19}, which was used as a prior in our spectral fits, is  $A_{V} = 0.582\text{ mag}$; this is close to the fit value of $A_{V}=0.61^{+0.34}_{-0.33}~\text{mag}$. Moreover, when using the same conversion to $A_{K}$ used by \citet{2014Boffin} ($A_{K} = 0.114 A_{V}$), we obtain $A_{K}=0.07^{+0.04}_{-0.04}~\text{mag}$, close to their value of $A_{K}=0.06~\text{mag}$.

The metallicity we obtained was not well constrained for this system, as can be seen in Figure \ref{fig:cornerstarfish} and no high-resolution spectroscopic [Fe/H] is available in the literature. So, when inferring the mass of the RG, we used a prior based on the Metallicity Distribution Functions (MDFs) in Table 2 of \citet{2015ApJ...808..132H}, selecting the column and row which corresponded to the location of V1472 Aql ($R_{\text{gal}}\approx8~\text{kpc}$, $z\approx16~\text{pc}$); this resulted in using a prior $\mathcal{N}(+0.02,\, 0.20)$ for [Fe/H] which we truncated to the MIST grid range $[-1.0, +0.5]$.

We found the masses of the RG and companion to be $M_{\text{G}}=1.24^{+0.64}_{-0.36}~M_{\odot}$ and $M_{2}=0.58^{+0.18}_{-0.12}~M_{\odot}$. Note that we used a uniform prior $\mathcal{U}(0.30,3.00)$ for $M_{2}$ because the nature of the secondary companion is unknown. This leads to a time-averaged Roche-lobe radius of $R_{L}=78^{+13}_{-10}~R_{\odot}$ and filling factor $f=0.83^{+0.11}_{-0.12}$. The model fits reported in Table \ref{tab:model_fits} also support a spherical shape for the star. The simulations suggest that we should see evidence for elongation by a filling factor of $f=0.90$, so it is not surprising that the filling factor is less than this. Our observations were obtained at orbital phase $\phi = 0.30$ from periastron (Table~\ref{tab:orbital}), which is near quadrature. Given the eccentricity of the orbit ($e = 0.048$), tidal elongation should still be visible at this phase. However, the position angle of the orbital axis is not known, and if the system were aligned unfavorably relative to our baseline coverage the elongation could be undetected regardless of phase.

There is a large spread in the inferred filling factor and Roche-lobe radius we obtained for this RG. We used an isotropic inclination prior truncated to
$50^\circ\leq i\leq90^\circ$, i.e., uniform in $\cos i$ over this
range, based on the presence of ellipsoidal variations in the light curve as suggested by \citet{2014Boffin}. However, tests of the sensitivity of our mass-inference technique to different inclinations for the system show that the adopted inclination ends up having a small impact on the filling factor. On the other hand, our technique is sensitive to the distance adopted. V1472 Aql has the second highest RUWE of our sample (RUWE=2.028), suggesting that the \textit{Gaia} astrometric solution is perturbed by the binary orbit. A test of the sensitivity of our mass-inference technique to distance for this system showed that decreasing the distance by 20\% would lead to a filling factor of  $f=0.72^{+0.04}_{-0.09}$ and an increase of 20\% in distance up to $f=0.90^{+0.17}_{-0.12}$. So, it is possible that the RG is close to or filling its Roche lobe if the distance we used is too small. 

Although V1472 Aql is classified as an SS Lep type system in the  \textit{New Online Database of Symbiotic Variables} \citep{2019Merc}, the companion mass we derived ($M_{2} = 0.58^{+0.18}_{-0.12}~M_{\odot}$) is lower than the typical masses of an A/F-type dwarf. Note that higher-mass solutions do exist in the posterior tail, as can be seen in Figure \ref{fig:masscorner}. UV observations could help distinguish between a low-mass MS companion and a quiescent WD.

\subsection{EG And} \label{subsubsec:And params}
EG And is an accretion-powered symbiotic system. Although \citet{2001MNRAS.326..553S} detected weak optical flickering, \citet{2004AJ....128.1790K} ruled out an accretion disk as the dominant source of far-UV flux. The RG in the system is classified as an M2-M3III star \citep{egandtype,egandwd1} and based on fits to \textit{Far Ultraviolet Spectroscopic Explorer} (FUSE) spectra, the hot component is a bare WD with a surface temperature of 80000-95000 K and log(\textit{g})=7.5 \citep{2017Sion}. There has not been a recorded outburst in the WD.

Two studies have used eclipses in the UV to determine the ratio of the radius of the giant to orbital separation, $R_{\text{G}}/A$, which when combined with system masses and orbital properties, allows an estimate of the radius of the giant (as well as a distance, when combined with flux and temperature measurements). \citet{1992Vogel} used \textit{International Ultraviolet Explorer} (IUE) spectra to time the eclipse of the WD by the RG, finding $R_{\text{G}} = 74\pm10~R_{\odot}$ and $d=400~\text{pc}$. On the other hand, \citet{2005Skopal} used IUE observations of the contact times of the continuum minimum at $\lambda~1320~\AA$ \citep{2001A&A...366..157S,1996Ap&SS.235..305P}, determining $R_{\text{G}} = 124 \pm19~R_{\odot}$ and $d=590\pm92~\text{pc}$, close to the $d = 593.52^{+11.87}_{-13.35}$ pc obtained by \textit{Gaia} EDR3 parallaxes in \citet{gaiadist}. The use of different criteria for the eclipse led to different measurements of the ratio of the RG to the orbital separation $R_{\text{G}}/A$, and the use of different total masses led to different distance determinations. \citet{2001A&A...366..157S} notes that their $R_{\text{G}}/A$ measurement should be considered an upper limit, whereas the value from \citet{1992Vogel} is generally adopted in literature \citep[e.g.][]{2016AJ....152....1K,shagatova_egand}. The angular diameters corresponding to each result are $\theta_{\text{G}}=1.75\pm0.23~\text{mas}$ for \citet{1992Vogel} and 1.95 mas for \citet{2005Skopal}. Crucially, the RG is underfilling its Roche lobe for both results ($f=0.479 \pm 0.064$ for \citet{1992Vogel}, and $f=0.668 \pm 0.102$ for \citet{2005Skopal}).

We measured an LD angular diameter of $\theta_{\text{LD}}=1.77^{+0.08}_{-0.05}$ mas for the RG in EG And, which is quite close to that of \cite{1992Vogel}. Using a distance of $d = 593.52^{+11.87}_{-13.35}$ pc from \citet{gaiadist}, we derive a radius of $R = 113^{+6}_{-4}~R_{\odot}$.  This is similar to the value that \citet{2023MNRAS.526..918G} derived using \textit{Gaia} DR3 parallaxes ($\sim110~R_{\odot}$). If we use the bolometric distance derived by \citet{1992Vogel} and updated by \citet{2016AJ....152....1K}, $d=400\pm20$ pc, the resulting radius is $R=77^{+5}_{-5}~R_{\odot}$. This is also in line with the radii reported for the stars of the spectral type of the RG in \citet{2021ApJ...922..163V}: $79.3\pm14.15~R_{\odot}$ for M2III and  $75.87\pm20.28~R_{\odot}$ for M3III. Thus, it is possible that the distance derived from the \textit{Gaia} EDR3 parallax is too large. The \textit{Gaia} RUWE for the system is 1.814, so the parallax is possibly impacted by photocenter motion. Note that the distance derived from the Hipparcos parallax,  $513\pm169~\text{pc}$ \citep{van2007hipparcos} encompasses both the smaller and larger distances discussed here.

The color excess we derive $E(B-V)=0.04^{+0.03}_{-0.02}$ is close to the value from the Bayestar19 map we used as a prior \citep{bayestar19} ($E(B-V)=0.044$) and in line with literature values. \citet{2023MNRAS.526..918G} reported $E(B-V)<0.07\pm 0.01$ and \citet{1992Vogel} used $E(B-V)=0.05$ from \citet{1991A&A...248..458M}. 

The effective temperature we derived, $T_{\mathrm{eff}}=3630^{+100}_{-100}$ K, is close to the value of \citet{2016AJ....152....1K} ($T_{\mathrm{eff}}=3730\pm100$ K), which was determined using the color-temperature calibration of \citet{2011ApJS..193....1W}. It is also possible to compare our result to the effective temperature that can be obtained from the bolometric flux reported in \citet{1992Vogel} ($1.9\times10^{-7}~\text{erg}~\text{cm}^{-2}~\text{s}^{-1}$) using
\begin{equation}
    T_{\mathrm{eff}} = \left( \frac{4 F_{\mathrm{bol}}}{\sigma \theta_{\mathrm{Ross}}^2} \right)^{1/4}\, .
\end{equation}
 \citet{1992Vogel} obtained this value using UV spectra they collected, optical spectra obtained from a collaborator, and literature IR magnitudes, noting that they did not observe changes in UV and optical flux (apart from when the WD was eclipsed) in different epochs. They de-reddened using $E(B-V)=0.05~\text{mag}$ \citep{1991A&A...248..458M}. Using this bolometric flux, de-reddening using $E(B-V)=0.04^{+0.03}_{-0.02}~\text{mag}$ instead, and using $\theta_{\text{LD}}=1.77^{+0.08}_{-0.05}~\text{mas}$ yields $T_{\mathrm{eff}}=3668^{+57}_{-86}~\textrm{K}$. Note that effective temperature is formally defined as the temperature at the Rosseland mean diameter, that is the location where the Rosseland optical depth is one, but \citet{1987A&A...186..200S} note that in continuum wavelengths, the LD diameter can be used as an approximation of the Rosseland mean diameter. This is similar to the approach taken in many optical interferometry papers \citep[e.g.][]{2021ApJ...922..163V}, in which a scaling factor is multiplied by the UD angular diameter to obtain an LD angular diameter and used with bolometric flux to get an effective temperature. For a star of spectral type M2-M3III, the calibration of \citet{2021ApJ...922..163V} suggests an effective temperature range between $T_{\mathrm{eff}}=3439-3709~\text{K}$, when taking the uncertainty limits into account.

With the effective temperature and radius, one can determine the luminosity, which can then be used with the effective temperature and metallicity to infer the masses of the stars in the system using the method described in Section \ref{sec:masses}. We found the luminosity of the RG to be $L=920^{+160}_{-140}~L_{\odot}$ when using the radius determined using a distance of $d=400\pm20~\text{pc}$. As seen in Figure \ref{fig:cornerstarfish}, the metallicity we obtained was poorly constrained. Instead, we adopt [Fe/H]=$-0.540\pm0.100$ which \citet{2023MNRAS.526..918G} obtained from high-resolution IR spectra ($R\sim50,000)$ using MARCS models and stellar parameters close to what we used  ($T_{\text{eff}}=3600$,$\log g=0.5$). After testing various metallicities, we found that while the inferred mass of the giant is moderately sensitive to metallicity, the filling factors remained well below unity.

A final component of the mass-inference process is the selection of a prior for the companion mass. The numerical model of \citet{1992Vogel}  suggests a mass of $M_{\mathrm{hot}} = 0.4~M_{\odot}$ for the WD (note they describe the hot companion as a pre-white dwarf based on their derived luminosity and radius).  \citet{2016AJ....152....1K} used cooling curves to limit the possible range of masses for the WD to be $M_{\mathrm{WD}} = 0.35\text{--}0.55~M_{\odot}$, although this range was influenced by a preference for a lower luminosity and temperature than was found by  \citet{2017Sion}  using FUSE spectra.  \citet{2005Skopal}  uses $M_{\mathrm{WD}} = 0.6~M_{\odot}$. Finally,  \citet{2017Sion} found a mass range of $M_{\mathrm{WD}} \approx 0.4\text{--}0.6~M_{\odot}$. Taking all of these into account, we selected a uniform prior over $M_{\mathrm{WD}} = [0.35, 0.6]~M_{\odot}$. Note that although this encompasses both helium (He) and carbon-oxygen (C-O) WDs, the lack of evidence for \textit{S}-process enhancement in the RG supports the presence of a WD under $0.55~M_{\odot}$  \citep{2017Sion, 2023MNRAS.526..918G}.

We found the masses of the RG and companion to be $M_{\text{G}}=1.29^{+0.23}_{-0.13}~M_{\odot}$ and $M_{WD}=0.38^{+0.04}_{-0.02}~M_{\odot}$, using distance $d=400\pm20~\text{pc}$. These values are within the range found by \citet{2016AJ....152....1K} and agree roughly with Table 2 in \citet{1992Vogel}. The resulting filling factor is $f=0.53^{+0.04}_{-0.04}$. If we instead use the distance from \citet{gaiadist}, we end up with $M_{\text{G}}=1.51^{+0.46}_{-0.26}~M_{\odot}$, $M_{WD}=0.42^{+0.07}_{-0.04}~M_{\odot}$ and $f=0.72^{+0.06}_{-0.06}$ thanks to the larger radius, $R=115^{+5}_{-5}~R_{\odot}$, found using the larger distance. Clearly, an improved solution for the \textit{Gaia} parallax would help resolve the question of the actual radius of the RG. However, regardless of these variations, the result is always that the RG is far from filling its Roche lobe. Even adding 20\% to the \citet{gaiadist} distance results in $f=0.81^{+0.08}_{-0.07}$. These values are also in agreement with the analysis presented in Section \ref{sec:fits}, which supports a spherical shape for the RG.

\citet{1997Wilson} reported ellipsoidal variations in the light curve of the system, suggesting that the RG was tidally distorted and filling $82-100\%$  of its Roche lobe.  As noted in Section \ref{sec:intro}, \citet{2021shagatova} and \citet{skopal_windtransfer} suggested that a focused wind could instead be the source of an ellipsoidal variation and, indeed, \citet{2016shagatova,shagatova_egand} found evidence of a wind focused toward the orbital plane in EG And using hydrogen column densities and radial velocities of \ion{Fe}{1} absorption lines, respectively.

We observed the system at phase $\phi=0.21$ from the time of spectroscopic conjunction as defined in  \citet{2016AJ....152....1K}, when any elongation would be most easily observed, particularly when the inclination, $i=80^{\circ}$, is also taken into account. Given the evidence for wind-directed overflow in \citet{shagatova_egand}, observations of this system at different phases and in multiple band passes would be beneficial to understanding this process of mass transfer. A full analysis of the system using observations at different phases will be described in a forthcoming work by this team.

\subsection{BD Cam} \label{subsubsec:Cam params} 
BD Cam is an accretion-powered system \citep{2024A&A...689A..86L} consisting of an S3.5/2 RG \citep{bdcamtype}, with the symbiotic nature of the system and presence of a hot companion first identified via IUE spectra \citep{bdcamwd2,bdcamwd1}.  An orbital solution was determined by \citet{1984Obs...104..224G}, who found a period of around 596 days and a nearly circular orbit. The spectrum of the RG presents no detectable Technetium (Tc) \citep{1991ApJ...381..278V,1993A&A...271..180G,1952merrill,1968ApJ...152L..13D,1988ApJ...333..219S,1987AJ.....94..981L}, which has a half-life on the order of the lifetime of the thermally-pulsing
asymptotic giant branch (TPAGB) \citep{2000A&A...360..196V}. Therefore the RG belongs to the class of extrinsic S stars, in which the higher abundance of S-process elements originates in mass exchange from an AGB companion that has since evolved to a WD, rather than from nucleosynthesis processes within the RG \citep[e.g.][]{1988A&A...198..187J,1993A&A...271..463J,1998A&A...332..877J,2000A&A...360..196V,1990AJ.....99.1930B}. Orbital modulation in UV flux and \ion{He}{1} $\lambda1083.0~\text{nm}$ suggest that the wind of the RG is funneled through its inner Lagrangian point \citep{1991AJ....101.1483D,1992A&A...255..215S,1994ASPC...56..413A}.

We measured an LD angular diameter of $\theta_{\text{LD}}=5.36^{+0.08}_{-0.06}$ mas for the RG in BD Cam. This, and the uniform-disk angular diameter we measured, $\theta_{\text{UD}}=5.22\pm 0.04~\text{mas}$, are less than the values determined by \citet{npoidiam} using observations from the Navy Precision Optical Interferometer (NPOI) that spanned 430 to 860 nm and were recorded at various orbital phases from 2002 to 2006: $\theta_{\mathrm{UD}}=5.314\pm0.014$ mas and $\theta_{\mathrm{LD}}=5.857\pm0.016$. These authors used a different limb-darkening law than we did, selecting a coefficient from \citet{claretld}. The larger angular diameters measured by NPOI could be the result of the different wavelength range used, which includes portions with strong molecular bands in stars of BD Cam's spectral type, as opposed to the \textit{H}-band data reported in this paper, which covers mainly continuum regions.

From the spectral fits, we determined an effective temperature of $T_{\text{eff}}=3600^{+120}_{-120}~\text{K}$. Using the distance from \citet{gaiadist} ($235.4\pm13.4$ pc), the radius is $R=136^{+8}_{-8}~R_{\odot}$ and luminosity $L=2780^{+530}_{-450}~L_{\odot}$. \citet{npoidiam} used a lower luminosity, resulting in a much cooler derived effective temperature: $T_{\text{eff}} = 2807\pm43$ K given $L = 1219.6\pm 0.1L_\odot$ from \citet{2017MNRAS.471..770M}. However, if we take the luminosity from \citet{2017MNRAS.471..770M}, which the authors of that paper derived using a \textit{Gaia} DR1 distance that is lower than the one we used, and update it with the distance we used, we find  $T_{\text{eff}}=3557 \pm 126~\text{K}$. This effective temperature is also close to that found by \citet{2026A&A...709A.143A} using fits of COMARCS stellar atmosphere models to a spectral energy distribution (SED): $T_{\text{eff}}=3500^{+100}_{-200}~\text{K}$.

The extinction we derive from the fit ($A_{V}=0.13^{+0.15}_{-0.09}~\text{mag}$) yields a color correction $E(B-V)=0.04^{+0.03}_{-0.03}~\text{mag}$, in line with the Bayestar19 map ($E(B-V)=0.082$) and the value assumed by \citet{ortiz2019} ($E(B-V)=0.1~\text{mag}$), who noted that the \textit{GALEX} value $(E(B-V) = 0.987~\text{mag}$) was likely an anomaly. Because the metallicity we derived from the \textit{Starfish} fits was not well constrained, in the following mass-inference process, we adopted [Fe/H]=$-0.200\pm0.100$ from \citet{2020AandA...633A..34C}.

We found $M_{\text{G}}=2.34^{+0.85}_{-0.67}~M_{\odot}$ and $M_{WD}=0.75^{+0.16}_{-0.14}~M_{\odot}$ leading to a time-averaged filling factor $f=0.65^{+0.09}_{-0.07}$. The RUWE for the \textit{Gaia} EDR3 parallax of BD Cam is the largest of all the stars analyzed in this paper. The mass of the RG ranges from $M_{\text{G}}=1.77^{+0.64}_{-0.57}~M_{\odot}$ for a $20\%$ reduction in the distance to $M_{\text{G}}=2.85^{+1.06}_{-0.70}~M_{\odot}$ for an increase in distance by $20\%$ and the filling factor ranges from $f=0.581^{+0.09}_{-0.07}$ to $f=0.72^{+0.09}_{-0.08}$ for the same range of distances, so it is not likely that the RG is filling its Roche lobe. This is also supported by the fits done in Section \ref{sec:fits}, which favored a more spherical shape. The derived mass for the RG is also in line with prior literature values. \citet{2019lebzelter} estimated a lower bound for the giant mass of $M \geq 1.5 \, M_\odot$ using line ratios and \citet{2024A&A...691A..98K} inferred a mass of $M_{\text{G}}=2.68~M_{\odot}$ from Gaia DR3 photometry.

Taking the orbital elements in Table \ref{tab:orbital} and the time of observation into account, this system was observed at phase $\phi=0.65$, between quadrature and conjunction, so elongation would only be partially visible. Imaging does suggest surface inhomogeneities, but these are expected in stars of BD Cam's evolutionary state. 

\subsection{SU Lyn}
SU Lyn was first identified as an accretion-powered symbiotic star by \citet{2016mukai} via \textit{Swift} X-Ray observations and interpretation of UV and optical spectra. This was later confirmed by additional UV, optical and near-IR spectra reported by \citet{2021kumar}. \citet{2022Ilkiewicz} noted that observational phenomena indicate interacting behaviors in the system are not persistent and that it appears as a classical symbiotic system at times and as one of the ``hidden'' symbiotic systems proposed by \citet{2016mukai} that lack detectable emission lines at other times.

We measured an LD angular diameter of $\theta_{\text{LD}}=3.39^{+0.06}_{-0.05}~\text{mas}$, which corresponds to a radius $R=263^{+14}_{-12}~R_{\odot}$ using a distance $719.74^{+38.18}_{-30.32}~\text{pc}$ from \citet{gaiadist}. The RUWE for the system is the lowest of those in this sample. The spectral type of the RG is M5.8III \citep{2016mukai}. \citet{2021ApJ...922..163V} found average radii between $R=145.59\pm 35.54~R_{\odot}$ for M5.5III and $R=162.74\pm24.84~R_{\odot}$ for M6III.  Using an H-R diagram from \citet{2007AJ....134.1089S}, \citet{2016mukai} found the absolute magnitude of the RG as $M_{V} = -0.83$ mag, leading to a distance $d= 640\pm100~\text{pc}$, which leads to $R=235 \pm 37~R_\odot$ if using the LD angular diameter we measured. \citet{2022Ilkiewicz} found a radius $R=280\pm40~R_{\odot}$ from SED fitting and use of the \citet{gaiadist} distance. Thus, the radius we measured falls between those currently reported in the literature and within the error bounds of those radii.

The extinction we derived $E(B-V)=0.05^{+0.04}_{-0.03}~\text{mag}$ is close to the Bayestar19 map value $E(B-V)=0.044~\text{mag}$ \citep{bayestar19}. \citet{2016mukai} found $ E(B-V) = 0.07~\text{mag}$ around JD 2457409 ($\approx$ 2016-01-22) but \citet{2022Ilkiewicz} note a variability due to changing circumstellar extinction with $E(B-V)=0.01~\text{mag}$ at JD = 2456000 ($\approx$ 2012-03-14) and a maximum $E(B-V)=0.18~\text{mag}$ at JD = 2457841 ($\approx$ 2017-03-28). The value we measure suggests the system had returned to the baseline value found by \citet{2016mukai}.
 
The temperature we measured is $T_{\text{eff}}=3230^{+60}_{-50}~{K}$. This value is less than that found by \citet{2019Akras} ($T_{\text{eff}}= 3565~\text{K}$) but close to that found by \citet{2022Ilkiewicz} using SED fitting ( $T_{\text{eff}}=3200\pm100~\text{K}$) and close to the $T_{\text{eff}}=3242\pm117$ suggested by the spectral type using the scale of \citet{2021ApJ...922..163V}. The luminosity we measure $L=6800^{+890}_{-750}~L_{\odot}$ is within the range found by \citet{2022Ilkiewicz} from SED fitting ($L=7700 \pm 1000 ~L_{\odot}$) and corresponds to the report by those authors that the RG is an early-stage AGB star going through its thermal pulsation period.

Because we lack an orbit for the system, we were unable to apply the same mass-inference approach as we used for the other stars in this paper. Instead, we ran an inference with only the measured parameters as priors for fits to MIST. Because the metallicity we measured was not well constrained and was high ([Fe/H]=$0.35^{+0.11}_{-0.14}$), we used the same approach as for V1472 Aql, using a prior based on the Metallicity Distribution Function (MDFs) in Table 2 of \citet{2015ApJ...808..132H}, for the location of SU Lyn ($R_{\text{gal}}\approx9~\text{kpc}$, $z\approx278~\text{pc}$). This resulted in using a normal distribution $\mathcal{N}(+0.02,0.20)$, truncated to the MIST grid [-1.0, +0.5].

We infer a mass of $M_{\text{G}}=2.69^{+0.80}_{-0.58}~M_{\odot}$ for the RG. \citet{2022Ilkiewicz} used MIST and parameters derived via an SED fit to estimate the mass of the giant as $M_{RG} = 2.78^{+0.62}_{-0.64} \; M_\odot$. Without an orbit we are unable to determine the mass of the WD, although we note that \citet{2018ApJ...864...46L} placed a lower limit of $M_{\text{WD}}=0.7~M_{\odot}$ on the mass of the WD, and \citet{2021kumar} reported $M_{\text{WD}} = 0.85~M_\odot$. Using the full range of possible WD masses, the mass ratio of the system ranges $q=\frac{M_{\text{G}}}{M_{\text{WD}}} = 3.2-3.84$  and the Roche lobe of the giant ranges $R_{l}=0.481-0.498~\text{a}$ for a circular orbit, where $\text{a}$ is the orbital separation. \citet{2022Ilkiewicz} suggest that the orbit of the system is highly eccentric, in which case RLOF, if it happens, would most likely occur during periastron. Model fitting favored a spherical object but this could also be the result of the phase during which the data were collected, rather than evidence of a lack of Roche-lobe filling throughout the orbit.

\citet{2022Ilkiewicz} analyzed existing literature photometry, finding evidence for a weak quasi-periodicity, in line with \citet{2021MNRAS.505.6121M}. \citet{2021MNRAS.505.6121M} suggest the variability could be linked to convection, and the image reconstruction of the RG in Figure \ref{fig:imagessphere} does show evidence of the large features that have been linked to convection in other AGBs like $\pi^{1}~\text{Gruis}$ \citep{2018Natur.553..310P}. Both \citet{2022Ilkiewicz} and \citet{2021MNRAS.505.6121M} find a very low radial velocity variability ($<2~\text{km s}^{-1}$) and \citet{2021MNRAS.505.6121M} note that the orbital plane is likely close to face-on. Thus, little is known about the orbit of the RG and like \citet{2022Ilkiewicz}, we stress the need for photometric and spectroscopic monitoring of this system.

\section{Discussion} \label{sec:discu}

Two of our targets have been the subject of much research on Roche filling factors in symbiotic stars. EG And \citep{2007Rutkowski, 2007Mikolajewska} and V1472 Aql \citep{1997samus} have both been observed to have strong ellipsoidal variations in their light curve. This has been interpreted by many including \citet{1997Wilson} and \citet{2014Boffin} to be a result of the giant filling its Roche lobe. However, recent analysis by \citet{2015skopal} and \citet{2016shagatova, 2021shagatova} indicates that the ellipsoidal variations in EG And are an indicator of wind-related mass transfer instead of RLOF. In addition \citet{Merc2025I} and \citet{Merc2025II} have shown that other symbiotic systems with ellipsoidal variations in their light curves are not filling their Roche lobes. \citet{Merc2025I} also suggested that slow dense winds could be the source of ellipsoidal variations and noted that \citet{2009A&A...507..891D} found that radiation or pulsation-driven winds can decrease the effective Roche lobe.

For the three systems with constrained orbits in our sample, we infer time-averaged filling factors below unity. Considering only confirmed symbiotic systems, EG And and BD Cam add two underfilling systems to the previously published interferometric sample from  \citet{2014Boffin}, \citet{Merc2025I}, \citet{Merc2025II}, and \citet{2026ApJ..1004L..13N}. Thus, 14 of the 17 confirmed symbiotic stars observed with optical interferometers have been found to underfill their Roche lobes; note that SU Lyn currently lacks the constraints required to determine its Roche-lobe radius.

\citet{Merc2025I} and \citet{Merc2025II} found that many RGs in symbiotic systems are larger and more luminous than typical luminosity class III field giants. Of the systems reported here, SU~Lyn most clearly corroborates this finding: 
 we find a radius and luminosity that exceed the calibration values for M5III and M6III stars from \citet{2021ApJ...922..163V}, consistent with its classification as a TP-AGB star \citep{2022Ilkiewicz} as shown in Figure~\ref{fig:hrdiag}. EG~And, by contrast, has parameters consistent with values given for an M2--M3III giant in \citet{2021ApJ...922..163V}. V1472~Aql falls below the typical M2III values reported in \citet{2021ApJ...922..163V}, although this may reflect a systematically low distance given its high \textit{Gaia} RUWE. This heterogeneity leads us to agree with \citet{Merc2025I} that comparisons between symbiotic RGs and standard luminosity class III stars should be made with care.
 
Our data represent only one epoch of observations on systems with long orbital periods. Observations at different phases  can further investigate the mechanism of mass transfer in symbiotic systems and test for the presence of either RLOF or WRLOF. In addition, multi-wavelength observations would offer information about the frequency-dependence of any geometric distortions that are uncovered through imaging. Distortions that are stronger or more pronounced at a longer wavelength would indicate that cooler gas is filling that space and could be a strong indicator in favor of a wind-driven mass transfer mechanism. A forthcoming paper will present observations of these systems in the \textit{K} band.

\section{Conclusion} \label{sec:conc}

In this paper, we present angular diameters and high-resolution images of giants and AGBs in symbiotic and related systems. Combining the interferometric radii with stellar and orbital constraints, we infer time-averaged Roche-lobe filling factors below unity for the three systems with constrained orbits. In addition, although a Roche-lobe radius cannot be determined for SU Lyn at the moment due to the lack of a measured orbital period, model fitting favors an LD disk over an elongated shape. Closure phases of these objects present evidence of deviation from centrosymmetry and imaging supports the presence of surface features in at least three of these stars: V1472 Aql, BD Cam, and SU Lyn.

These results provide evidence against Roche-lobe filling at the observed epochs in the systems with constrained Roche lobes. However, they also motivate future study of these systems with optical interferometry at different phases. Additionally, these observations fail to rule out the presence of stronger distortions in a band such as the \textit{K} band utilized by both the CHARA MYSTIC and VLTI GRAVITY beam combiners. Different radii at different wavelength bands could be indicative of a temperature gradient across the face of the star that could indicate the presence of a focused wind. Finally, the degree to which these four stars typify standard symbiotic stars in terms of mass-transfer mechanism is unclear. The three confirmed symbiotic systems in our sample are accretion-powered, whereas many symbiotic systems are shell-burning \citep{2019Merc}. 

These results build upon those of past work, showing that many symbiotic systems are not experiencing mass transfer via RLOF. There are several steps we recommend in future applications of optical interferometry to the study of symbiotic stars. The first is long-term interferometric monitoring of previously observed systems. In particular, by observing objects at several different phases we can better constrain the dependence of the shape of the giant on orbital phase. Variability in the shape of a symbiotic star over time could be an indication of the presence of a Roche lobe or other distortion pointing towards the WD. Secondly, simultaneous multi-wavelength interferometric imaging across different bands is crucial for helping to characterize the nature of any distortions found and identifying potential wind-related features. Finally, a broader population of symbiotic stars should be observed with optical interferometers. Of the systems listed as confirmed Galactic symbiotics in the New Online Database of Symbiotic Variables, there are roughly 32 symbiotic systems that can be observed and resolved by the CHARA Array and 17 by the VLTI. This number is based on declination and magnitude limits, as well as  angular diameter estimates from $V$ and $K$ photometry using the surface brightness relation of \citet{2021ApJ...922..163V} with distances from \citet{gaiadist}. Forthcoming work will undertake some of these steps, including analysis of the systems reported in this paper in the $K$ band and analysis of observations of several additional systems. Understanding symbiotic stars and their evolution depends on determining the mechanism of mass transfer in these systems and optical interferometry is demonstrating its strong potential to help answer that question.

\begin{acknowledgements}
We thank the anonymous referee for their helpful review and comments. This material is based upon work supported by the National Science Foundation under Grant Number 2213518. MOH was supported by the Polish National Science Centre grant
2019/32/C/ST9/00577. TG acknowledges funding in support of this project that was received via an award from the New Mexico Space Grant Consortium. TG obtained data as a Visiting Astronomer at the Infrared Telescope Facility, which is operated by the University of Hawaii under contract 80HQTR19D0030 with the National Aeronautics and Space Administration. This work is based upon observations obtained with the Georgia State University Center for High Angular Resolution Astronomy Array at Mount Wilson Observatory. RMR acknowledges support from the Heising-Simons Foundation's 51 Pegasi b Fellowship Program. The CHARA Array is supported by the National Science Foundation under Grant No. AST-1636624, AST-2034336, and AST 2407956.  Institutional support has been provided from the GSU College of Arts and Sciences, Office of the Provost, and Office of the Vice President for Research and Economic Development. SK acknowledges funding for MIRC-X from the European Research Council (ERC) under the European Union's Horizon 2020 research and innovation programme (Starting Grant No. 639889 and Consolidated Grant No. 101003096). JDM acknowledges funding for the development of MIRC-X (NASA-XRP NNX16AD43G, NSF-AST 1909165) and MYSTIC (NSF-ATI 1506540, NSF-AST 1909165).
Time at the CHARA Array was granted through the NOIRLab community access program (NOIRLab 2021B-0254; PI T. Gaudin).
 This research has made use of the Jean-Marie Mariotti Center \texttt{SearchCal} service, which involves the JSDC and JMDC catalogues
\footnote{Available at https://www.jmmc.fr/searchcal}. This research has made use of the Jean-Marie Mariotti Center \texttt{Aspro}
service \footnote{Available at http://www.jmmc.fr/aspro}.
 RN acknowledges the use of Claude Opus 4.5 (Anthropic) and Gemini 3.1 Pro (Google) for assistance with development of Julia scripts, improving plot formatting,  table formatting, and proofreading. No part of the technical analysis or data interpretation was performed by these models.

\end{acknowledgements}




%
\facilities{CHARA(MIRCX), IRTF (SpeX)}

\software{OITOOLS, PMOIRED, ROTIR, Starfish, SURFING}

\bibliography{symbiotic}{}
\bibliographystyle{aasjournal}
\appendix
\restartappendixnumbering 
\section{Method for Inferring Mass and Filling Factors}\label{app:mass}
\subsection{Algorithm}
For each target we proceed as follows.

\begin{enumerate}

\item We draw $\theta_{\rm LD}$, $d$, and $T_{\rm eff}$ from asymmetric two-sided Gaussian approximations,
$\mathcal{A}(x,\sigma_{-},\sigma_{+})$, constructed from their reported median and 16th/84th-percentile uncertainties. The metallicity is drawn from the target-specific prior $P_{\rm [Fe/H]}$ listed in Table~\ref{tab:priors}. For systems with orbital solutions, we also draw $M_2$, $i$, $P$, and $e$ from the target-specific prior or measurement distributions
listed in Table~\ref{tab:priors}; the measured mass functions and their uncertainties are taken from Table~\ref{tab:orbital}.

\item We compute $R$ and $L$ from the angular diameters, adopted distances, and effective temperatures from the spectral fits. We then infer $M_G$ using a kernel-weighted interpolation over the
$k=16$ nearest MIST giant-phase grid points in
$(\log T_{\rm eff},\log L,[\mathrm{Fe/H}])$ space. We use the MIST current stellar mass at each evolutionary grid point rather than the MIST initial mass so that mass loss along the evolutionary track is included in the inferred present-day giant mass.

\item For systems with measured spectroscopic mass functions, we apply the mass-function constraint as an importance weight. For each sample,

\begin{equation}
f(m)_{\rm pred}^{(i)} =
\frac{\left(M_2^{(i)}\sin i^{(i)}\right)^3}
{\left(M_G^{(i)}+M_2^{(i)}\right)^2},
\end{equation}

and the importance weight is

\begin{equation}
W^{(i)} =
\exp\left[
-\frac{1}{2}
\left(
\frac{f(m)_{\rm pred}^{(i)}-f(m)_{\rm obs}}
{\sigma_{f(m)}}
\right)^2
\right].
\end{equation}

\item We resample the complete set of Monte Carlo quantities using probabilities proportional to $W^{(i)}$. The resulting equal-weight
samples preserve the covariance among $M_G$, $M_2$, $R$, $d$, $T_{\rm eff}$, $[\mathrm{Fe/H}]$, and inclination. The radius carried through this resampling is therefore conditioned on the MIST and mass-function constraints, and is used only internally when computing
the correlated Roche-lobe quantities. The stellar radii and luminosities quoted in Table~\ref{tab:stellarparams} are instead the
direct observational values obtained before application of the mass-function constraint.

\item We compute the Roche-lobe quantities from each matched posterior sample. The Keplerian semimajor axis is 
\begin{equation}
a_{\rm sma}^{(i)} =
\left[
\frac{G\left(M_G^{(i)}+M_2^{(i)}\right)P^2}
{4\pi^2}
\right]^{1/3}.
\end{equation}

For the eccentric systems we use the orbit-averaged separation

\begin{equation}
\langle r\rangle^{(i)} =
a_{\rm sma}^{(i)}
\left(1+\frac{\left(e^{(i)}\right)^2}{2}\right),
\end{equation}

which reduces to $a_{\rm sma}$ for a circular orbit. The
orbit-averaged Roche-lobe radius is then

\begin{equation}
R_L^{(i)} =
\langle r\rangle^{(i)}
\frac{0.49q^{2/3}}
{0.6q^{2/3}+\ln\left(1+q^{1/3}\right)},
\qquad
q=\frac{M_G^{(i)}}{M_2^{(i)}}.
\end{equation}

The filling factor is calculated from the matched posterior samples,

\begin{equation}
f^{(i)} =
\frac{R_{\rm joint}^{(i)}}{R_L^{(i)}}.
\end{equation}

\end{enumerate}

For SU Lyn, which lacks an orbital solution, the mass-function and Roche-lobe steps are omitted. Its giant mass is obtained directly from the MIST inference described above. Reported central values and
asymmetric uncertainties are the 50th percentile and the 16th/84th percentile intervals, respectively.

\section{Additional Plots}\label{sec:all plots}
The following appendix provides additional plots referenced in the paper. Figure \ref{fig:uv_coverage} presents plots of the (\textit{u},\textit{v}) coverage for the observations reported here. Figure \ref{fig:v2all} presents plots of the squared visibilities, V$^{2}$, for the observations. Figure \ref{fig:t3phiall} presents plots of the closure phase for the observations. Figures \ref{fig:sept_spectra} and \ref{fig:nov_spectra} present the IRTF spectra used in this paper. Figure \ref{fig:cornerstarfish} presents the corner plots for the Starfish parameter determinations. Figure \ref{fig:spectrastarfish} presents a comparison of the best fitting model spectra determined with Starfish to the observed spectra for each star. Figure \ref{fig:masscorner} presents corner plots from the mass-inference procedure. Figure \ref{fig:hrdiag} presents HR diagrams from the mass-inference procedure.
\input{uvplotsall}
\input{v2all}
\input{t3phiall}

\input{allspectra}

\input{cornerstarfish}
\input{spectrastarfish}
\input{masscorner}
\input{hrdiagrams}



\end{document}

%% file: observations.tex
\begin{deluxetable}{lcccc}[h]
    \tablewidth{0pt}
    \tablecaption{CHARA Array Observation Log \label{tab:int_log}}
     
        \tablehead{
       \colhead{Date (UT)} & \colhead{Target} & \colhead{N$_{telescopes}$} & \colhead{N$_{brackets}$} & \colhead{Calibrator(s)}}
     \startdata
        2021-09-20 & V1472 Aql (HD 190658) & 6 & 2 & HD 201601 \\
        2021-09-21 & V1472 Aql (HD 190658) & 6 & 1 & HD 195810, HD 201601 \\
        2021-09-21 & EG And (HD 4174) & 6 & 2 & HD 219080, HD 571 \\
        2021-09-21 & BD Cam (HD 22649) & 5 & 2 & HD 26553, HD 29826 \\
        2021-09-22 & BD Cam (HD 22649) & 5 & 1 & HD 20509, HD 29826 \\
        2021-09-22 & SU Lyn (HD 47648) & 6 & 2 & HD 43352, HD 55575 \\
    \enddata
\tablecomments{The number of brackets reported refers to ``cal-target-cal" brackets that are standard to observations with CHARA and involve observation of small calibrator targets both before and after observing the science target. }
\end{deluxetable}

%% file: cals.tex
\begin{deluxetable}{lccccc}[ht]
    \tablewidth{0pt}
     \tablecaption{Interferometric Calibrators  \label{tab:int_log_cal}}
     \tablehead{   \colhead{Calibrator} & \colhead{Target} & \colhead{RA (J2000)} & \colhead{Dec (J2000)} & \colhead{\textit{H} Band UD (mas)} & \colhead{Source }}
       \startdata
       HD 43352 & SU Lyn ( HD 47648) & 06 19 10.0 & $+$56 31 35.7 & 0.735 $\pm$ 0.047 & SearchCal \citep{searchcal} \\
       HD 29826 & BD Cam (HD 22649) & 04 46 24.5 & $+$64 48 37.1 & 0.681 $\pm$ 0.054 & SearchCal \citep{searchcal} \\
       HD 219080 & EG And (HD 4174) & 23 12 33.0 & $+$49 24 22.3 & 0.676 $\pm$ 0.047 & SearchCal \citep{searchcal} \\
       HD 20509 & BD Cam (HD 22649) & 03 20 22.4 & $+$55 32 13.5 & 0.658 $\pm$ 0.059 & SearchCal \citep{searchcal} \\
       HD 55575 & SU Lyn ( HD 47648) & 07 15 50.1 & $+$47 14 23.9 & 0.593 $\pm$ 0.050 & SearchCal \citep{searchcal}\\
       HD 571 & EG And (HD 4174) & 00 10 19.2 & $+$46 04 20.2 & 0.591 $\pm$ 0.041 & SearchCal \citep{searchcal} \\
       HD 201601 & V1472 Aql (HD 190658) & 21 10 20.5 & $+$10 07 53.6 & 0.560 $\pm$ 0.016 & \citep{2011Perrault} \\
       HD 26553 & BD Cam (HD 22649) & 04 15 01.8 & $+$57 27 37.3 & 0.500 $\pm$ 0.045 & SearchCal \citep{searchcal} \\
       HD 195810 & V1472 Aql (HD 190658)  & 20 33 12.8 & $+$11 18 11.7 & 0.350 $\pm$ 0.050 & \citep{2018Gardner} \\               
    \enddata
\end{deluxetable}

%% file: sulynobs.tex
\begin{figure}[!h]
    \gridline{\fig{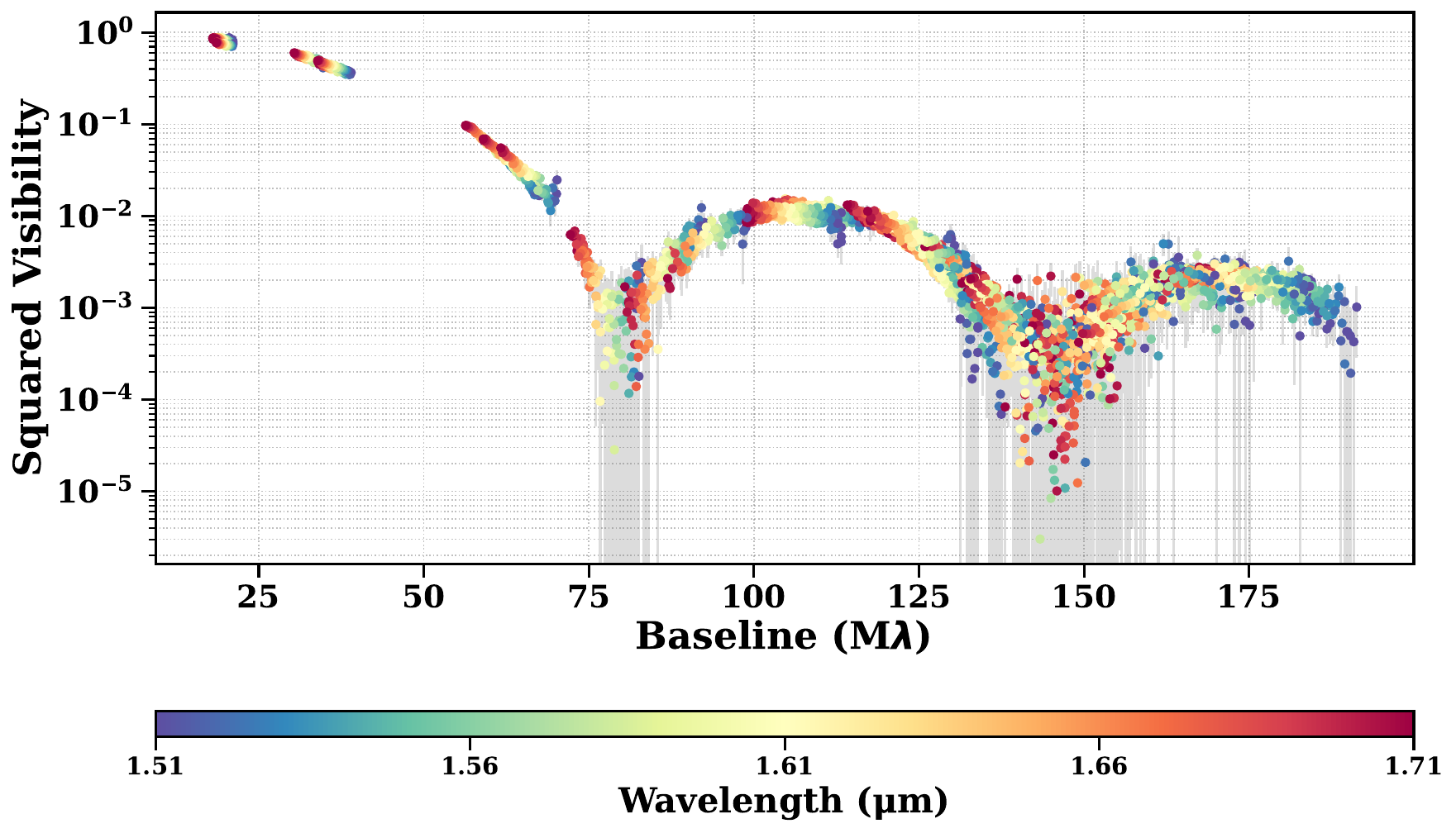}{0.5\textwidth}{a) Squared visibility data for target SU Lyn}}
        \gridline{\fig{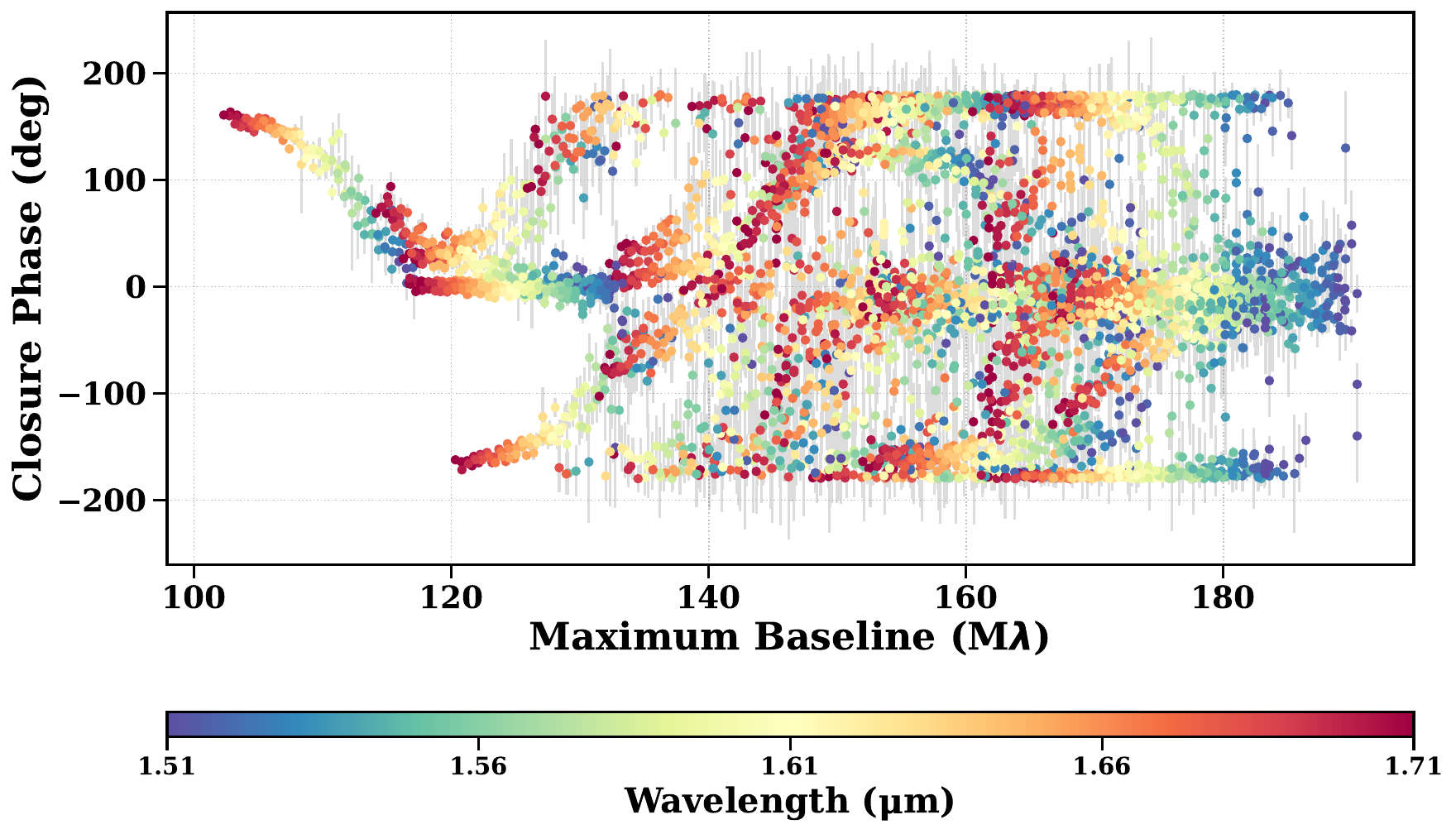}{0.5\textwidth}{b) Closure phase data for target SU Lyn}}
     \caption{An example of the processed data used in the analysis of interferometry data from CHARA. Both plots depict data collected on SU Lyn. a) shows the squared visibility amplitude data b) shows the closure phase data. Data was collected with MIRC-X on the CHARA Array. Plots for all targets are available in the Appendix.}
        \label{fig:su lyn}
\end{figure}

%% file: irtfobs.tex
\begin{deluxetable}{lcccc}
     \tablewidth{0pt}
      \tablecaption{NASA IRTF Observation Log \label{tab:spec log}}
        \tablehead{\colhead{Date (UTC)} & \colhead{Target} & \colhead{Mode} & \colhead{AB Pairs} & \colhead{Standard (Sp. Type)}}
        \startdata
        2021-09-02 & V1472 Aql (HD 190658) & SXD & 4 & HD 182919 (A0V) \\
        2021-09-02 & V1472 Aql (HD 190658) & LXD & 4 & HD 182919 (A0V) \\
        2021-09-02 & EG And (HD 4174) & SXD & 4 & HD 1561 (A0Vs) \\
        2021-09-02 & EG And (HD 4174)& LXD & 4 & HD 1561 (A0Vs) \\
        2021-09-02 & BD Cam (HD 22649) & SXD & 8 & HD 19844 (A0V) \\
        2021-09-02 & BD Cam (HD 22649) & LXD & 4 & HD 19844 (A0V) \\
        2021-11-26 & EG And (HD 4174) & SXD & 8 & HD 1561 (A0Vs) \\
        2021-11-26 & EG And (HD 4174) & LXD & 4 & HD 1561 (A0Vs) \\
        2021-11-26 & BD Cam (HD 22649) & SXD & 8 & HD 19844 (A0V) \\
        2021-11-26 & BD Cam (HD 22649) & LXD & 4 & HD 19844 (A0V) \\
        2021-11-26 & SU Lyn (HD 47648) & SXD & 16 & HD 38831 (A0Vs) \\
        2021-11-26 & SU Lyn (HD 47648) & LXD & 8 & HD 38831 (A0Vs) \\
 \enddata
\end{deluxetable}

%% file: modelfits.tex
\begin{deluxetable*}{ccccccc}
\tablecaption{Model Fits \label{tab:model_fits}}
\tablehead{
    \colhead{Model} & \colhead{Diameter (mas)} & \colhead{$\alpha$} & \colhead{Axis-Ratio} & \colhead{Position Angle (degrees)} & \colhead{$\chi_{\nu}^2$} & \colhead{log(Z)}
}
\startdata
\multicolumn{7}{c}{V1472 Aql} \\
\hline
uniform disk & $2.36^{+0.04}_{-0.03}$ & -- & -- & -- & 15.61 & -14.00 \\
limb darkened disk & $2.43^{+0.05}_{-0.05}$ & $0.38^{+0.17}_{-0.15}$ & -- & -- & 8.24 & -11.18 \\
elongated disk & $2.41^{+0.06}_{-0.05}$ & -- & $0.96^{+0.03}_{-0.04}$ & $132^{+35}_{-115}$ & 13.91 & -16.21 \\
hybrid disk & $3.10^{+0.18}_{-0.11}$ & -- & $0.92^{+0.05}_{-0.07}$ & $33^{+45}_{-14}$ & 28.07 & -26.36 \\
\hline
\multicolumn{7}{c}{EG And} \\
\hline
uniform disk & $1.69^{+0.03}_{-0.03}$ & -- & -- & -- & 2.10 & -7.69 \\
limb darkened disk & $1.77^{+0.08}_{-0.05}$ & $0.25^{+0.29}_{-0.17}$ & -- & -- & 2.00 & -7.91 \\
elongated disk & $1.72^{+0.05}_{-0.04}$ & -- & $0.96^{+0.02}_{-0.03}$ & $90^{+46}_{-50}$ & 1.80 & -10.16 \\
hybrid disk & $2.26^{+0.08}_{-0.06}$ & -- & $0.95^{+0.03}_{-0.05}$ & $46^{+84}_{-23}$ & 28.16 & -27.24 \\
\hline
\multicolumn{7}{c}{BD Cam} \\
\hline
uniform disk & $5.22^{+0.04}_{-0.04}$ & -- & -- & -- & 12.64 & -12.39 \\
limb darkened disk & $5.36^{+0.08}_{-0.06}$ & $0.21^{+0.09}_{-0.08}$ & -- & -- & 5.78 & -10.89 \\
elongated disk & $5.54^{+0.26}_{-0.22}$ & -- & $0.93^{+0.04}_{-0.04}$ & $159^{+13}_{-131}$ & 8.05 & -14.12 \\
hybrid disk & $7.56^{+0.27}_{-0.25}$ & -- & $0.90^{+0.04}_{-0.04}$ & $33^{+18}_{-14}$ & 25.71 & -24.04 \\
\hline
\multicolumn{7}{c}{SU Lyn} \\
\hline
uniform disk & $3.20^{+0.02}_{-0.02}$ & -- & -- & -- & 32.42 & -23.01 \\
limb darkened disk & $3.39^{+0.06}_{-0.05}$ & $0.45^{+0.12}_{-0.10}$ & -- & -- & 7.09 & -11.37 \\
elongated disk & $3.23^{+0.05}_{-0.03}$ & -- & $0.98^{+0.01}_{-0.02}$ & $95^{+39}_{-30}$ & 31.13 & -26.09 \\
hybrid disk & $4.78^{+0.10}_{-0.10}$ & -- & $0.68^{+0.03}_{-0.03}$ & $128^{+4}_{-4}$ & 76.21 & -51.06 \\
\hline
\enddata
\tablecomments{Model fit results for each target star based on visibility curve data. The log(Z) listed here is the Bayesian evidence. The complex visibility function that describes each model can be found in Section \ref{sec:models}}
\end{deluxetable*}

%% file: modelfits_sim.tex
\begin{deluxetable*}{lccccccc}
\tablecaption{Detectability of Roche lobe geometry.\label{tab:modelfitssim}}
\tablehead{
    \colhead{System} & \colhead{Type/Fillout} & \colhead{Axis Ratio} &
    \colhead{$\chi_{\nu}^2$ UD} & \colhead{$\chi_{\nu}^2$ LD} & \colhead{$\chi_{\nu}^2$ ellipse} &
    \colhead{$\Delta\log(Z)$} & \colhead{Interpretation}
}
\startdata
\multicolumn{8}{c}{Simulated Roche filling star} \\
\hline
Roche & $f = 0.50$ & 0.995 & 1.47 & 1.44 & 1.02 & 4.46 & Not detectable \\
Roche & $f = 0.60$ & 0.992 & 2.02 & 2.02 & 1.00 & 3.93 & Not detectable \\
Roche & $f = 0.70$ & 0.988 & 2.14 & 2.09 & 1.01 & 3.89 & Not detectable \\
Roche & $f = 0.80$ & 0.975 & 2.44 & 2.44 & 1.03 & 2.91 & Marginally detectable \\
Roche & $f = 0.90$ & 0.953 & 3.51 & 3.16 & 1.03 & 2.28 & Clearly detectable \\
Roche & $f = 0.95$ & 0.931 & 4.55 & 4.49 & 1.03 & 1.58 & Strongly detectable \\
Roche & $f = 1.00$ & 0.907 & 5.78 & 5.77 & 1.74 & 1.17 & Strongly detectable \\
\hline
\multicolumn{8}{c}{Simulated limb darkened star} \\
\hline
Sphere & $r = 1.00$ mas & 0.995 & 1.00 & 1.00 & 1.00 & 4.62 & Spherical \\
Sphere & $r = 1.50$ mas & 0.991 & 0.97 & 0.97 & 0.97 & 3.80 & Spherical \\
Sphere & $r = 1.85$ mas & 0.981 & 1.03 & 1.03 & 1.03 & 3.07 & Spherical \\
Sphere & $r = 1.98$ mas & 0.978 & 1.01 & 1.01 & 1.01 & 3.14 & Spherical \\
\enddata
\tablecomments{$\Delta \log(Z) = \log(Z_{\rm UD}) - \log(Z_{\rm ellipse})$. Larger values indicate greater preference for the simpler uniform disk model over the ellipse model. Interpretation thresholds are defined from the simulated limb darkened star control simulations: $\Delta\log(Z) > 3.0$ indicates the Roche geometry is not detectable as the model looks more like a sphere; $2.5 < \Delta\log(Z) \leq 3.0$: marginally detectable; $1.5 < \Delta\log(Z) \leq 2.5$: clearly detectable; $\Delta\log(Z) \leq 1.5$: strongly detectable.}
\end{deluxetable*}

%% file: simulations.tex
\begin{figure*}[!t]
    \gridline{
        \fig{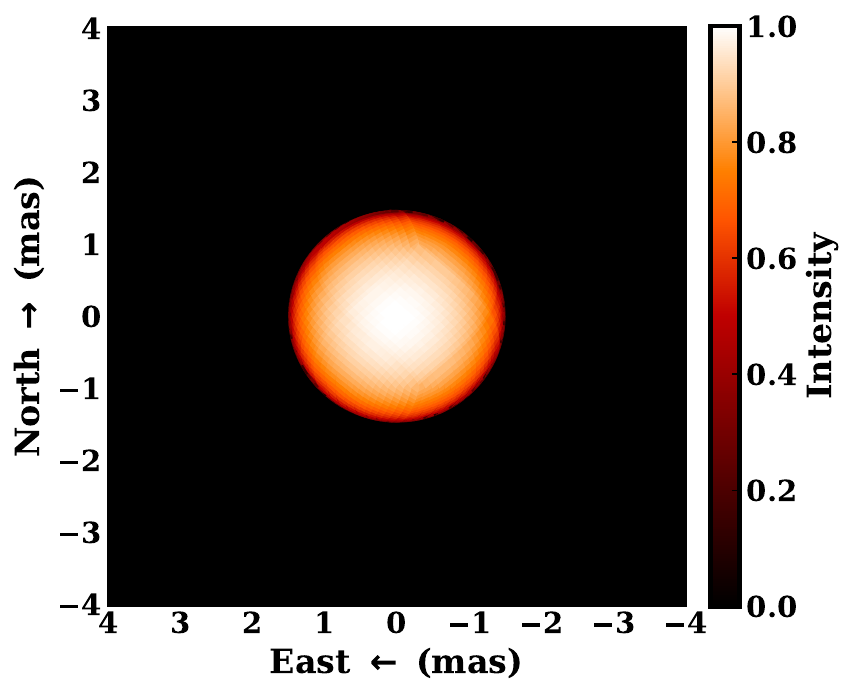}{0.33\textwidth}{(a) Fillout factor 0.7}
        \fig{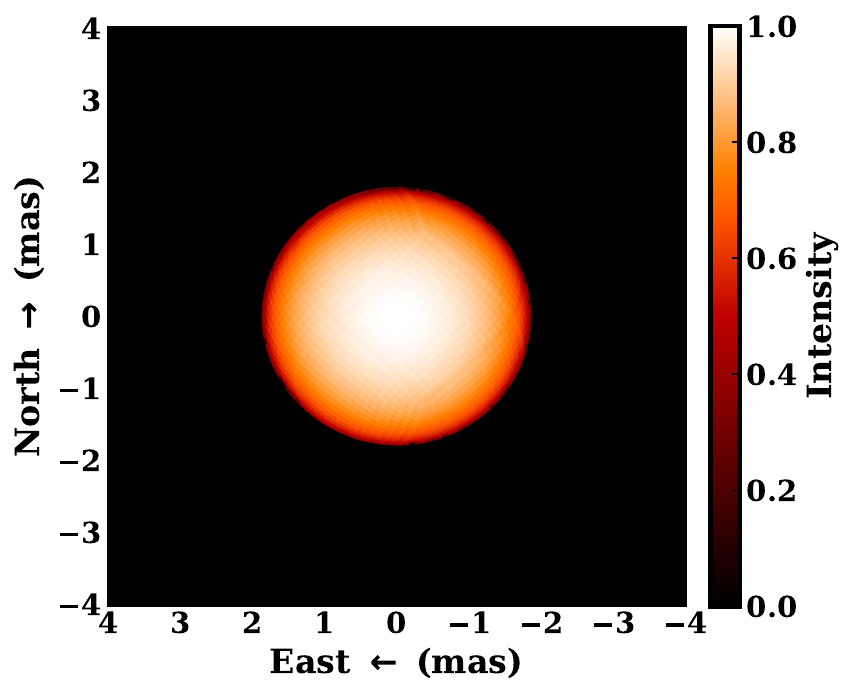}{0.33\textwidth}{(b) Fillout factor 0.8}
        \fig{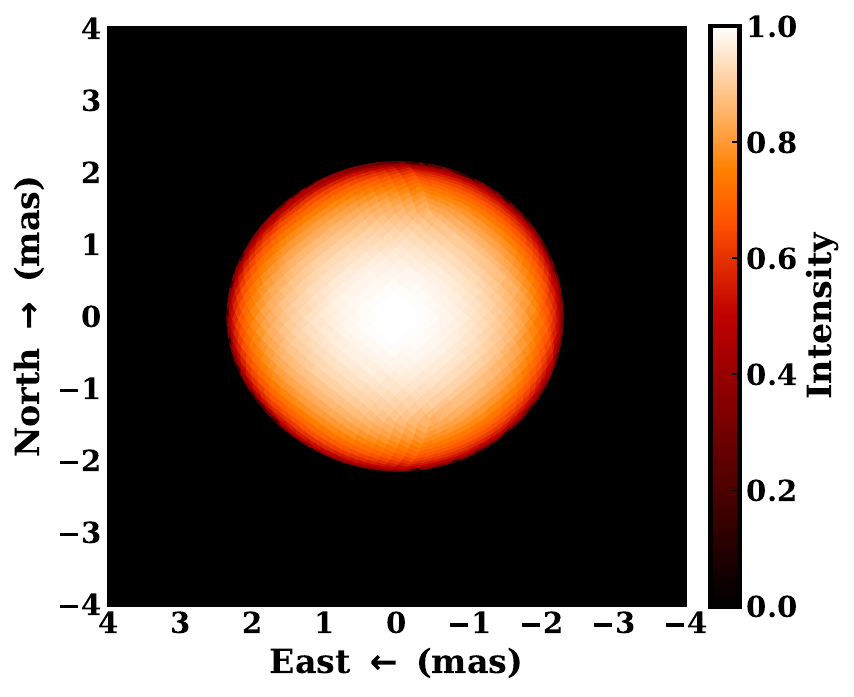}{0.33\textwidth}{(c) Fillout factor 0.9}}
    \gridline{
      \fig{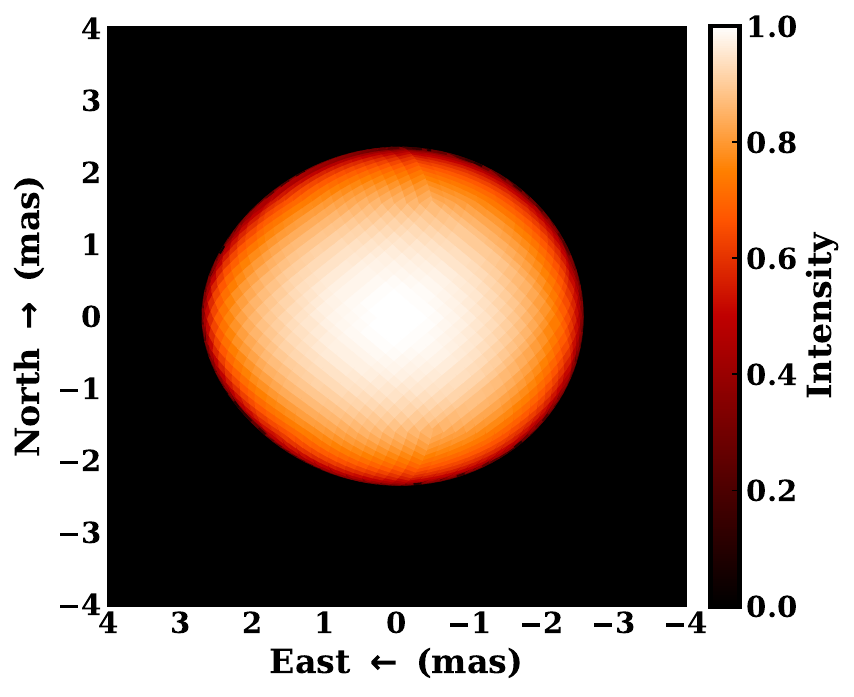}{0.33\textwidth}{(d) Fillout factor 0.95}
       \fig{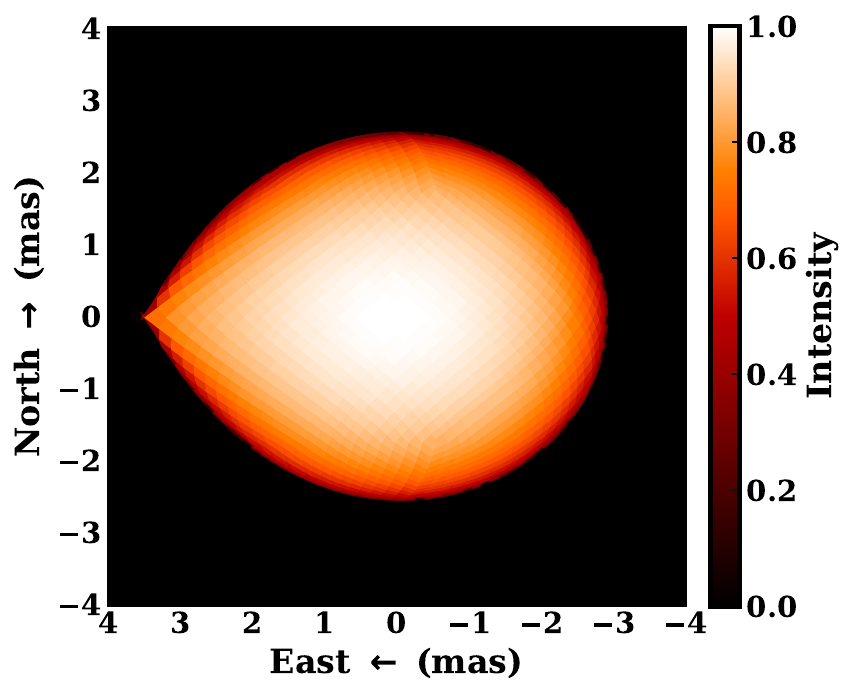}{0.33\textwidth}{(e) Fillout factor 1.00}
        \fig{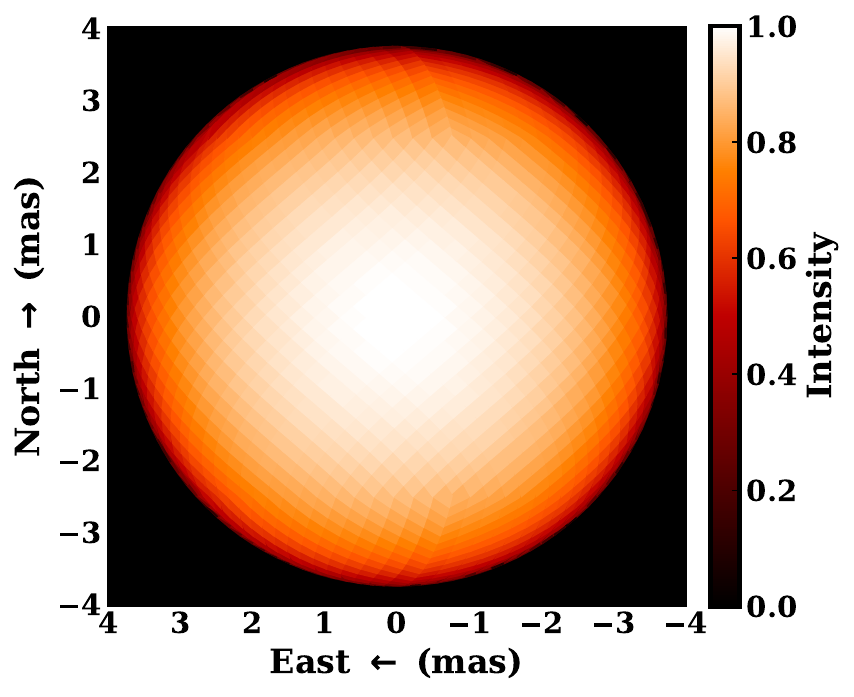}{0.33\textwidth}{(f) Sphere}}
    \caption{Images of the simulated stars used in the tests presented in Table \ref{tab:modelfitssim}. Simulations with fillout factors below 0.7 are not displayed as they were not easily distinguished from spheres.}
    \label{fig:simimages}
\end{figure*}

%% file: parameters.tex
\begin{deluxetable*}{ccccccc}
\tablecaption{Inferred Stellar Parameters
\label{tab:stellarparams}}
\tablewidth{0pt}
\tablehead{
\colhead{Target} &
\colhead{$A_{V}$ (mag)} &
\colhead{$E(B-V)$ (mag)} &
\colhead{$[\text{Fe/H}]$} &
\colhead{$T_{\text{eff}}$ (K)} &
\colhead{$R\ (R_{\odot})$} &
\colhead{$L\ (L_{\odot})$}
}
\startdata
V1472 Aql
    & $0.61_{-0.33}^{+0.34}$
    & $0.20_{-0.11}^{+0.11}$
    & $+0.02_{-0.30}^{+0.28}$
    & $3620_{-110}^{+120}$
    & $65_{-2}^{+2}$
    & $650_{-80}^{+100}$ \\
EG And\tablenotemark{a}
    & $0.11_{-0.07}^{+0.10}$
    & $0.04_{-0.02}^{+0.03}$
    & $-0.05_{-0.27}^{+0.28}$
    & $3630_{-100}^{+100}$
    & $77_{-5}^{+5}$
    & $920_{-140}^{+160}$ \\
BD Cam
    & $0.13_{-0.09}^{+0.15}$
    & $0.04_{-0.03}^{+0.05}$
    & $-0.10_{-0.26}^{+0.27}$
    & $3600_{-120}^{+120}$
    & $136_{-8}^{+8}$
    & $2780_{-450}^{+530}$ \\
SU Lyn
    & $0.16_{-0.10}^{+0.12}$
    & $0.05_{-0.03}^{+0.04}$
    & $+0.34_{-0.14}^{+0.11}$
    & $3230_{-50}^{+60}$
    & $263_{-12}^{+14}$
    & $6800_{-750}^{+890}$ \\
\enddata
\tablecomments{Effective temperature, metallicity, and
extinction were inferred from Starfish MCMC spectral fitting. Color excess was derived from the extinction with $R_{V}=3.1$ \citep{caradelli}. For some stars,  metallicities were adopted as priors for the MIST mass inference, as described in Section~\ref{sec:masses}.
Physical radii and luminosities were derived by Monte Carlo propagation of the LD angular diameter, adopted distance, and effective temperature measurement distributions. Radii were calculated from the angular diameters and distances, and luminosities from the resulting radii and effective temperatures using the Stefan--Boltzmann relation. Reported values are the medians, with uncertainties corresponding to the 16th and 84th percentiles.  Unless otherwise stated distances used came from \citet{gaiadist}
(a)Radius and luminosity for EG And used $d=400\pm20~\text{pc}$
\citep{1992Vogel,2016AJ....152....1K}.}
\end{deluxetable*}

%% file: orbitalparams.tex
\begin{deluxetable}{lccc}
\tabletypesize{\footnotesize}
\tablecaption{Orbital Parameters\label{tab:orbital}}
\tablehead{
\colhead{Parameter} & 
\colhead{V1472 Aql} &
\colhead{EG And} & 
\colhead{BD Cam}
}
\startdata
Period (d)     & $198.716 \pm 0.038$         & $482.5 \pm 1.3$        & $596.2 \pm 0.19$     \\
$T$ (JD$-2400000$)  & $43721.1 \pm 1.6$ & $50208.11 \pm 0.07$    & $42794 \pm 25$      \\
Eccentricity ($e$)  & $0.048 \pm 0.016$     & $0$ (fixed)            & $0.088 \pm 0.023$   \\
$\omega$ ($^{\circ}$)   & $99.3 \pm 29.7$               & \nodata      & $322 \pm 15$ \\
Inclination ($^{\circ}$) & $50$--$90$\tablenotemark{a} & $\approx 80$\tablenotemark{b} & $109\pm6.5$\tablenotemark{c}       \\
$\gamma$ (km s$^{-1}$)  & \nodata  & $-94.75 \pm 0.09$      & $-22.28 \pm 0.14$   \\
$K$ (km s$^{-1}$)  & $12.97 \pm 0.27$     & $7.30 \pm 0.13$        & $8.47 \pm 0.20$      \\
$a \sin i$ (au)  & $0.237 \pm 0.005$       & $0.323 \pm 0.003$      & $0.463 \pm 0.011$    \\
$f(M)$ ($M_{\odot}$) & $0.045 \pm 0.003$    & $0.019 \pm 0.001$      & $0.037 \pm 0.003$   \\
\hline
\textbf{Ref.}   & \citet{1982Lucke}   & \citet{2016AJ....152....1K} & \citet{1984Obs...104..224G}  \\
\enddata
\tablecomments{Orbital parameters for SU Lyn were not published as of the date of writing. The $T$ given by \citet{2016AJ....152....1K} for EG And is the time of conjunction whereas the $T$ given by \citet{1982Lucke} for V1472 Aql and \citet{1984Obs...104..224G} for BD Cam are times of periastron. (a) \citet{2014Boffin}. (b) \citet{1992Vogel}. (c) \citet{2000pourbaix}}
\end{deluxetable}

%% file: priors.tex
\begin{deluxetable*}{lllllll}
\tabletypesize{\footnotesize} 
\tablewidth{0pt} 
\tablecaption{Priors used in the mass inference procedure
\label{tab:priors}}
\tablehead{
  \colhead{Target} &
  \colhead{$[\mathrm{Fe/H}]$} &
  \colhead{$i$} &
  \colhead{$M_\mathrm{2}~(M_\odot)$} &
  \colhead{$P$ (d)} &
  \colhead{$e$} &
  \colhead{$d$ (pc)}
}
\startdata
V1472 Aql
& $\mathcal{N}(+0.02,0.20)$\tablenotemark{a}
& isotropic $50^\circ$--$90^\circ$
& $\mathcal{U}(0.30,3.00)$\tablenotemark{b}
& $198.716\pm0.038$
& $0.048\pm0.016$
& $247.6^{+5.0}_{-4.0}$\tablenotemark{c} \\
EG And
& $\mathcal{N}(-0.54,0.10)$\tablenotemark{d}
& isotropic $78^\circ$--$90^\circ$
& $\mathcal{U}(0.35,0.60)$\tablenotemark{e}
& $482.5\pm1.3$
& $0.0$ (fixed)
& $400\pm20$\tablenotemark{f} \\
BD Cam
& $\mathcal{N}(-0.20,0.10)$\tablenotemark{g}
& $\mathcal{N}(109^\circ,6.5^\circ)$\tablenotemark{h}
& $\mathcal{U}(0.30,1.44)$
& $596.21\pm0.19$
& $0.088\pm0.023$
& $235.4\pm13.4$ \\
SU Lyn
& $\mathcal{N}(+0.02,0.20)$\tablenotemark{a}
& \nodata
& \nodata
& \nodata
& \nodata
& $719.7^{+38.2}_{-30.3}$\tablenotemark{c} \\
\enddata
\tablecomments{
The mass function $f(m)$ and orbital period $P$, and eccentricity are treated as
Gaussian measurements when uncertainties are given; the eccentricity of EG And is fixed at zero. Inclination and $[\mathrm{Fe/H}]$ are sampled from the distributions listed; $M_\mathrm{2}$ is drawn from a uniform prior over the stated range. ``Isotropic'' denotes $p(i)\propto\sin i$, equivalently a prior uniform in $\cos i$ over the listed inclination interval. Distributions are defined as normal $\mathcal{N}(\mu, \sigma)$ and uniform $\mathcal{U}(a, b)$. \\
References: 
(a) \citet{2015ApJ...808..132H}, truncated to $[-1.0,+0.5]$; 
(b) Unknown companion spectral type; 
(c) \citet{gaiadist}; 
(d) \citet{2023MNRAS.526..918G}, truncated to $[-1.0,+0.5]$; 
(e) \citet{2017Sion,2016AJ....152....1K}; 
(f) \citet{1992Vogel,2016AJ....152....1K}; 
(g) \citet{2020AandA...633A..34C}, truncated to $[-1.0,+0.5]$; 
(h) \citet{2000pourbaix}.}
\end{deluxetable*}

%% file: masstab.tex
\begin{deluxetable*}{cccccc}
\tablecaption{Inferred masses, orbit-averaged separation ($\langle r\rangle$),
Roche-lobe radius, and time-averaged filling factor.
\label{tab:masstab}}
\tablewidth{0pt}
\tablehead{
\colhead{Target} &
\colhead{$M_{\text{G}}~(M_{\odot})$} &
\colhead{$M_{\text{2}}~(M_{\odot})$} &
\colhead{$\langle r\rangle$ ($R_\odot$)} &
\colhead{$R_{\text{L}}~(R_{\odot})$} &
\colhead{time-averaged $f$}
}
\startdata
V1472 Aql
    & $1.24_{-0.36}^{+0.64}$
    & $0.58_{-0.12}^{+0.18}$
    & $175_{-17}^{+23}$
    & $78_{-10}^{+13}$
    & $0.83^{+0.11}_{-0.12}$ \\
EG And
    & $1.29^{+0.23}_{-0.13}$
    & $0.38^{+0.04}_{-0.02}$
    & $307^{+16}_{-10}$
    & $150^{+10}_{-6}$
    & $0.53^{+0.04}_{-0.04}$ \\
BD Cam
    & $2.34^{+0.85}_{-0.67}$
    & $0.75^{+0.16}_{-0.14}$
    & $436^{+43}_{-42}$
    & $209^{+26}_{-25}$
    & $0.65^{+0.09}_{-0.07}$ \\
SU Lyn
    & $2.69^{+0.80}_{-0.58}$
    & \nodata
    & \nodata
    & \nodata
    & \nodata \\
\enddata
\end{deluxetable*}

%% file: surfing.tex
\begin{deluxetable*}{lccccc}
\tablecaption{Best fitting parameters determined by SURFING image reconstruction. \label{tab:surfing}}
\tablehead{
\colhead{Star Name} &
\colhead{Model} &
\colhead{$\theta$ (mas)} &  
\colhead{$\alpha$} &
\colhead{$V_0$} & 
\colhead{$\chi^{2}$}
}
\startdata
V1472 Aql & Sphere & 2.44 & 0.30 & 1.00 & 4.64\\
\hline
EG And & Sphere & 1.75 & 0.26 & 0.99 & 2.68\\
\hline
BD Cam & Sphere & 5.52& 0.17 & 1.08 & 3.65\\
\hline
SU Lyn & Sphere & 3.38  & 0.37 & 1.01 & 2.44\\
\enddata
\tablecomments{$V_{0}$ is the zero-baseline visibility.}
\end{deluxetable*}

%% file: images_sphere.tex
\begin{figure*}[!t]
    \gridline{
            \fig{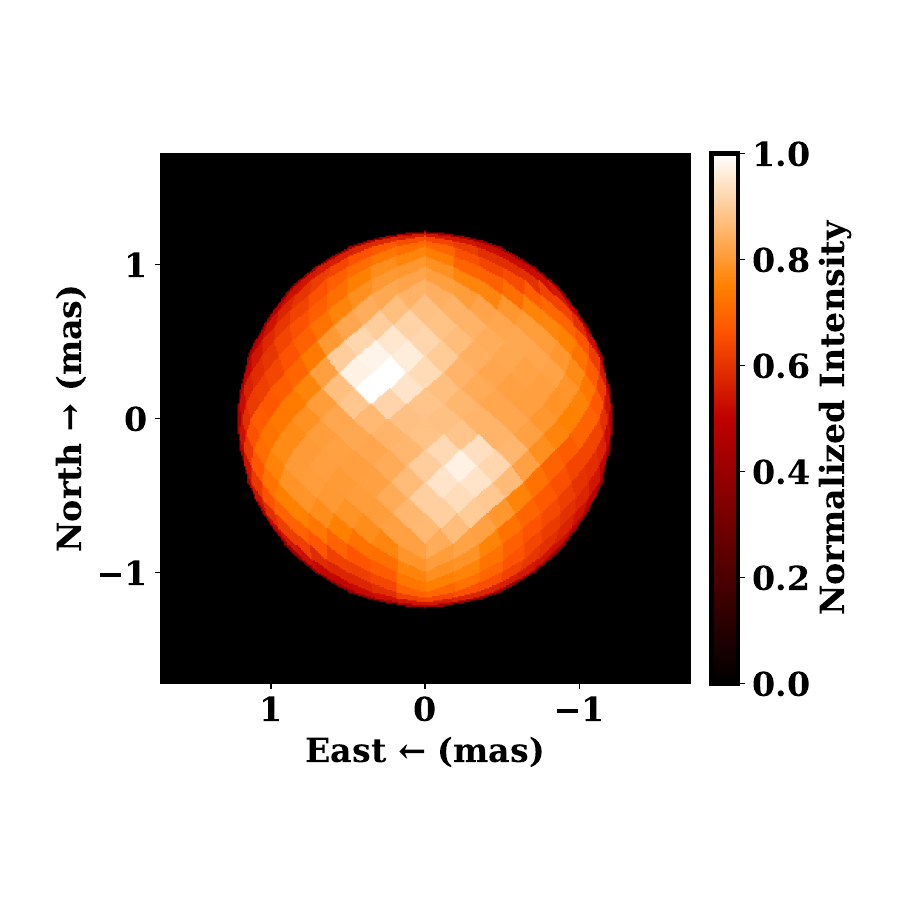}{0.25\textwidth}{V1472 Aql}
             \fig{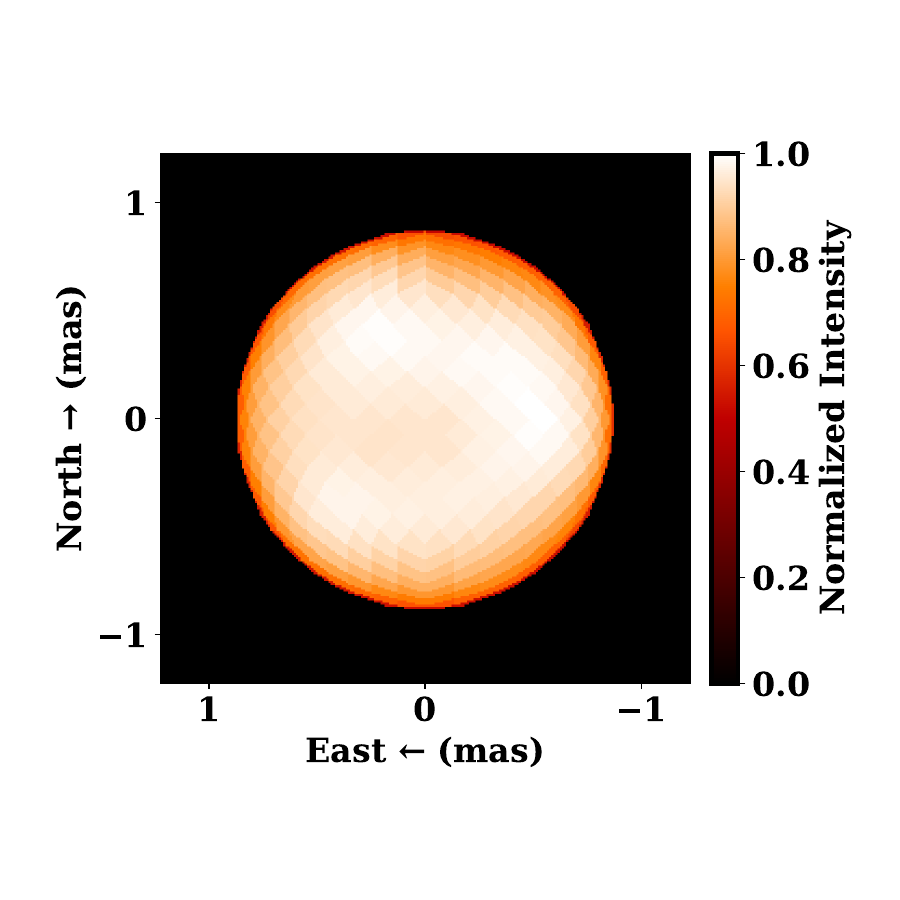}{0.25\textwidth}{EG And}
            \fig{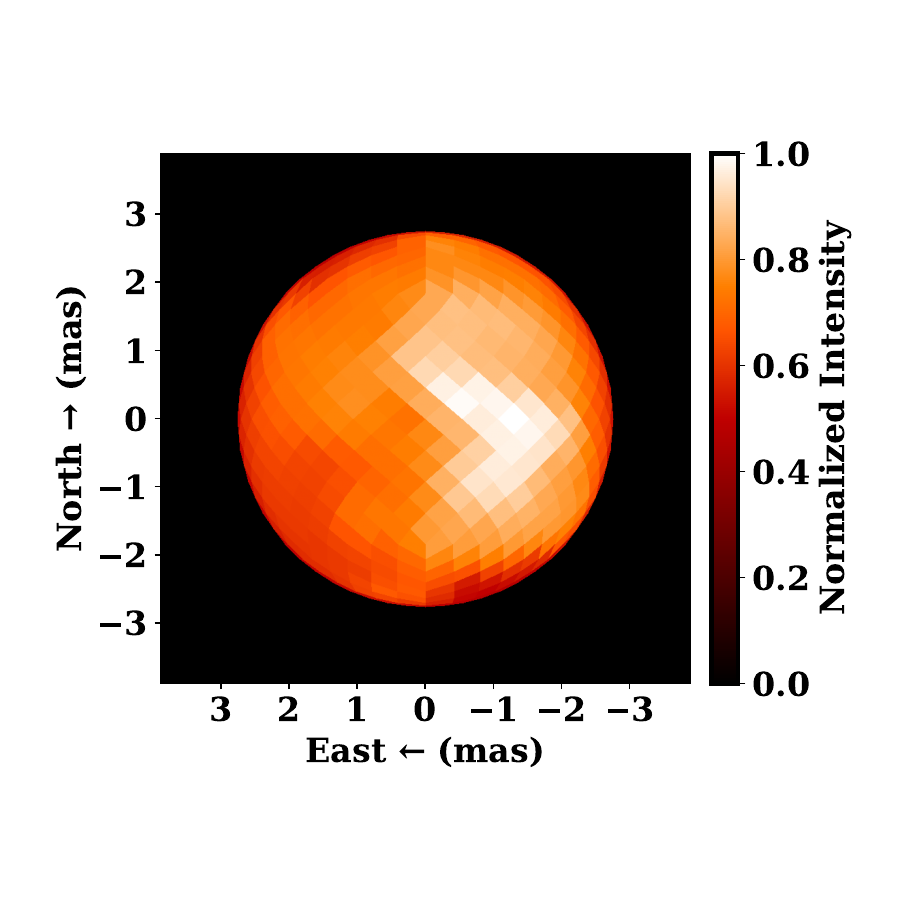}{0.25\textwidth}{BD Cam}
             \fig{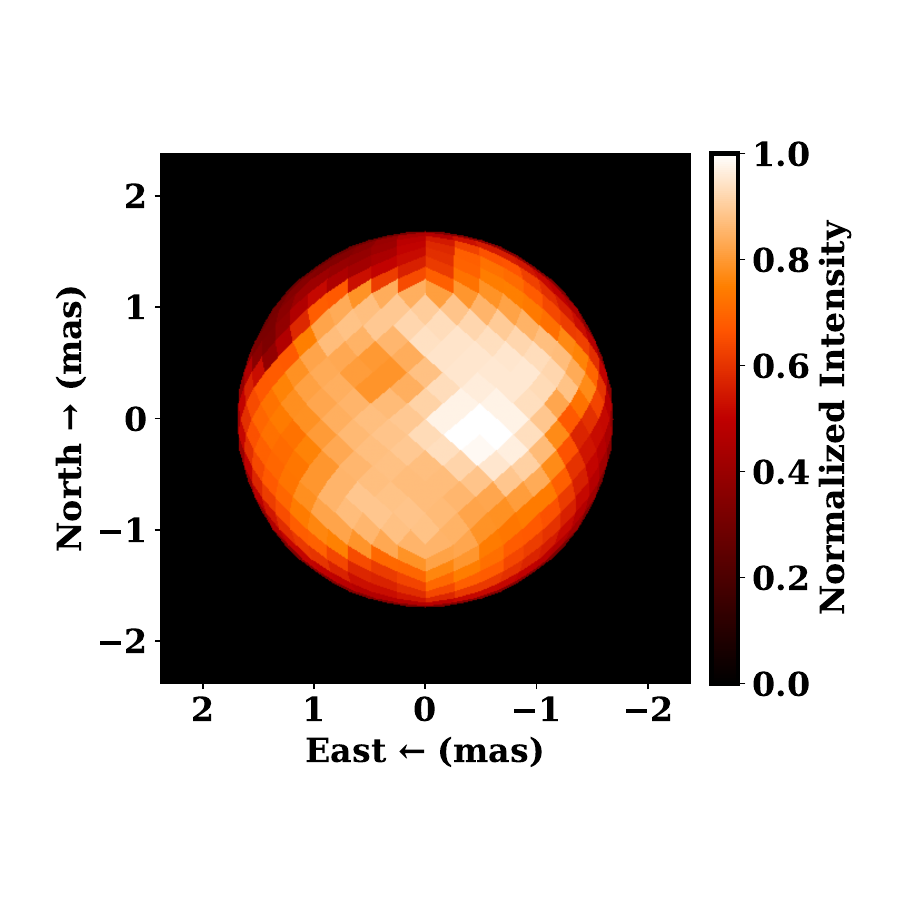}{0.25\textwidth}{SU Lyn}}


    \caption{Images of the targets reconstructed on spheres using SURFING. The images are normalized, with the sum of an image equal to the value of $V_{0}$ of  each star as listed in Table \ref{tab:surfing}. }
    \label{fig:imagessphere}
\end{figure*} 

%% file: uvplotsall.tex
\begin{figure*}[!h]
    \gridline{\fig{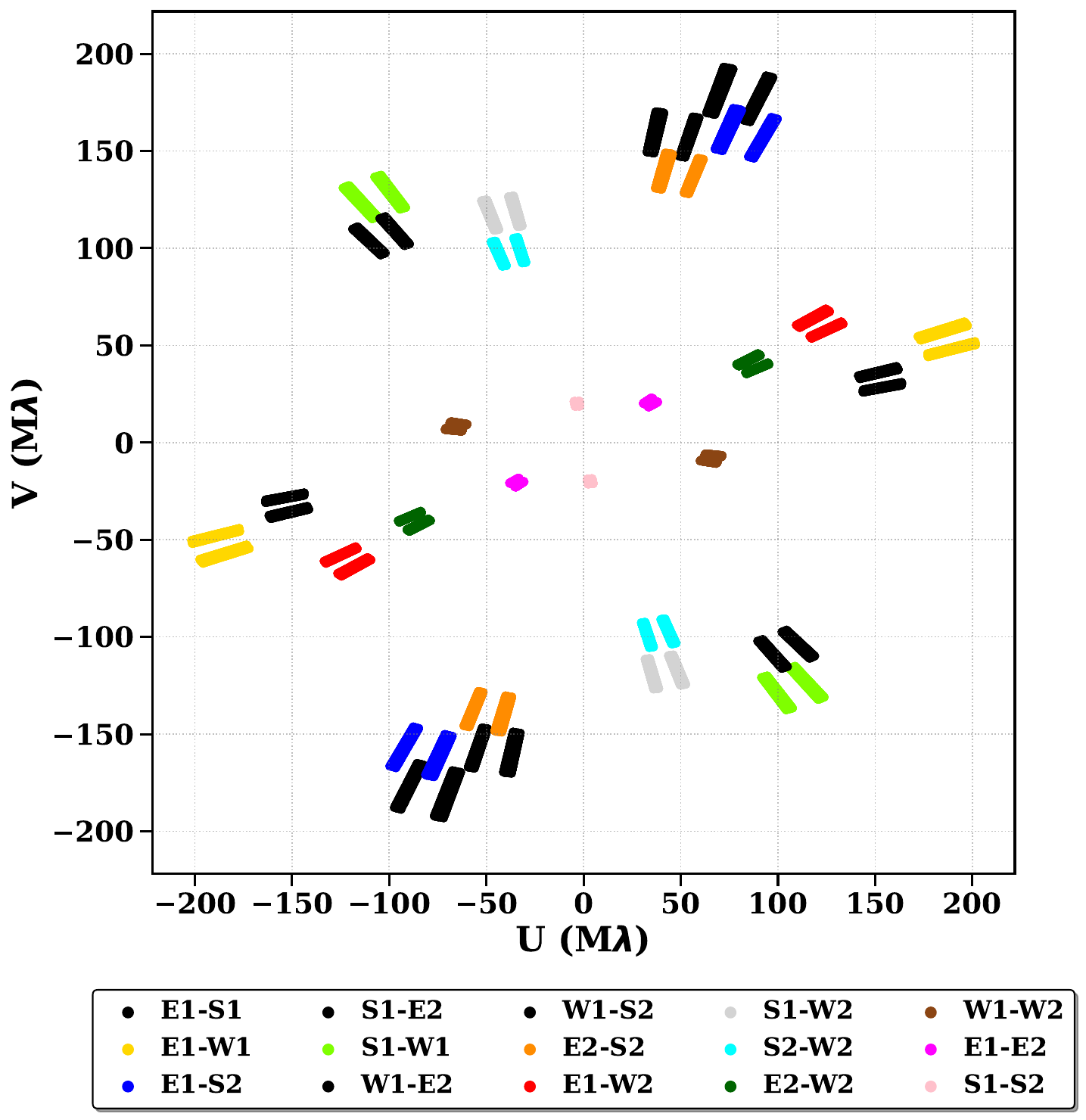}{0.45\textwidth}{(a) V1472 Aql}
              \fig{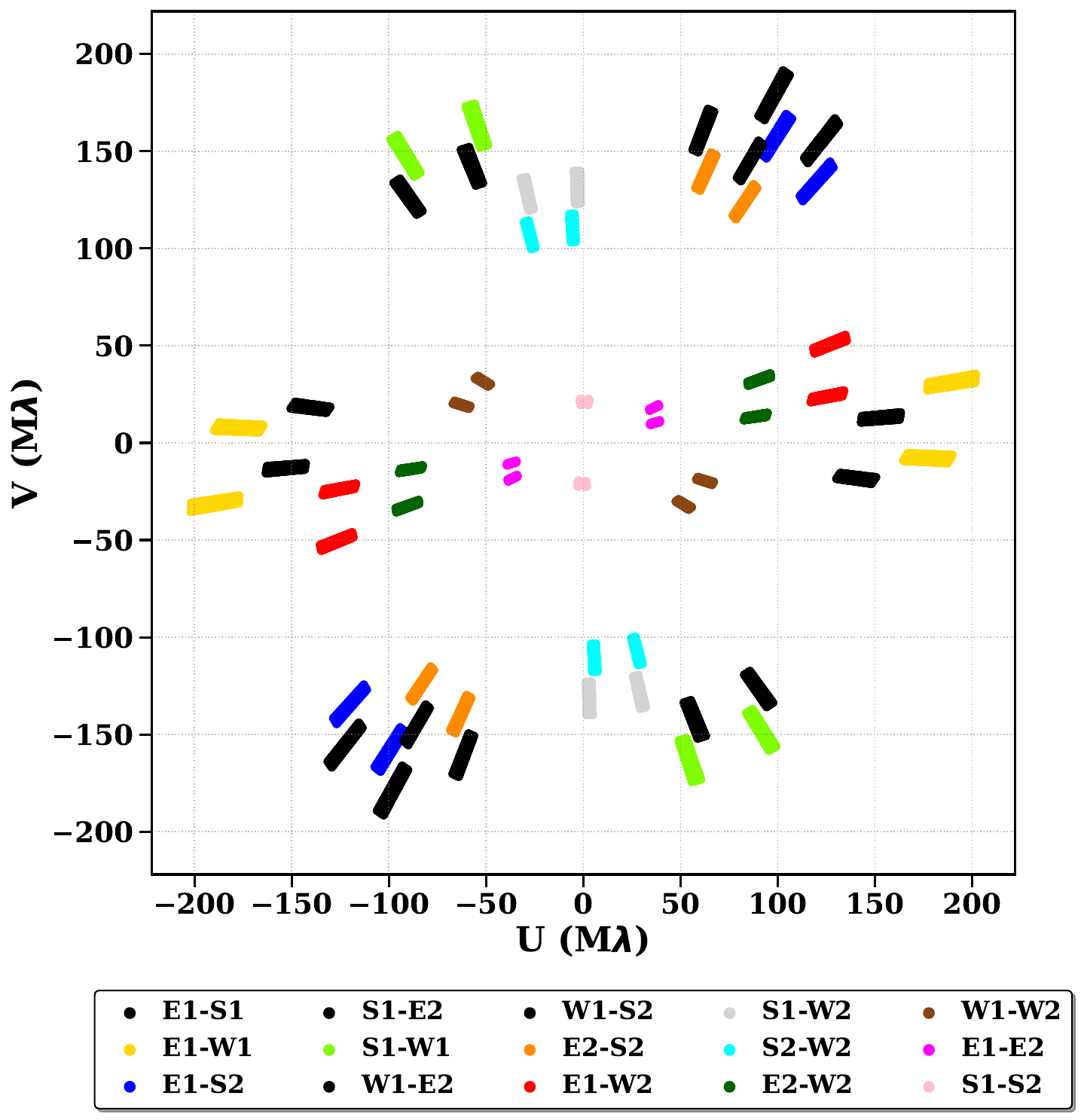}{0.45\textwidth}{(b) EG And}
             }
    \gridline{\fig{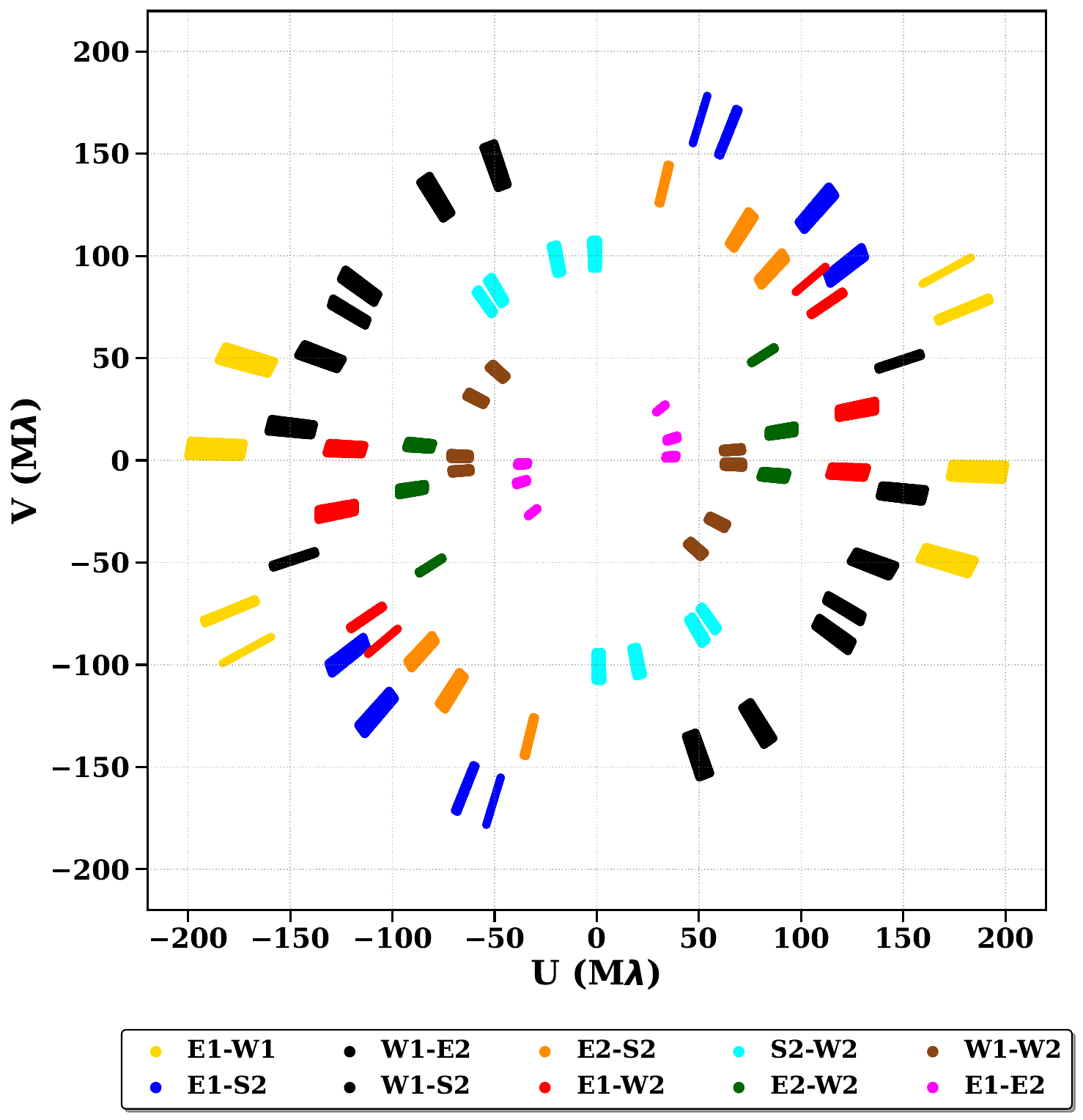}{0.45\textwidth}{(c) BD Cam}
              \fig{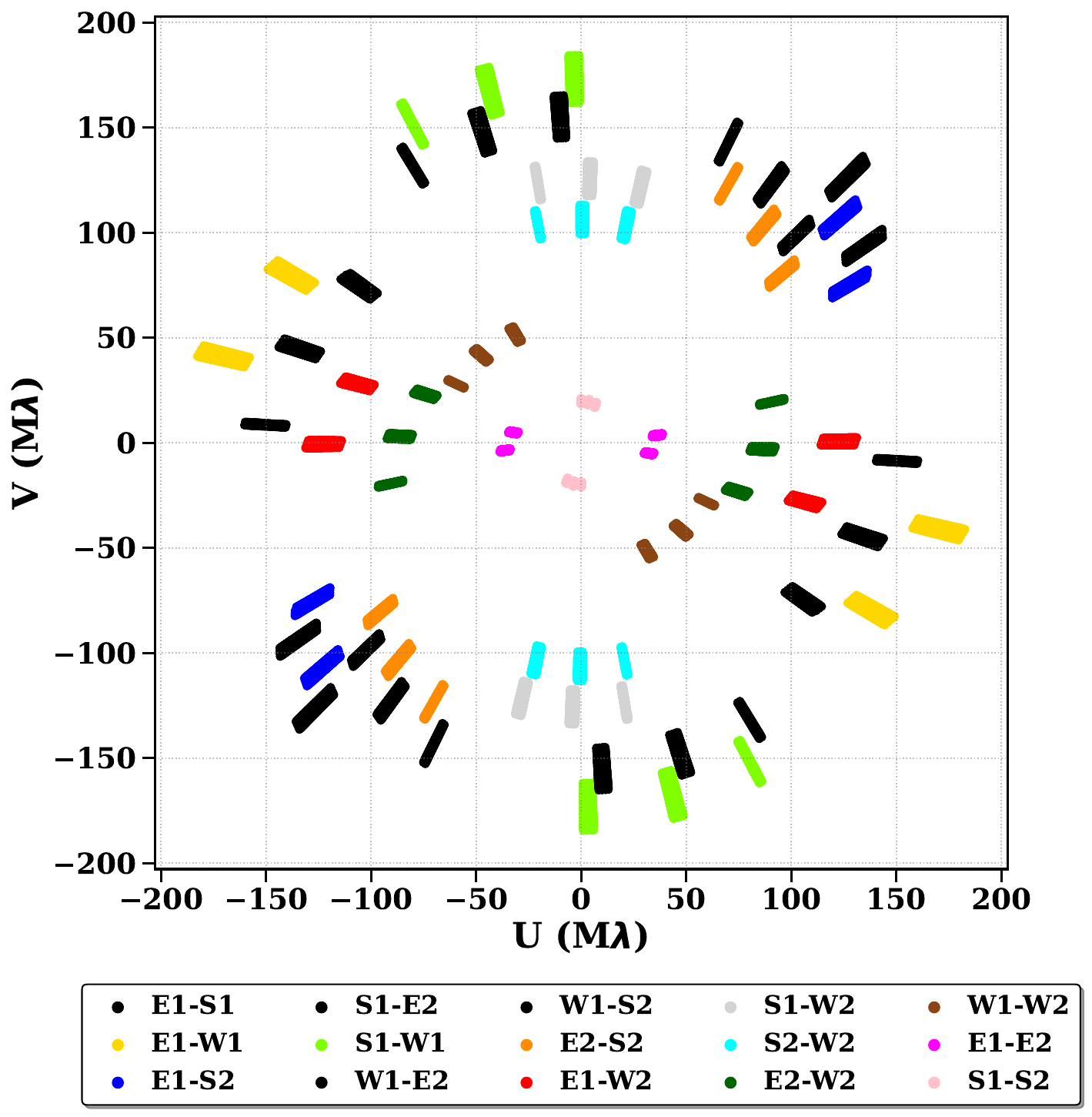}{0.45\textwidth}{(d) SU Lyn}
             }
    \caption{$(u,v)$ plane coverage for all four targets. The color of the points describes baseline as described by the legend. Data was obtained with MIRC-X at the CHARA Array.}
    \label{fig:uv_coverage}
\end{figure*}

%% file: v2all.tex
\begin{figure*}[!h]
    \gridline{\fig{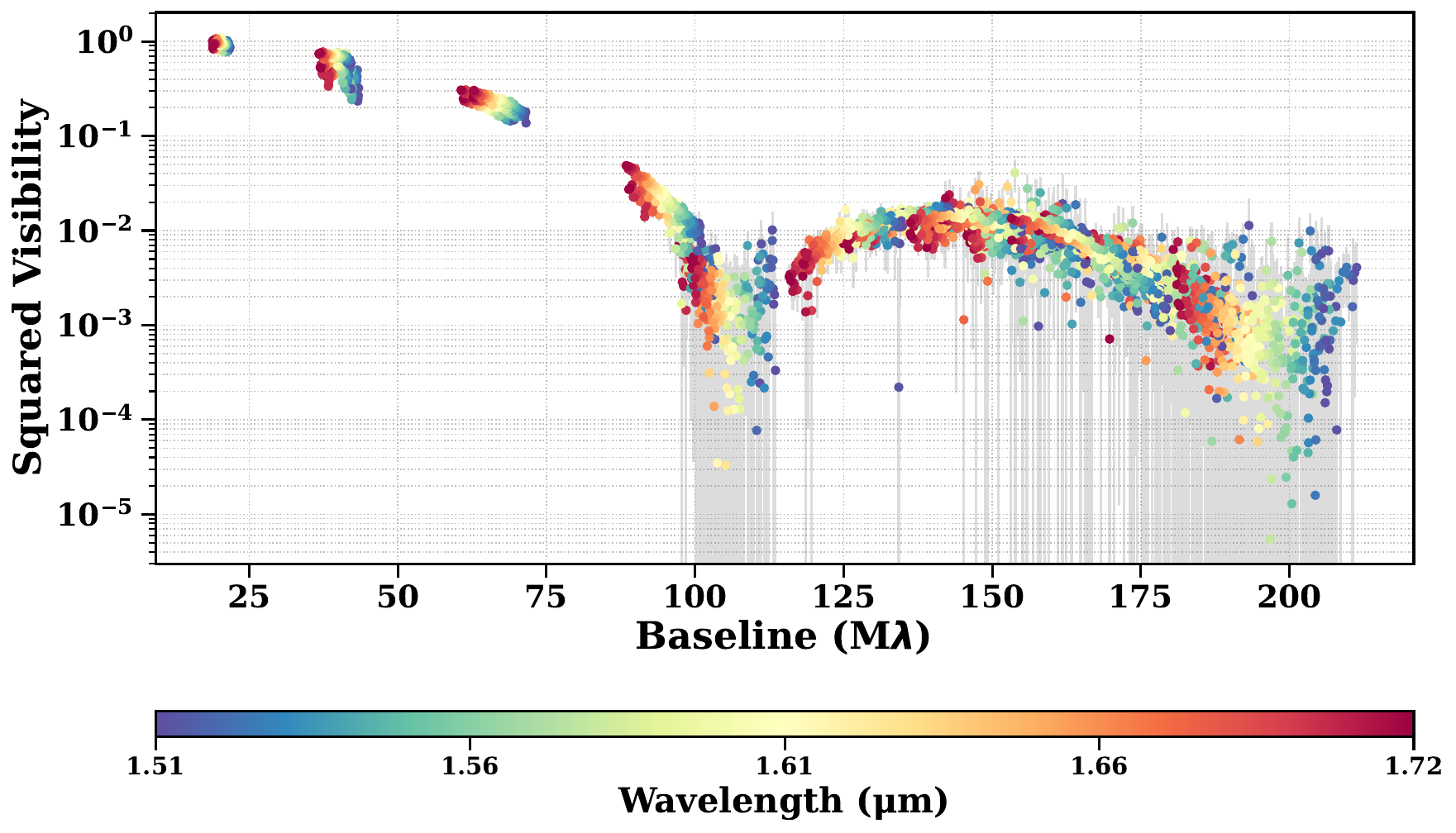}{0.5\textwidth}{(a) V1472 Aql}
              \fig{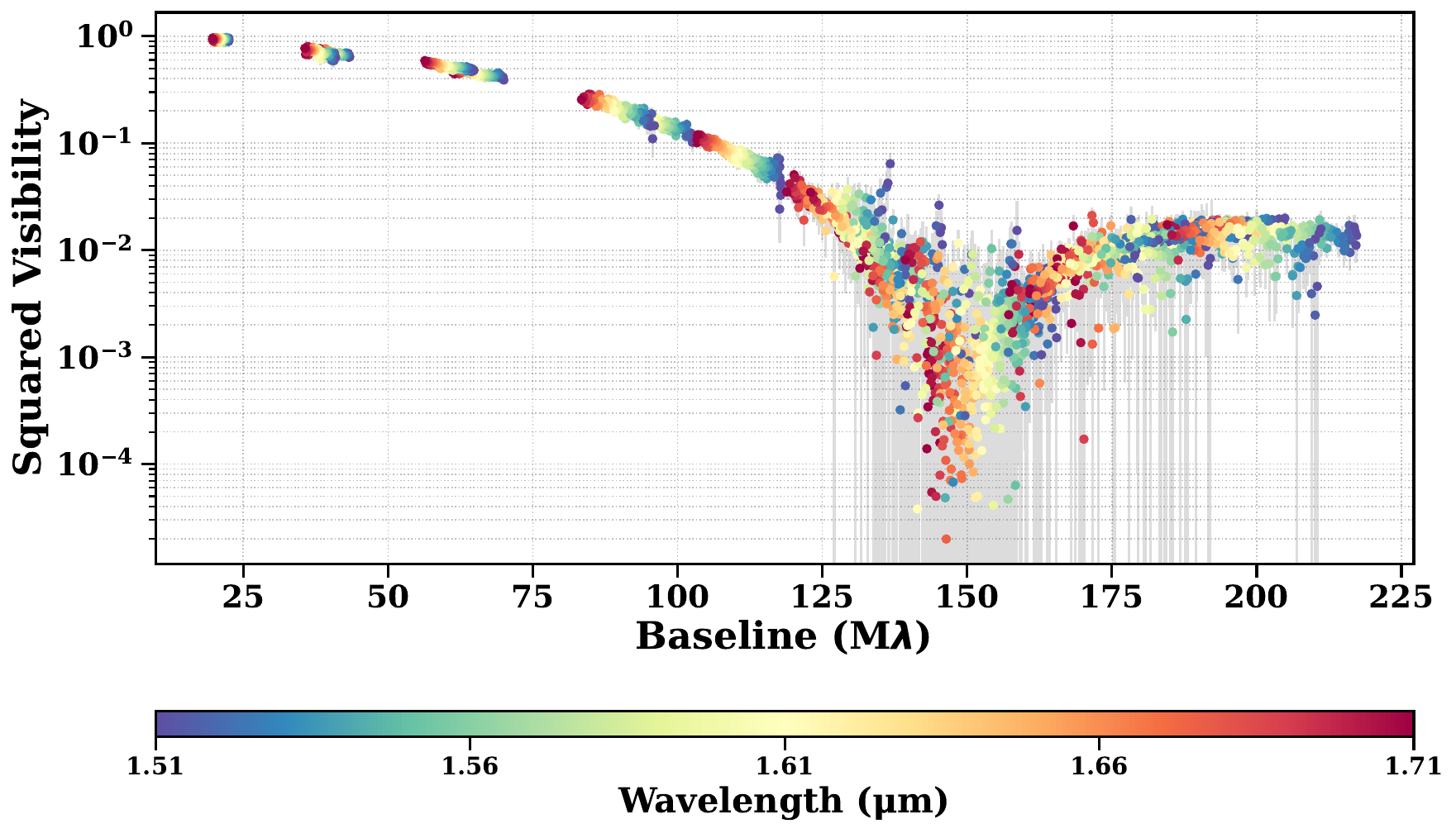}{0.5\textwidth}{(b) EG And}
             }
    \gridline{\fig{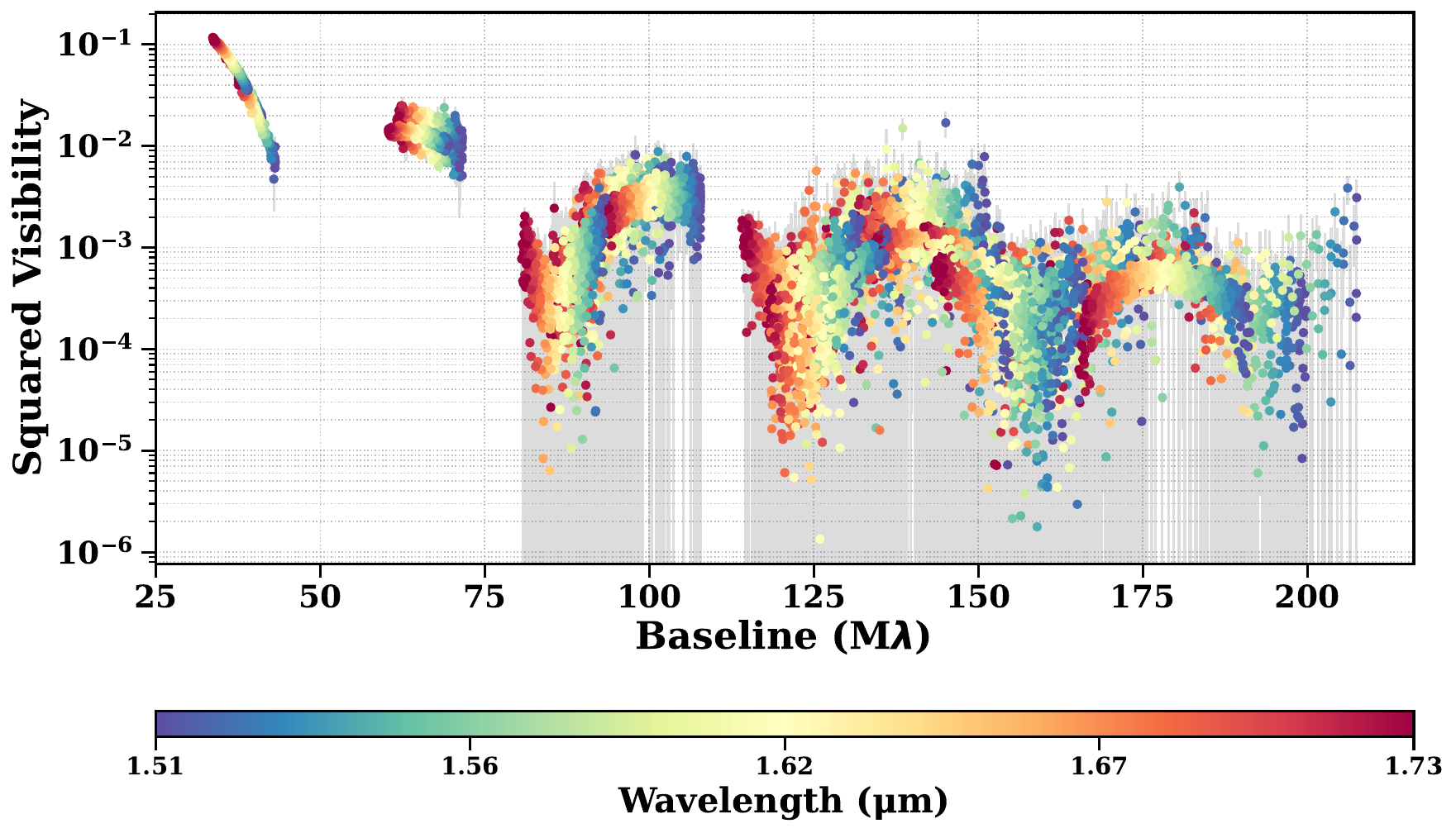}{0.5\textwidth}{(c) BD Cam}
              \fig{su_lyn_v2.pdf}{0.5\textwidth}{(d) SU Lyn}
             }
    \caption{V$^{2}$ data for all four targets. The color of the points describes wavelength as described by the colorbar. Data was obtained with MIRC-X at the CHARA Array.}
    \label{fig:v2all}
\end{figure*}

%% file: t3phiall.tex
\begin{figure*}[!h]
    \gridline{\fig{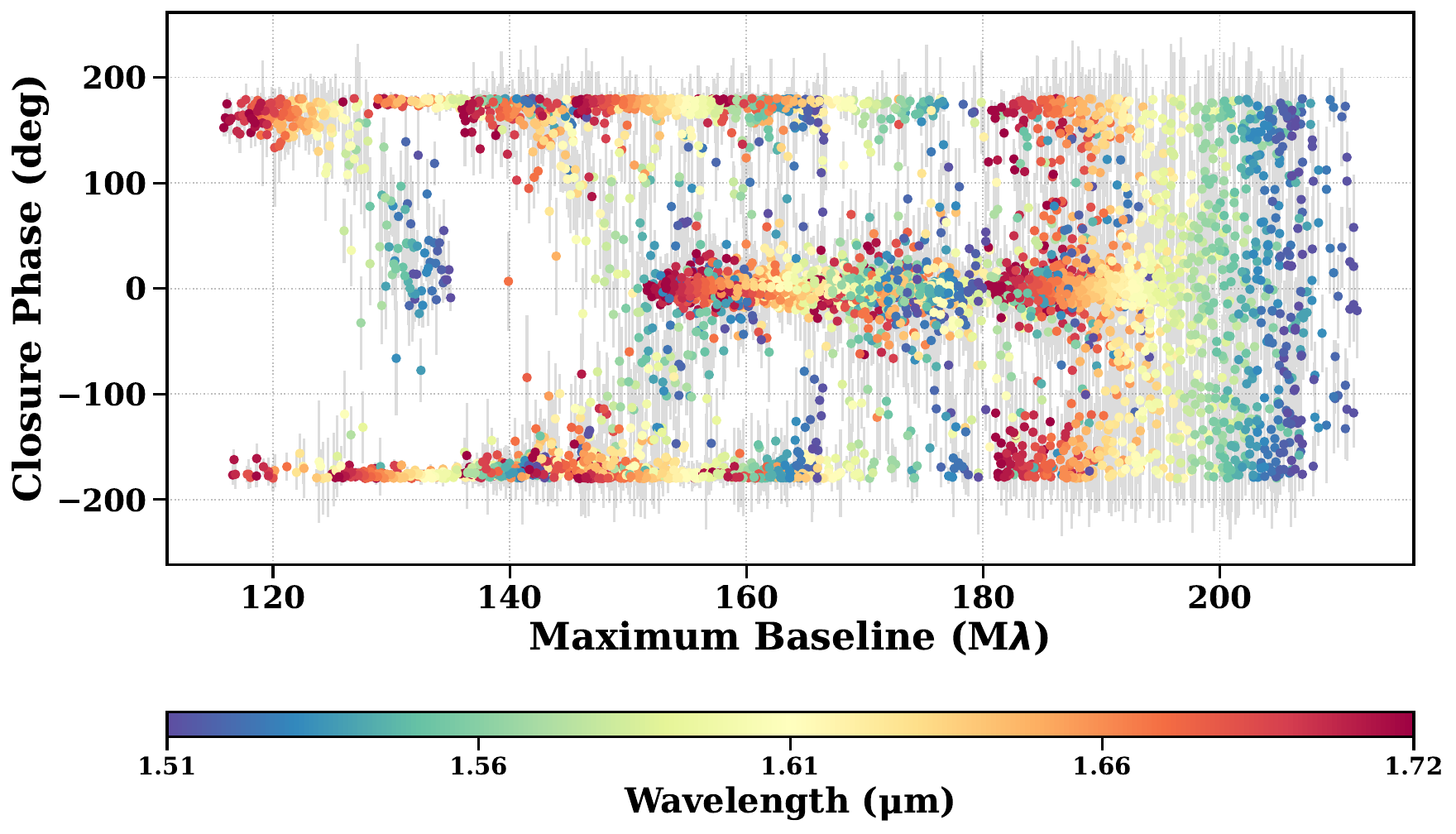}{0.5\textwidth}{(a) V1472 Aql}
              \fig{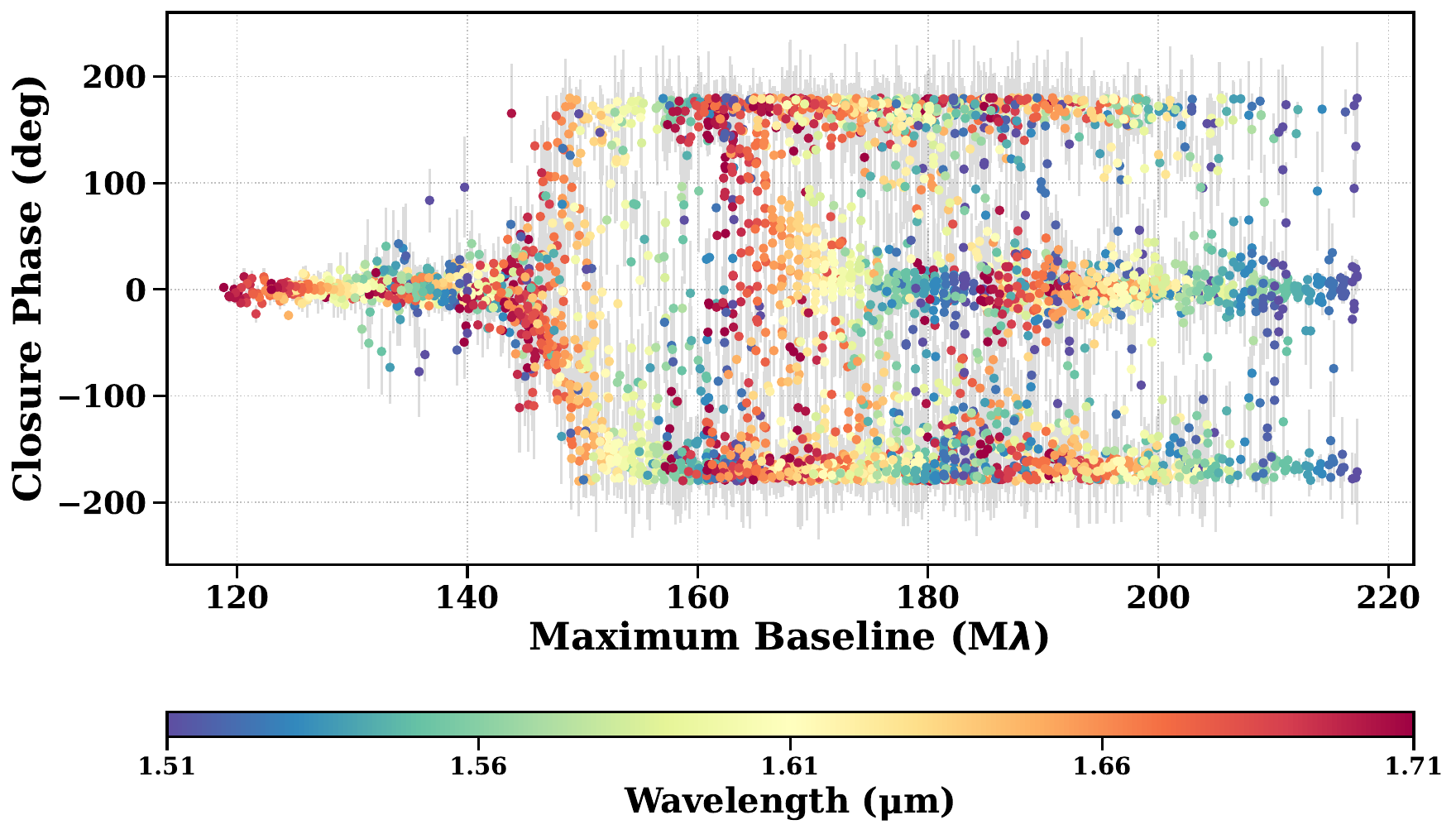}{0.5\textwidth}{(b) EG And}
             }
    \gridline{\fig{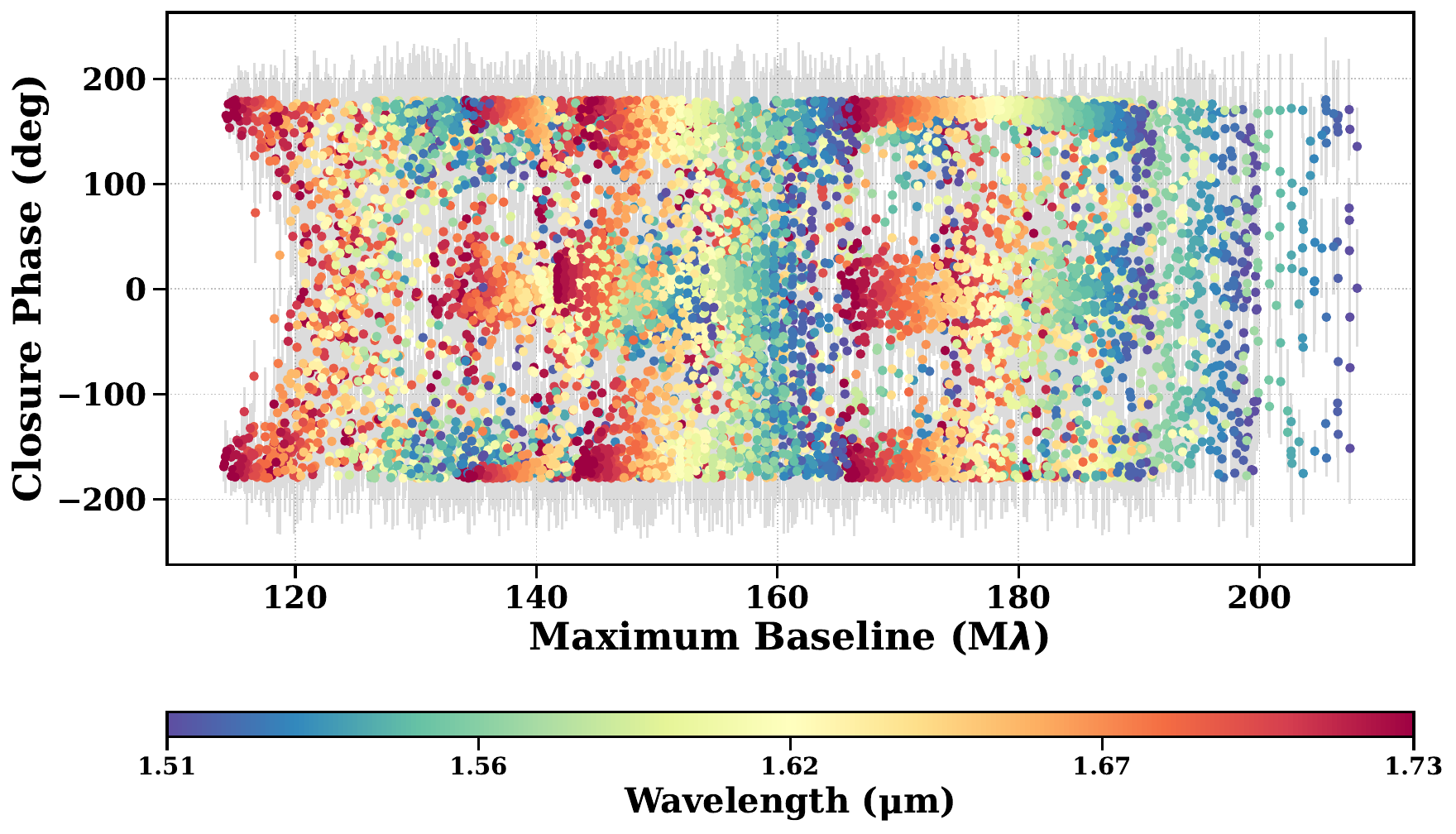}{0.5\textwidth}{(c) BD Cam}
              \fig{su_lyn_t3phi.pdf}{0.5\textwidth}{(d) SU Lyn}
             }
    \caption{Closure phase data for all four targets. The color of the points describes wavelength as described by the colorbar. Data was obtained with MIRC-X at the CHARA Array.}
    \label{fig:t3phiall}
\end{figure*}

%% file: allspectra.tex
\begin{figure*}[!ht]
    \centering
    \includegraphics[width=0.5\textwidth]{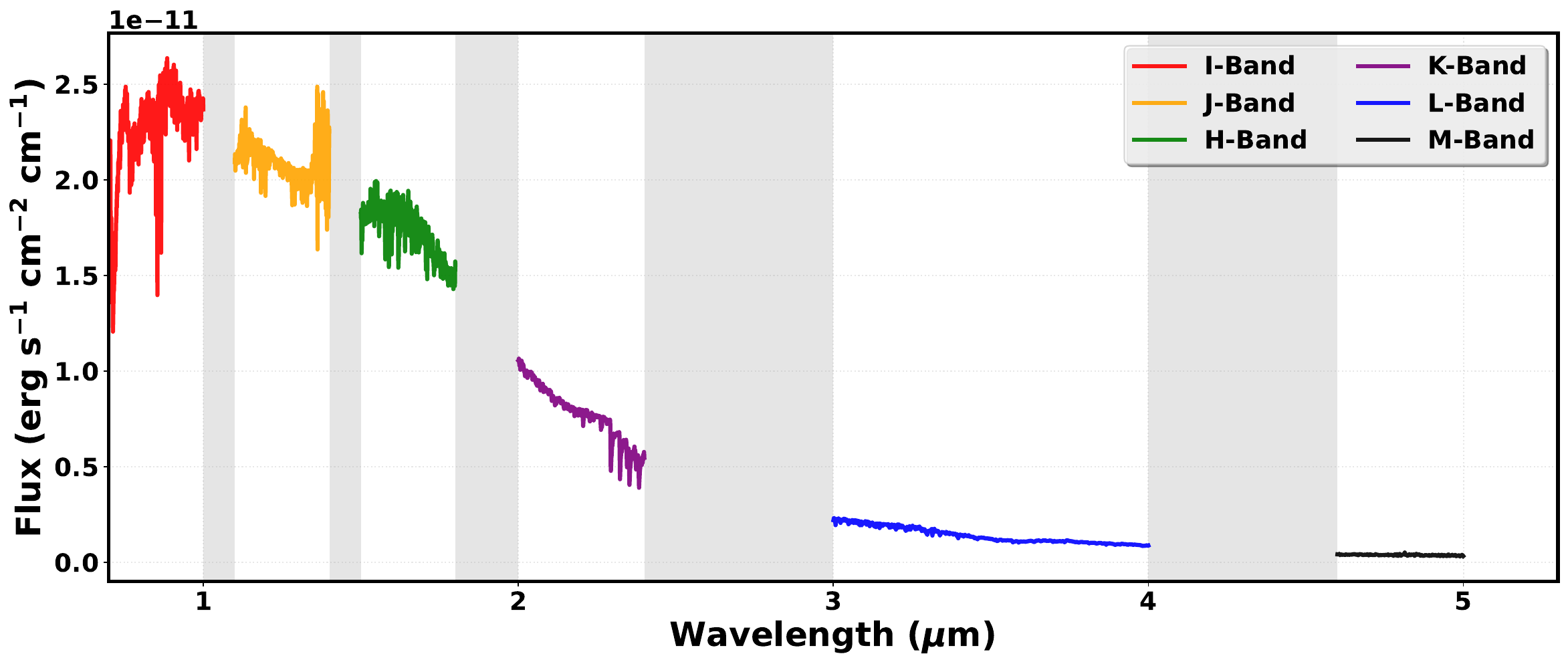}
    \includegraphics[width=0.5\textwidth]{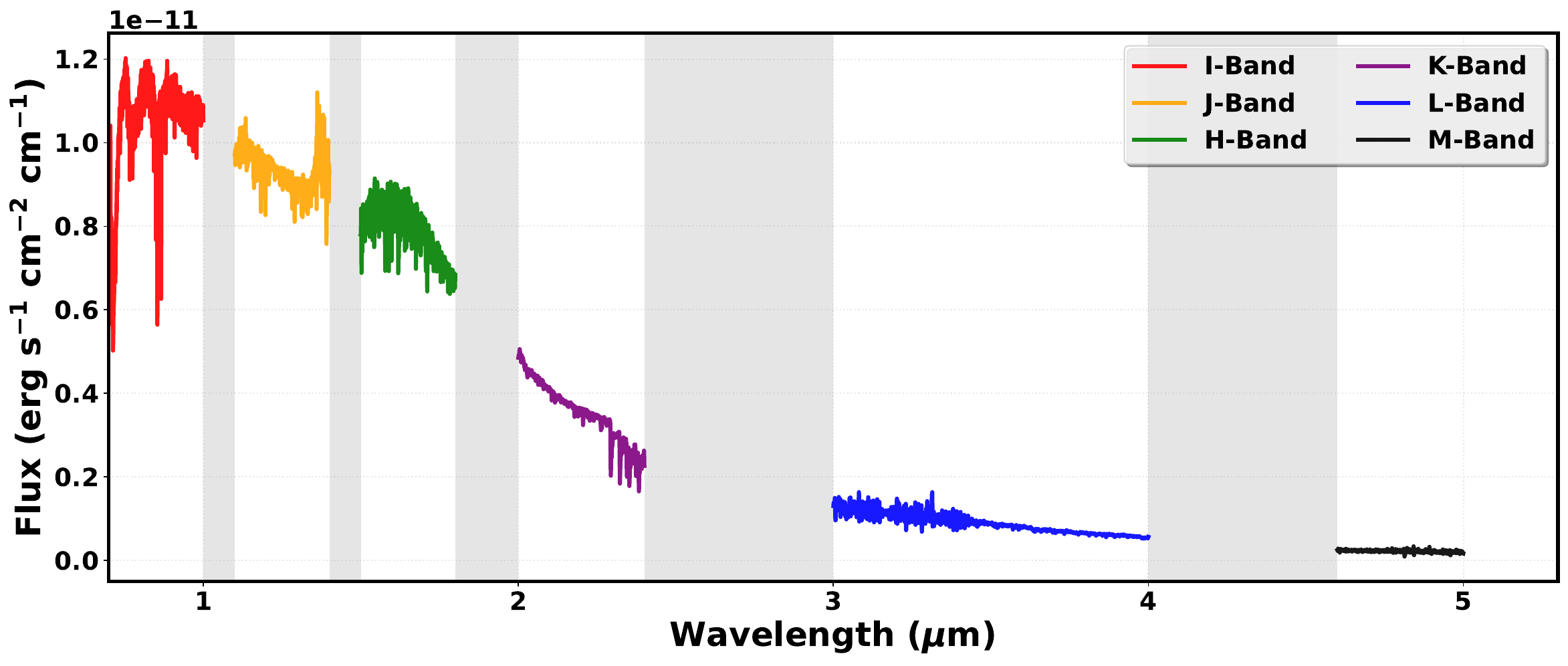}
    \includegraphics[width=0.5\textwidth]{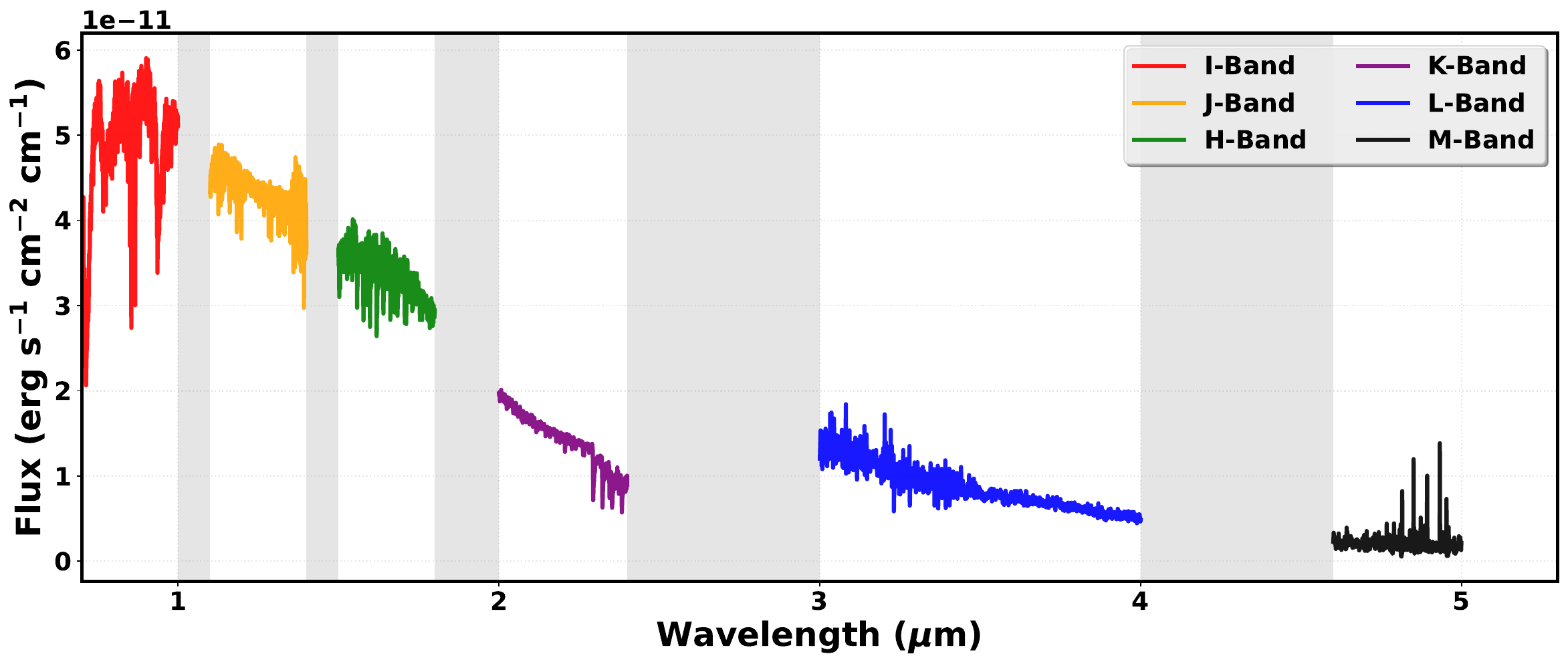}
    
    \caption{2021 September IRTF SpeX spectra of V1472 Aql (top), EG And (middle), and BD Cam (bottom) spanning 0.7--5.3~$\mu$m. Each color represents a different photometric band: I-band (red), J-band (orange), H-band (green), K-band (purple), L-band (blue), and M-band (black). Gray shaded regions indicate telluric absorption bands where atmospheric water vapor and other molecules prevent transmission of stellar flux. Spectra are shown in flux-calibrated units.}
    \label{fig:sept_spectra}
\end{figure*}

\begin{figure*}[!ht]
    \centering
    \includegraphics[width=0.5\textwidth]{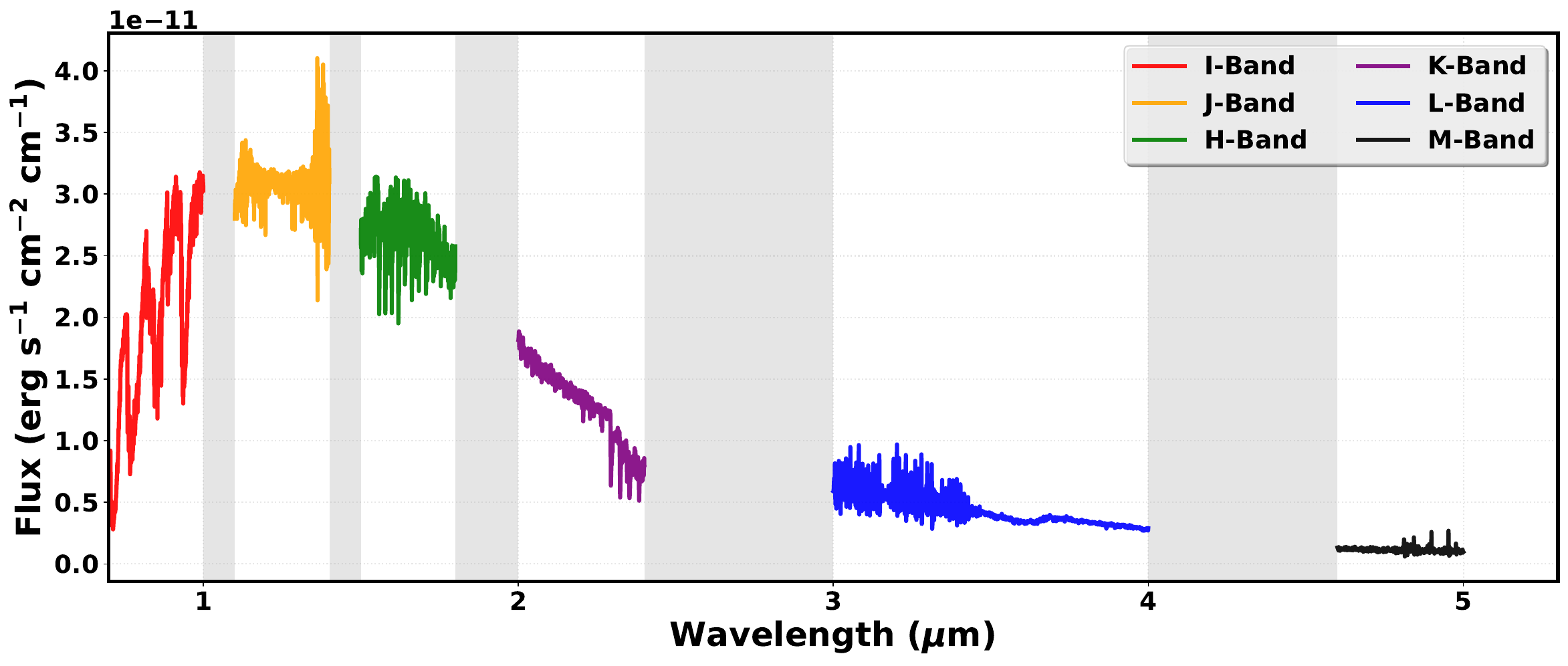}
    \includegraphics[width=0.5\textwidth]{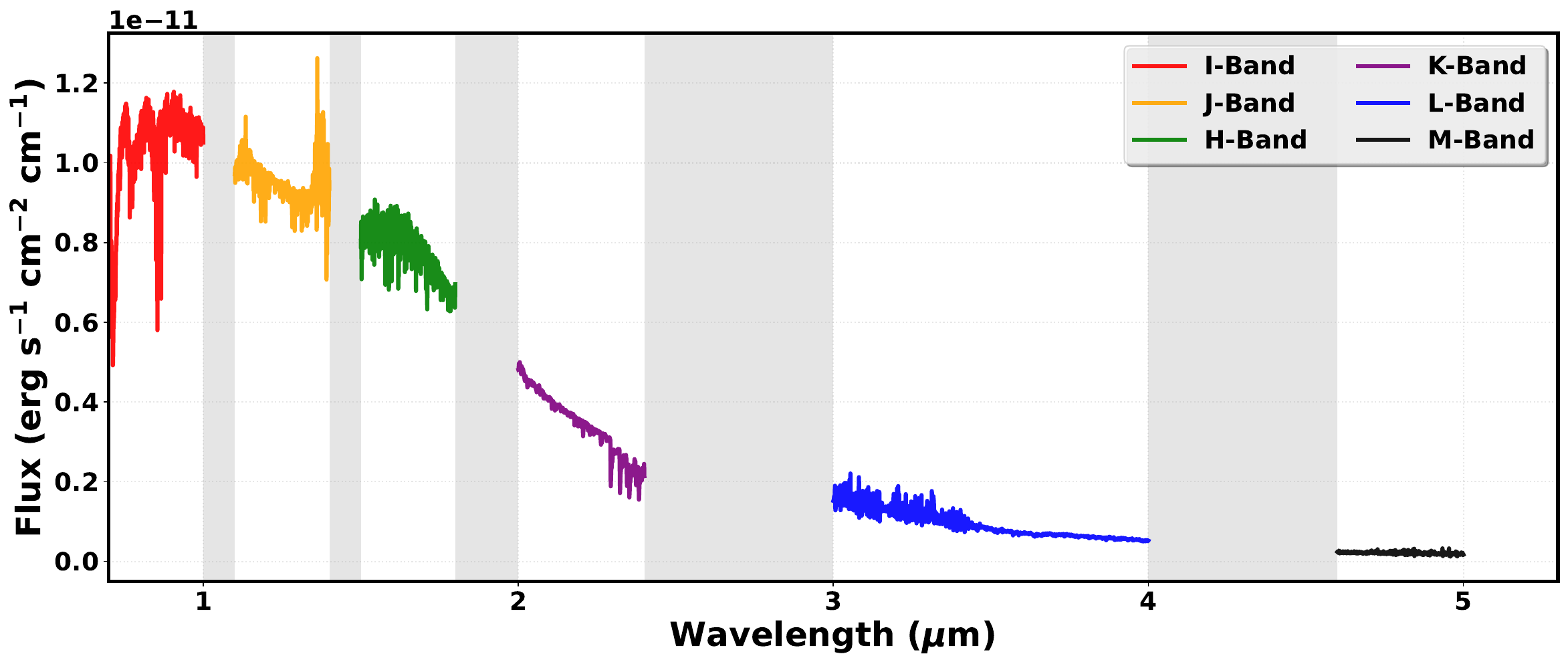}
    \includegraphics[width=0.5\textwidth]{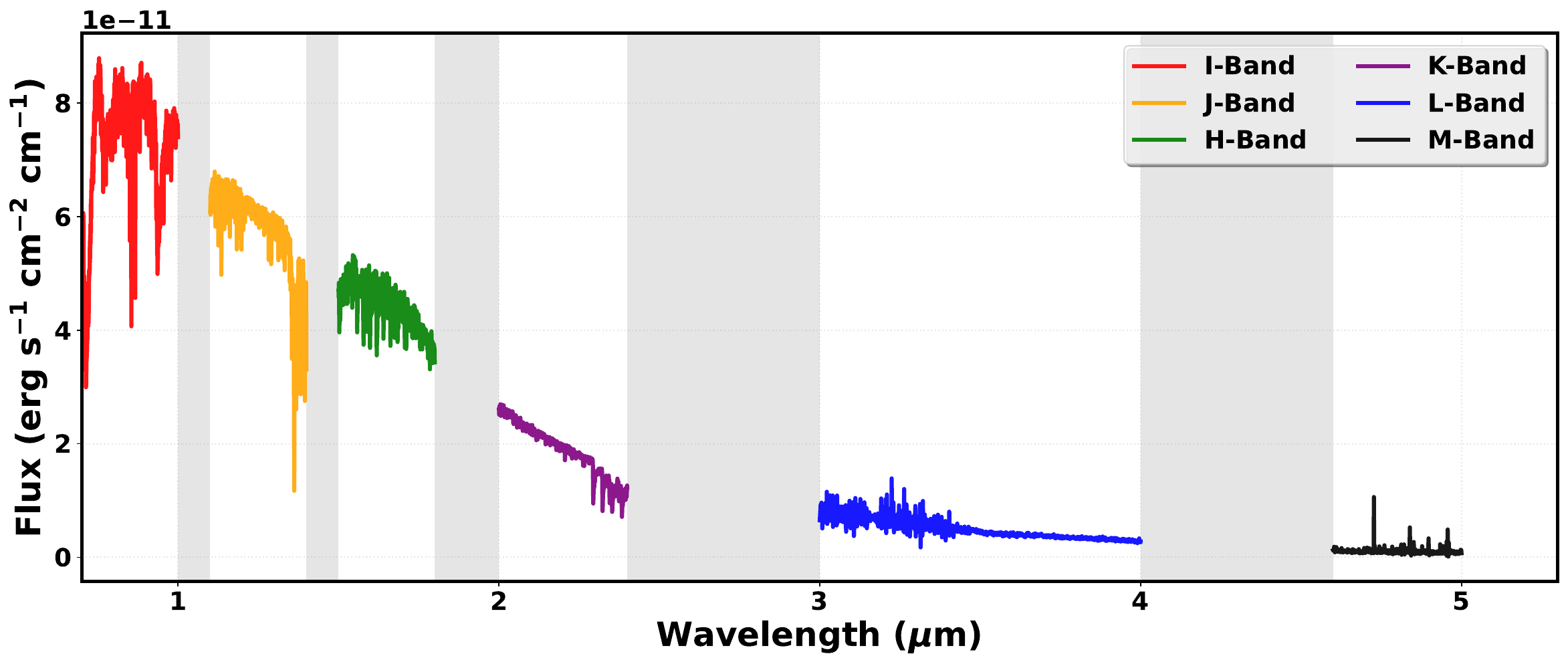}
    
    \caption{2021 November IRTF SpeX spectra of SU Lyn (top), EG And (middle), and BD Cam (bottom). Band colors and telluric regions as in Figure~\ref{fig:sept_spectra}. Two targets (EG And and BD Cam) were observed in both epochs, enabling direct comparison of spectral evolution over the $\sim$2 month baseline.}
    \label{fig:nov_spectra}
\end{figure*}

%% file: cornerstarfish.tex
\begin{figure*}[!h]
    \gridline{\fig{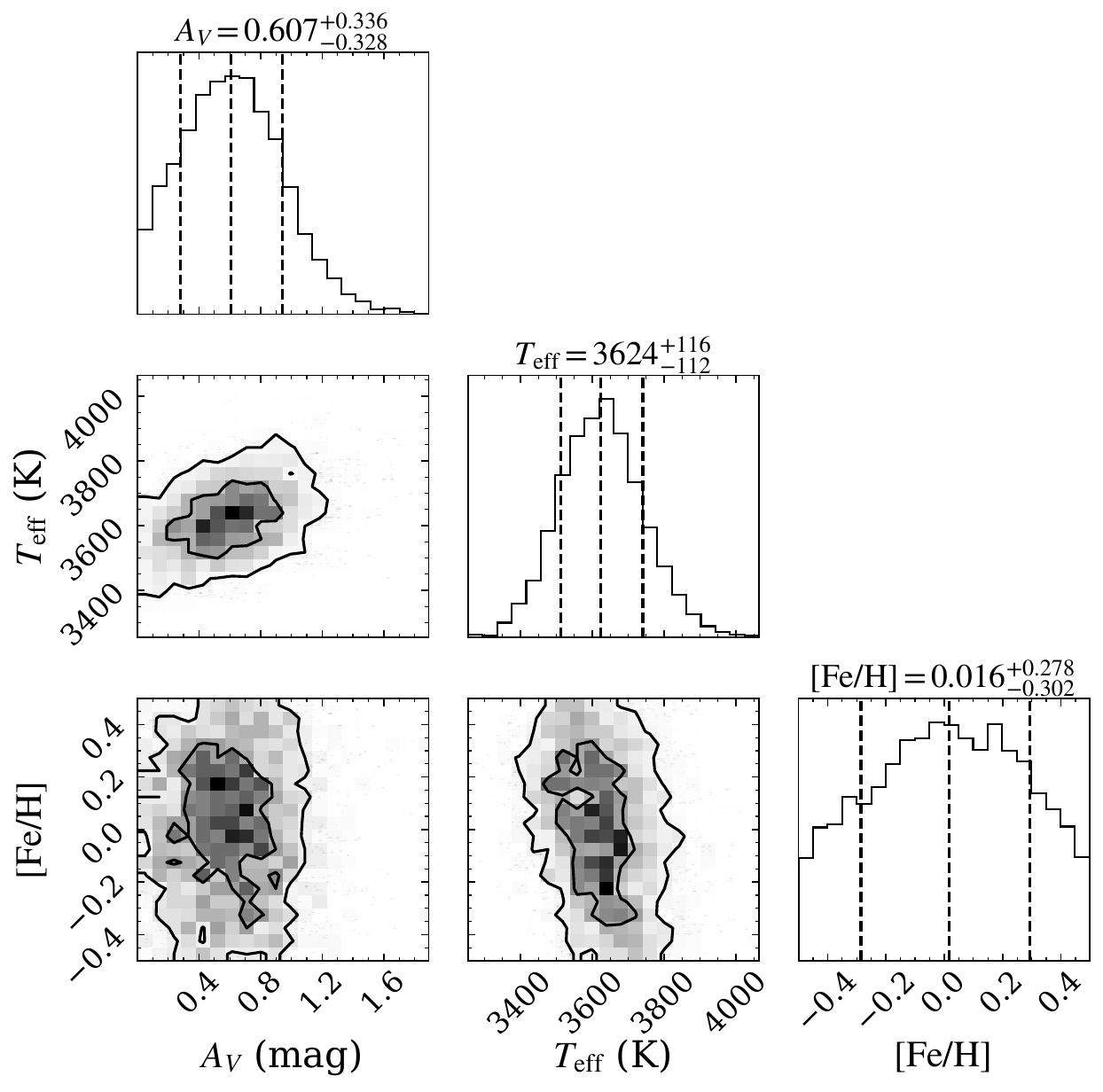}{0.45\textwidth}{(a) V1472 Aql}
              \fig{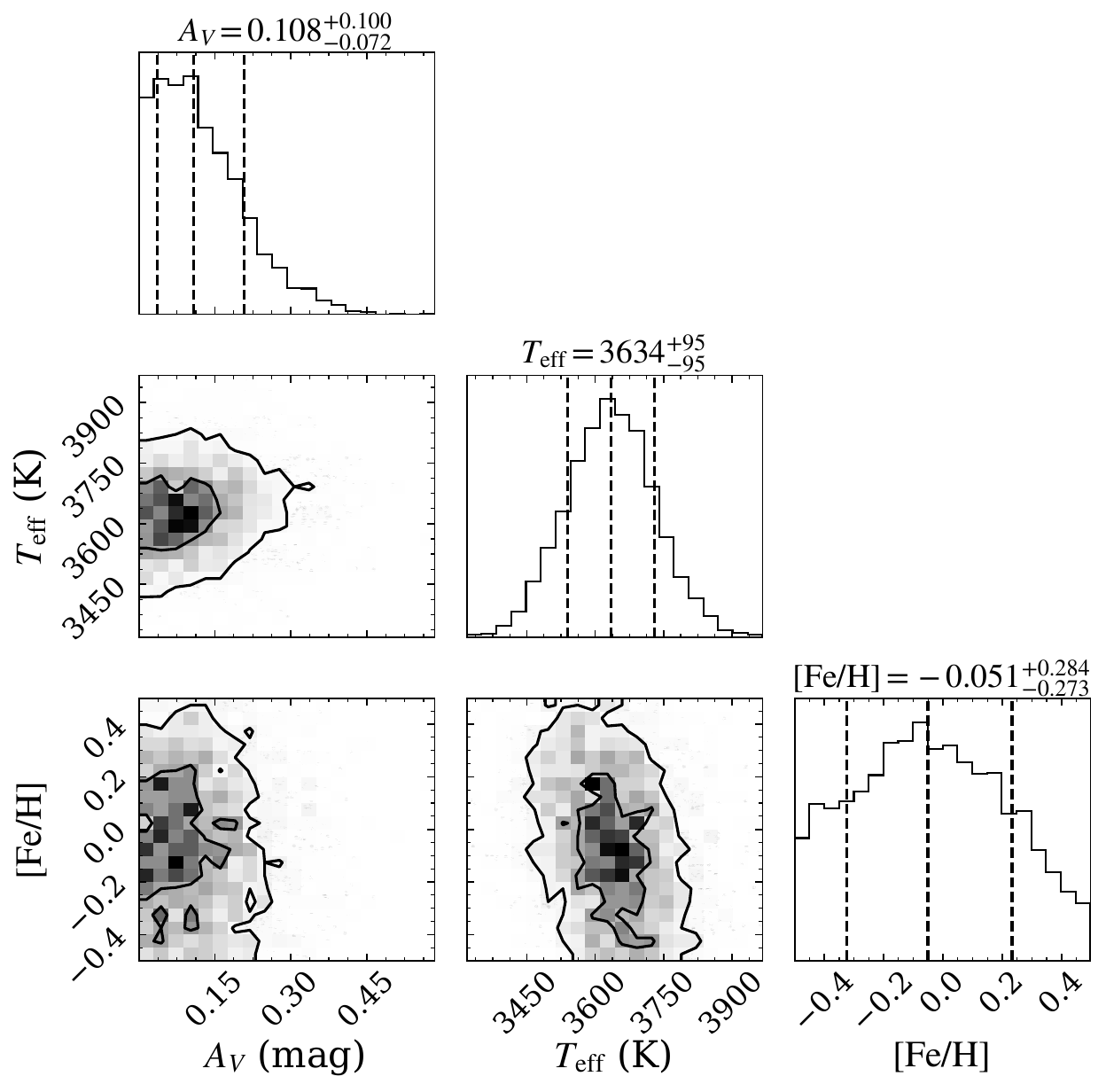}{0.45\textwidth}{(b) EG And}
             }
    \gridline{\fig{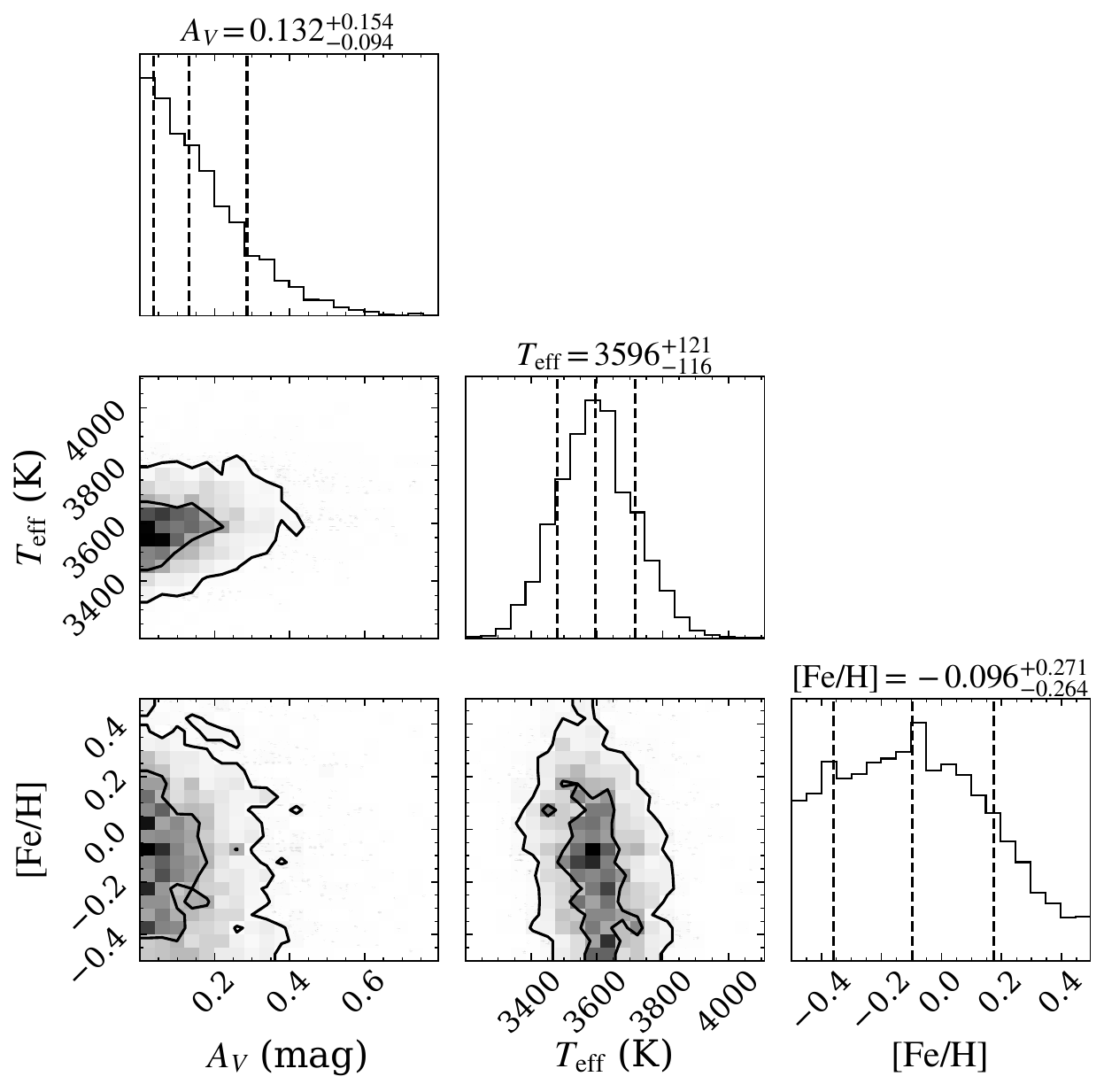}{0.45\textwidth}{(c) BD Cam}
              \fig{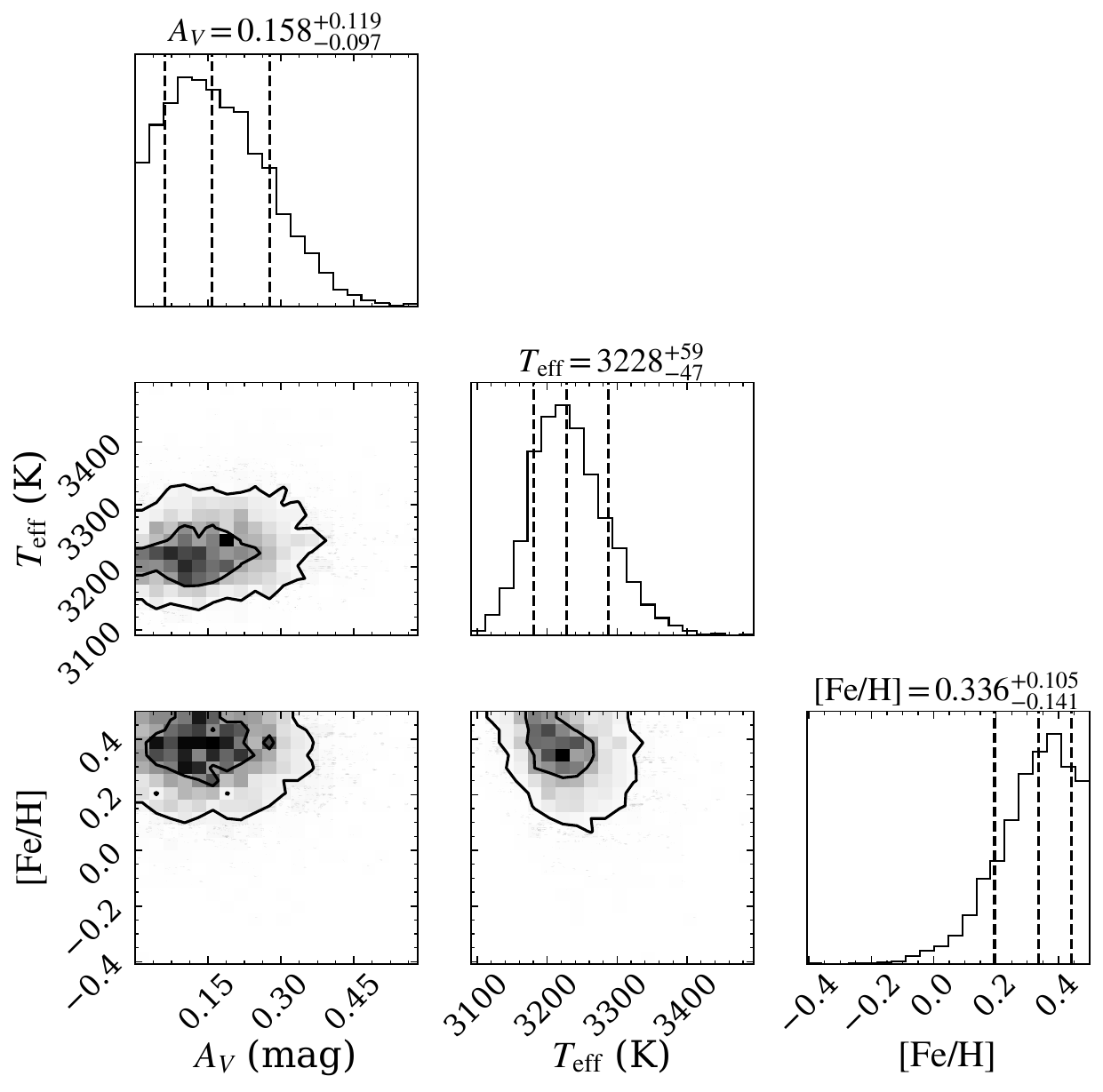}{0.45\textwidth}{(d) SU Lyn}
             }
    \caption{The corner plot for the fundamental stellar parameters of each target star as derived by running the Starfish model spectrum fitting code. Each box along the main diagonal represents the likelihood of values for interstellar extinction (Av), effective temperature (T), and metallicity (Z) as best-fitting the star's spectrum. The off-diagonal plots compare the likelihood of each parameter's value of fitting the stellar spectrum given a value of one of the other two parameters. More likely values for each parameter are colored darker.}
    \label{fig:cornerstarfish}
\end{figure*}

%% file: spectrastarfish.tex
\begin{figure*}[!h]
    \gridline{\fig{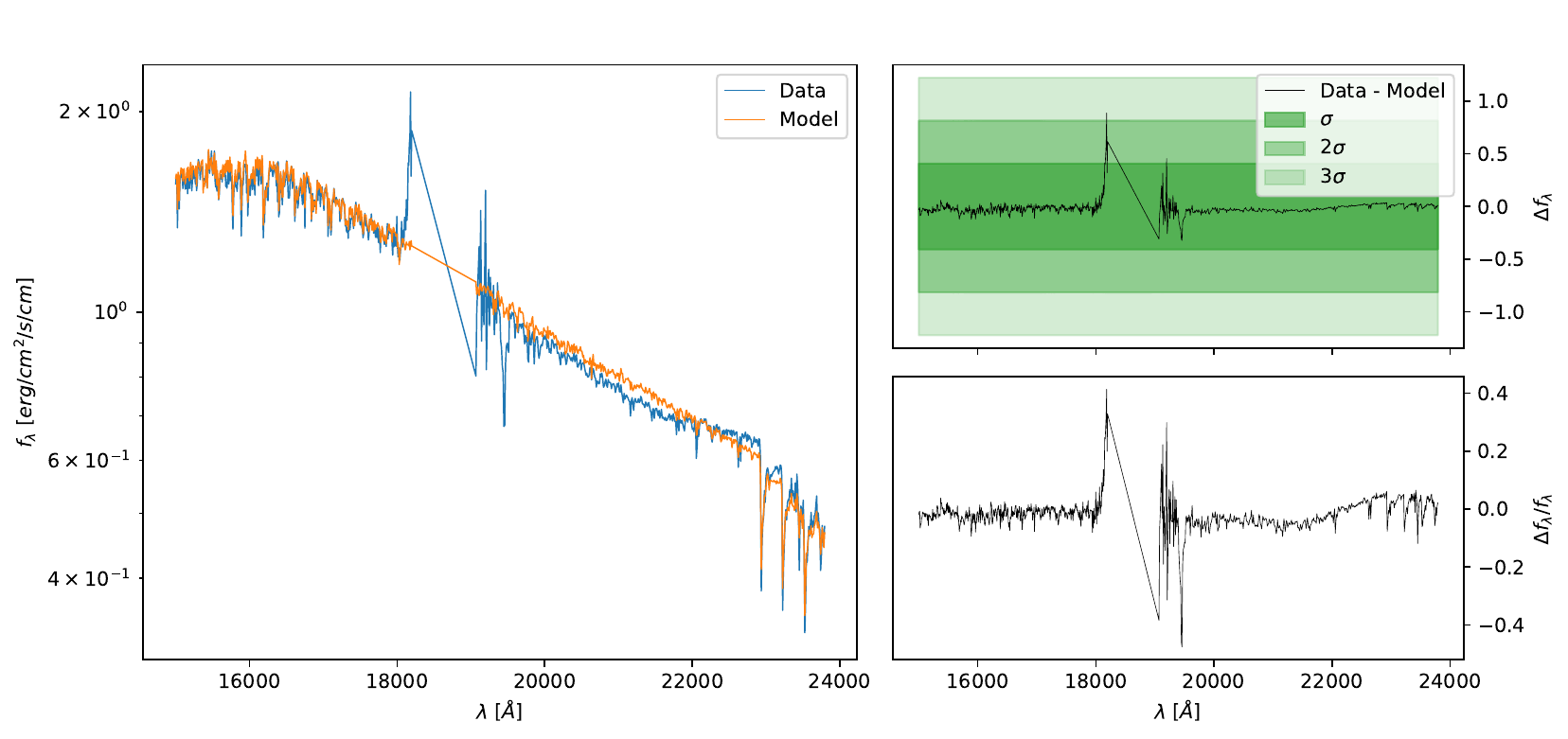}{0.5\textwidth}{(a) V1472 Aql}}
    \gridline{\fig{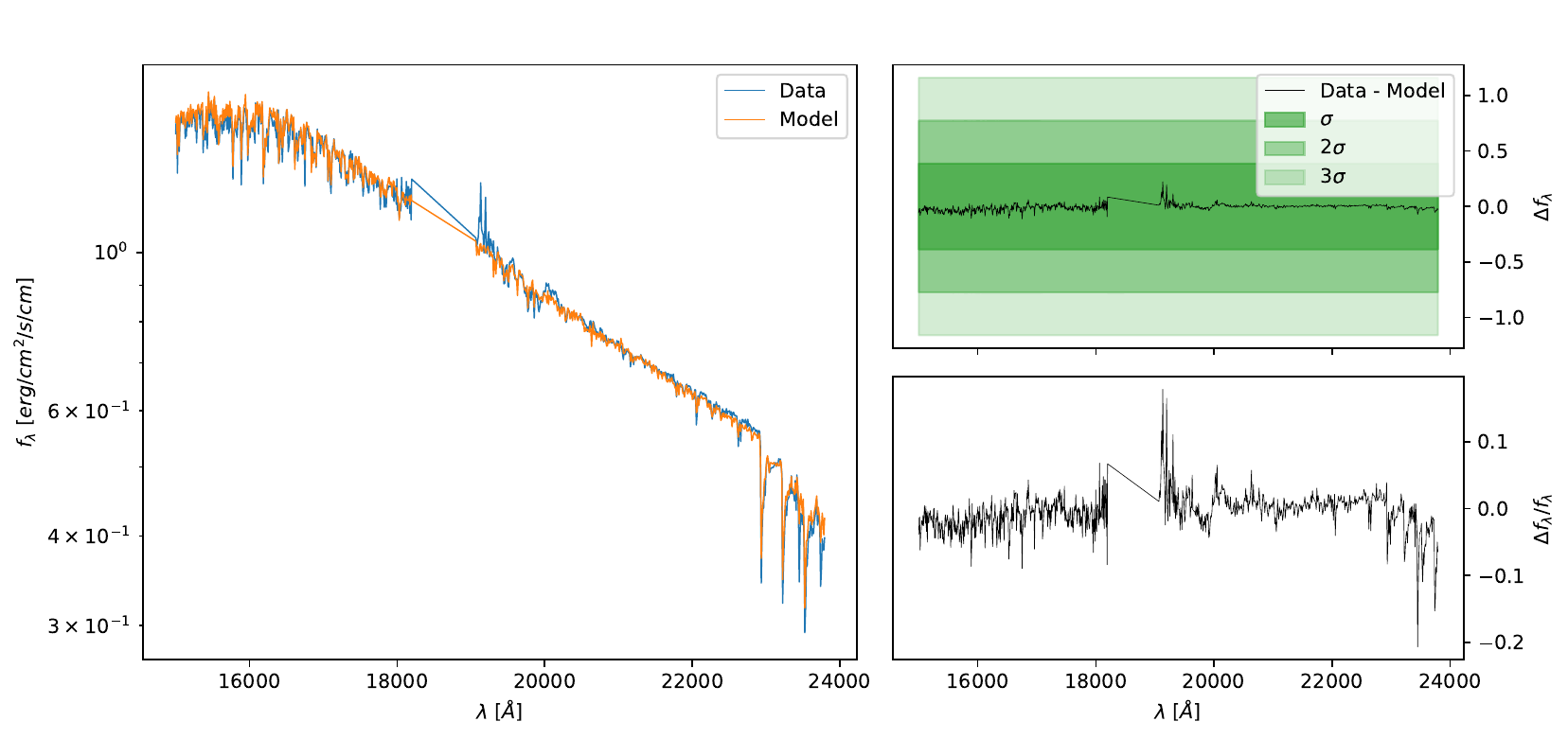}{0.5\textwidth}{(b) EG And}}
    \gridline{\fig{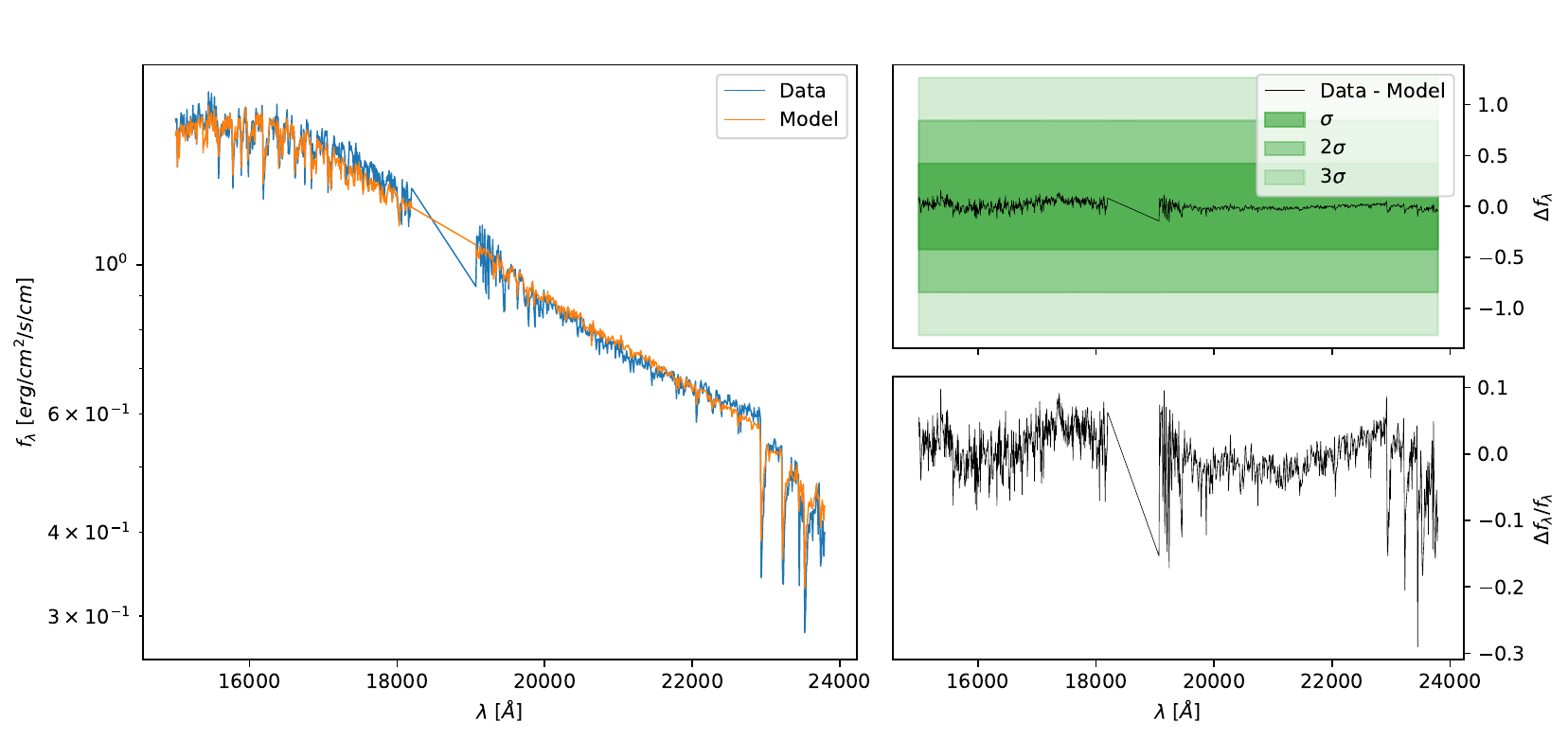}{0.5\textwidth}{(c) BD Cam}}
    \gridline{\fig{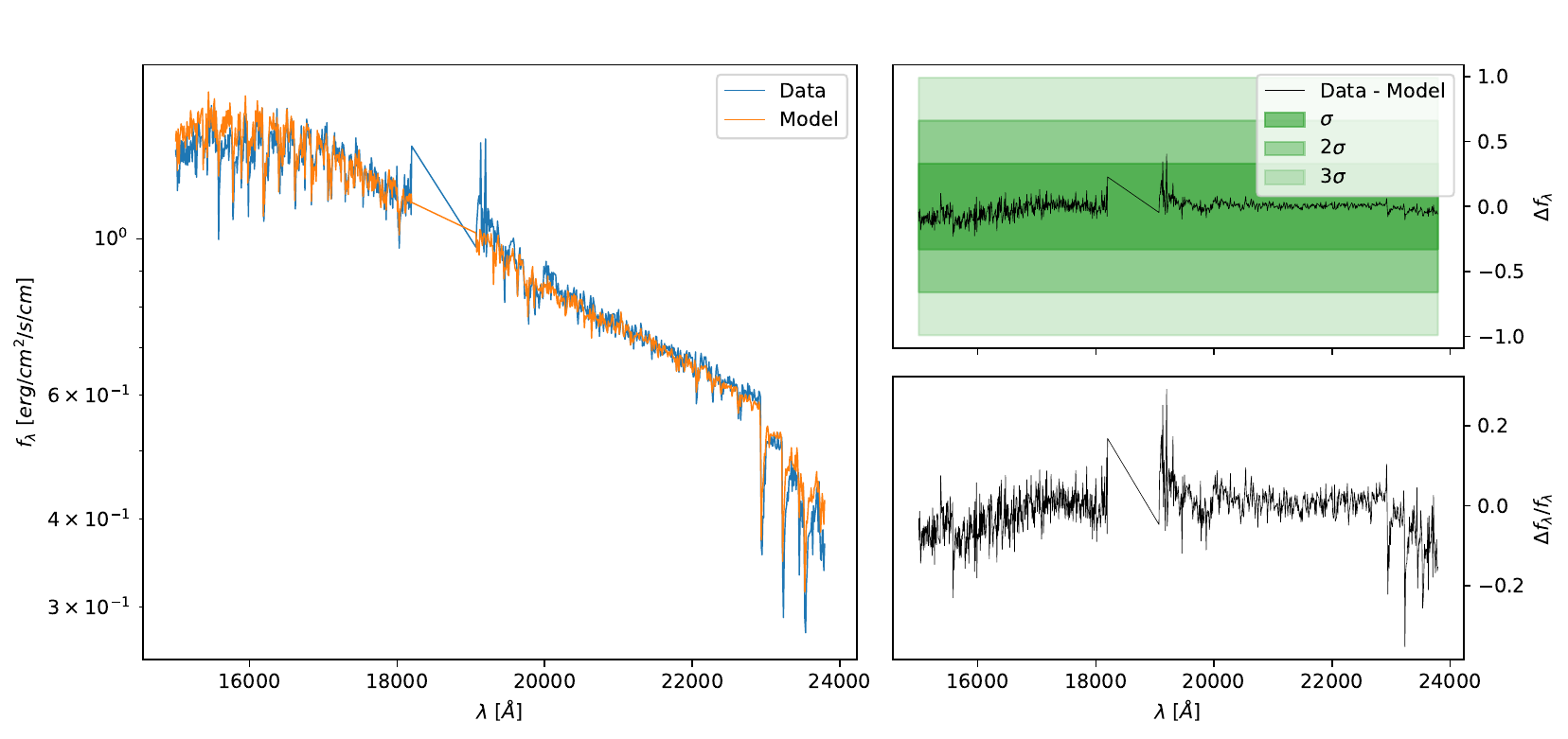}{0.5\textwidth}{(d) SU Lyn}}
    \caption{Plots of the best-fitting model Phoenix spectrum (orange) to the observed spectral data (blue) for each star in the H and K NIR wavelength bands. The top right figure shows the residuals between the two spectra with an overlay showing the 1, 2, and 3 standard deviation contours of the data. The bottom right graph shows the residuals without the standard deviation contours.}
    \label{fig:spectrastarfish}
\end{figure*}

%% file: masscorner.tex
\clearpage

\begin{figure*}[p]
    \centering
    \includegraphics[width=0.95\textwidth]{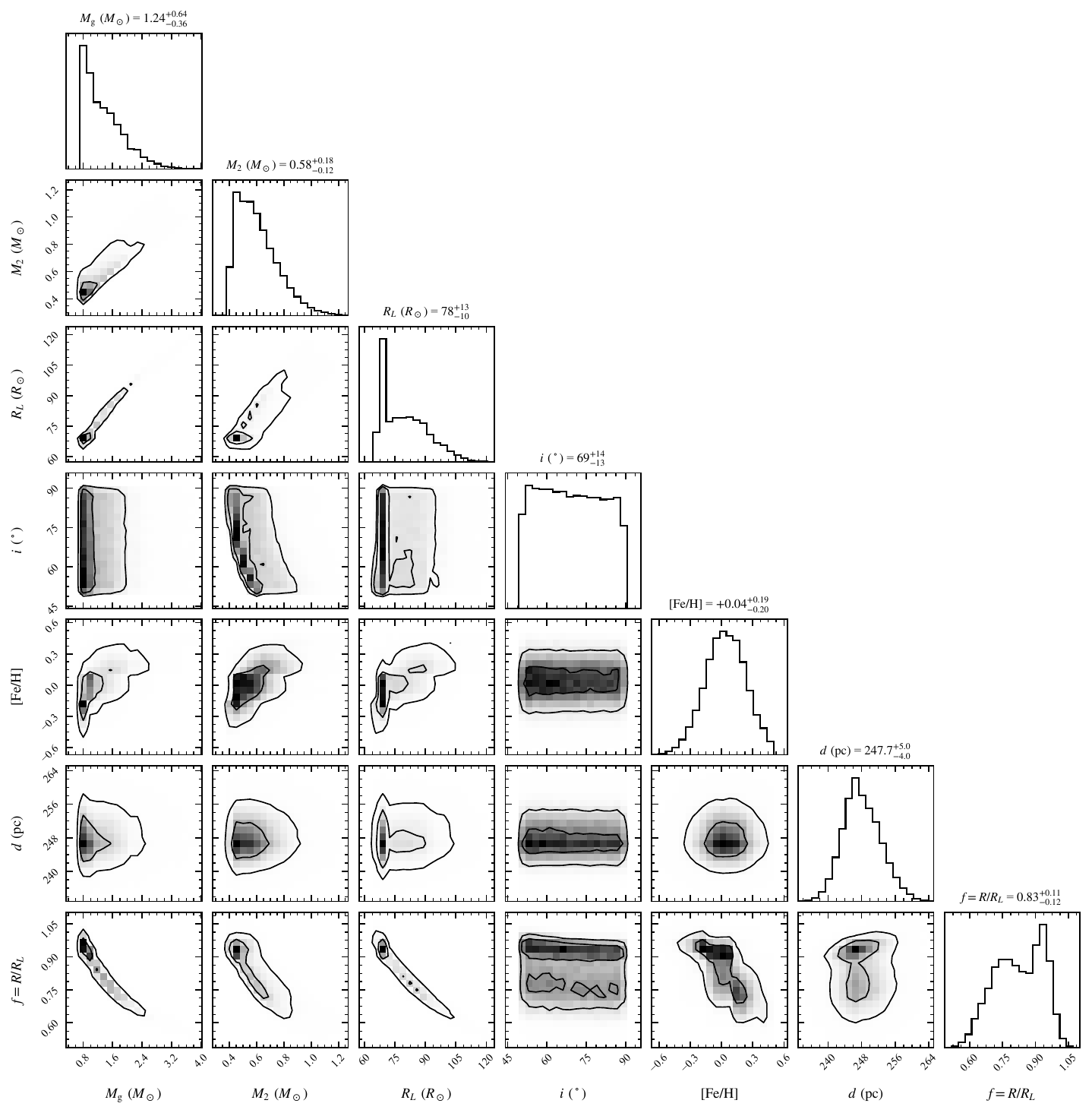}
    \caption{Joint distributions from the mass-inference procedure.
    For the three systems with orbital solutions, the plotted samples
    represent the equal-weight joint posterior after application of the
    mass-function constraint. Diagonal panels show marginalized
    one-dimensional distributions with median and 16th/84th-percentile
    intervals, while off-diagonal panels show two-dimensional covariances.
   The input priors are listed in Table~\ref{tab:priors}. (a) V1472 Aql.\label{fig:masscorner}}
\end{figure*}

\clearpage

\begin{figure*}[p]
    \addtocounter{figure}{-1} 
    \centering
    \includegraphics[width=0.95\textwidth]{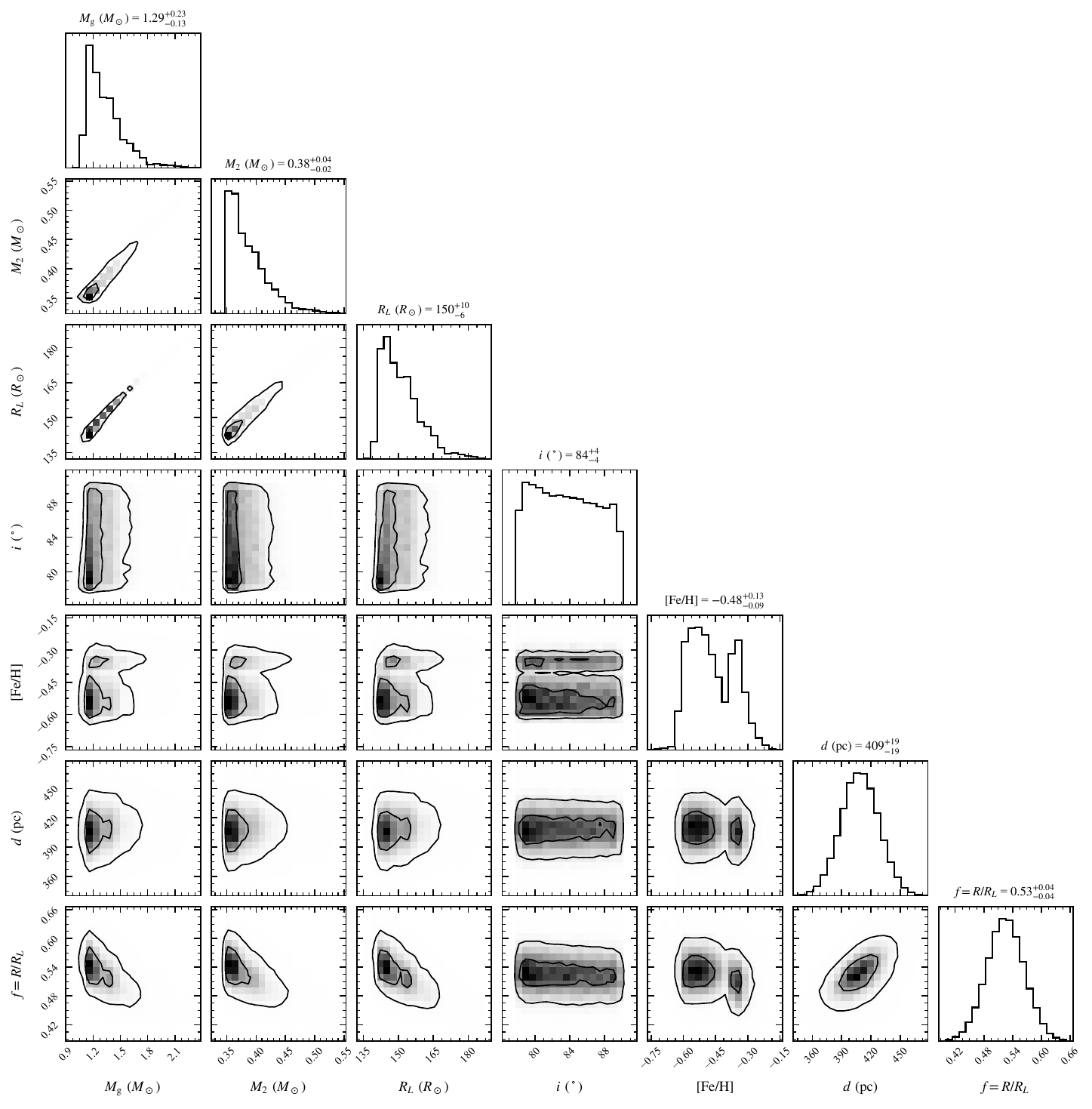}
    \caption{(Continued) (b) EG And.}
\end{figure*}

\clearpage

\begin{figure*}[p]
    \addtocounter{figure}{-1} 
    \centering
    \includegraphics[width=0.95\textwidth]{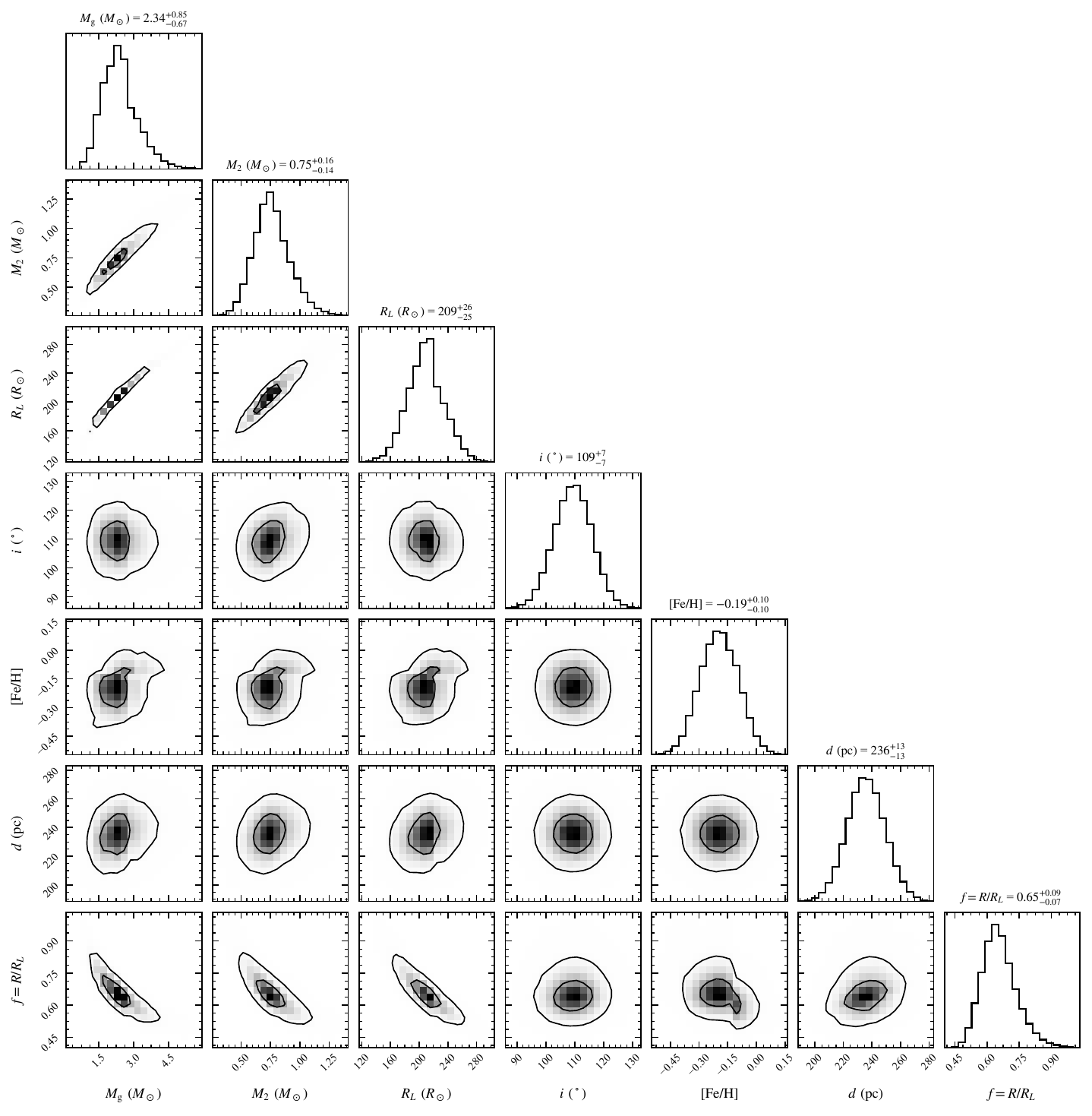}
    \caption{(Continued). (c) BD Cam.}
\end{figure*}

\clearpage

\begin{figure*}[p]
    \addtocounter{figure}{-1} 
    \centering
    \includegraphics[width=0.95\textwidth]{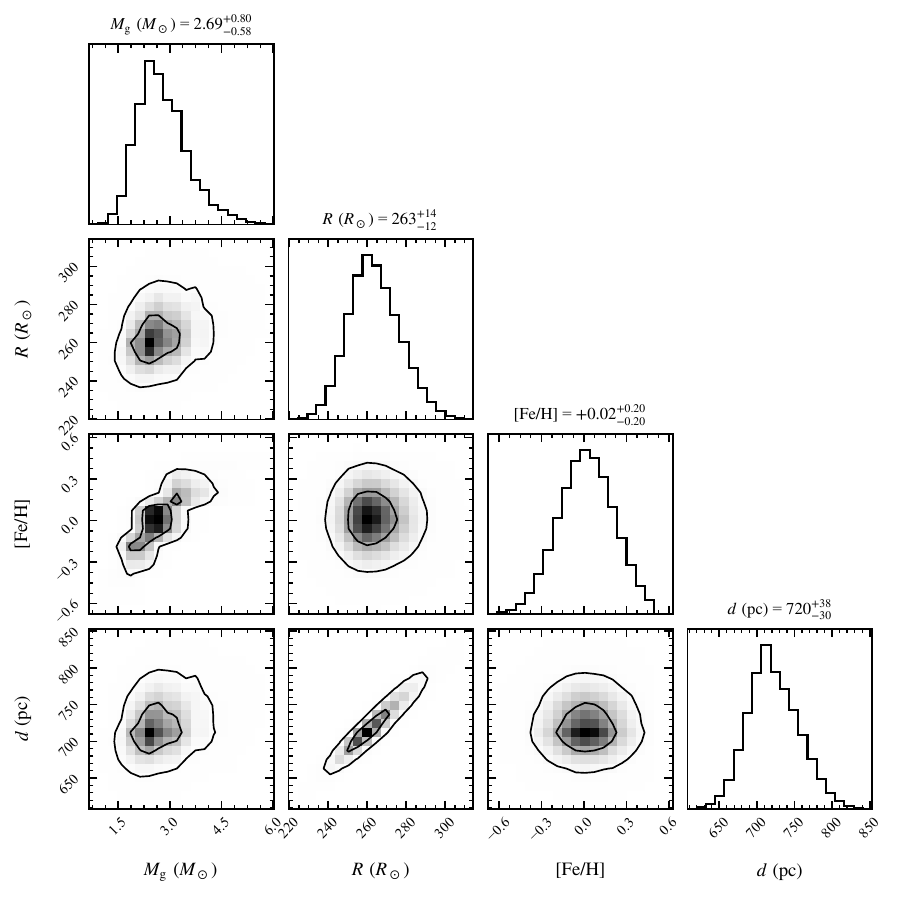}
    \caption{(Continued).  (d) SU Lyn. Note that because SU Lyn does not have an orbit, physical radius, rather than Roche-lobe radius is displayed.}
\end{figure*}

\clearpage

%% file: hrdiagrams.tex
\begin{figure*}[!h]
    \gridline{\fig{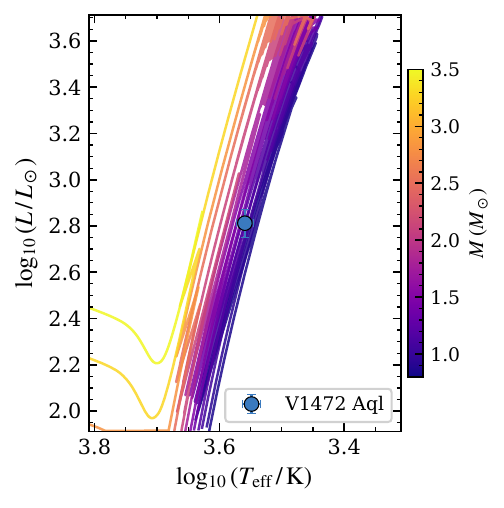}{0.45\textwidth}{(a) V1472 Aql}
              \fig{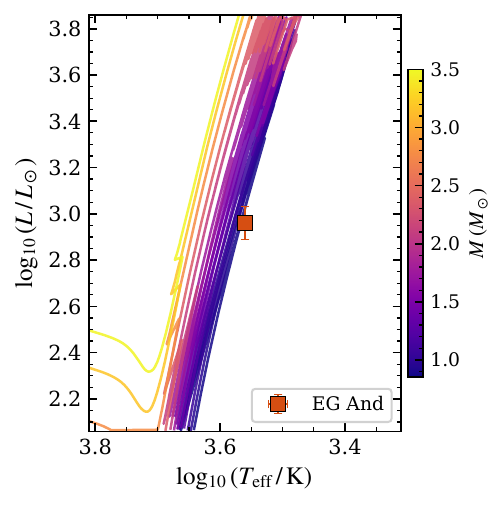}{0.45\textwidth}{(b) EG And}
             }
    \gridline{\fig{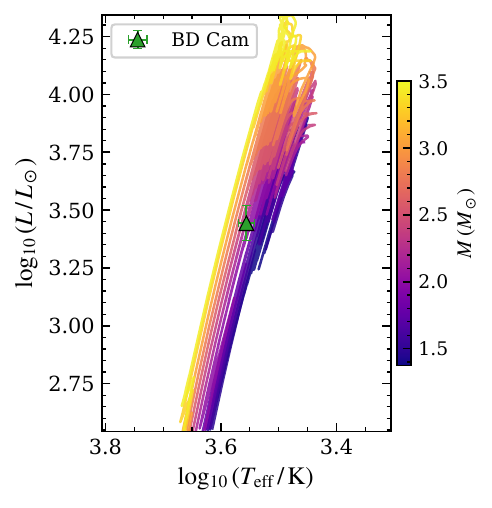}{0.45\textwidth}{(c) BD Cam}
              \fig{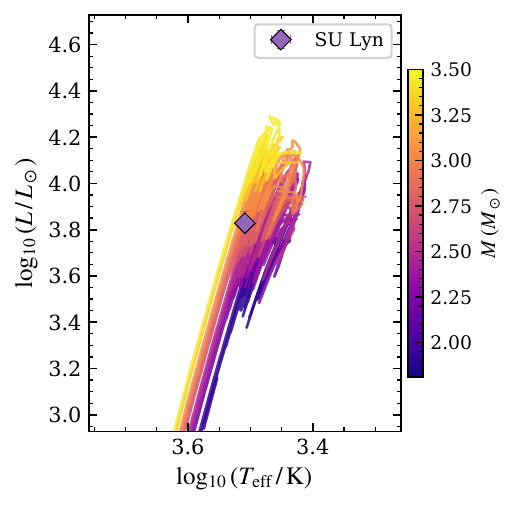}{0.45\textwidth}{(d) SU Lyn}
             }
    \caption{HR Diagrams from the fits to MIST. }
    \label{fig:hrdiag}
\end{figure*}